\documentclass[useAMS, usenatbib, a4paper]{mn2e}

\usepackage{mathrsfs}
\usepackage[space]{grffile}
\usepackage[T1]{fontenc}
\usepackage{aecompl}
\usepackage{times}
\usepackage{subfigure}
\usepackage{amsmath, amssymb}
\usepackage{amsfonts}
\usepackage{graphicx}
\usepackage{multirow}
\usepackage{hyperref}
\usepackage{url}
\usepackage{bm}
\usepackage{ulem} 
\usepackage[abs]{overpic}
\usepackage{tikz}
\usepackage{todonotes}
\usepackage{times}
\usepackage{amssymb}
\usepackage{amsmath}
\usepackage{graphicx,color}
\usepackage{bmpsize}
\usepackage{epstopdf}
\usepackage{adjustbox}
\usepackage{multicol}
\usepackage{rotate}

\renewcommand{\d}[1]{\ensuremath{\operatorname{d}\!{#1}}}
\newcommand{\vect}[1]{\mbox{\boldmath ${#1}$}}
\newcommand {\apgt} {\ {\raise-.5ex\hbox{$\buildrel>\over\sim$}}\ }
\newcommand {\aplt} {\ {\raise-.5ex\hbox{$\buildrel<\over\sim$}}\ }

\def\myputfigure#1#2#3#4#5%
{\vskip#5pt\makebox[0pt]{\hskip#2in
\includegraphics[width=#3\textwidth]{#1}}\vskip#4pt\hfill}

\newcommand{\cmfast}{\textsc{\small 21CMFAST }}

\newcommand{\ccalc}{\textsc{\small cosmocalc }}
\newcommand{\toolscm}{\textsc{\small tools21cm }}

\newcommand{\hii}{H\textsc{ii} } 
\newcommand{\xh}{x_{\mathrm{H}\textsc{i}} }
\newcommand{\Tcmb}{T_{\mathrm{cmb}}}
\newcommand{\Ts}{T_{\mathrm{s}}}

\newcommand{\Tb}{\delta T_{\mathrm{b}}}

\newcommand{\hmxb}{\textit{HMXB} }
\newcommand{\qso}{\textit{QSO} }

\title[21cm normalised bispectrum due to X-ray heating.]
{The 21cm bispectrum as a probe of non-Gaussianities due to X-ray heating.}
\author[C. A. Watkinson, S. K. Giri, H. E. Ross, K. L. Dixon, I. T. Iliev, G. Mellema, J. R. Pritchard]
{Catherine ~A.~Watkinson$^1$\thanks{Email: \href{mailto:catherine.watkinson@gmail.com}
{\protect\nolinkurl{catherine.watkinson@gmail.com}}}, Sambit ~K.~Giri$^2$,
Hannah ~E.~Ross$^{2\,,3}$, Keri L. Dixon$^{4\,,3}$, \newauthor
Ilian T. Iliev$^3$, Garrelt Mellema$^2$ and Jonathan ~R.~Pritchard$^1$\\
$^1$Department of Physics, Blackett Laboratory, Imperial College, London, SW7 2AZ, UK \\
$^2$Department of Astronomy and Oskar Klein Centre, Stockholm University, AlbaNova, SE-106 91 Stockholm, Sweden \\
$^3$Astronomy Centre, Department of Physics \& Astronomy, Pevensey III Building, University of Sussex, Falmer, Brighton, BN1 9QH, United Kingdom \\
$^4$New York University Abu Dhabi, PO Box 129188, Saadiyat Island, Abu Dhabi, UAE\\}
\date{\today}

\pubyear{\number\year}

\begin{document}
\maketitle

\begin{abstract}
We present analysis of the normalised 21-cm bispectrum from fully-numerical simulations of
intergalactic-medium heating by stellar sources and high-mass X-ray binaries (\textit{HMXB}) during the cosmic dawn.
Lyman-$\alpha$ coupling is assumed to be saturated,
we therefore probe the nature of non-Gaussianities produced by X-ray heating processes.
We find the evolution of the normalised bispectrum to be very different from that
of the power spectrum. It exhibits a turnover whose peak moves from large to small scales with decreasing redshift, and corresponds to the typical
separation of emission regions.
This characteristic scale reduces as more and more regions move into emission with time.
Ultimately, small-scale fluctuations within heated regions come to dominate the
normalised bispectrum, which at the end of the simulation is almost entirely driven
by fluctuations in the density field.
To establish how generic the qualitative evolution of the normalised bispectrum
we see in the stellar + \textit{HMXB} simulation is, we examine several other simulations -
two fully-numerical simulations that include QSO sources, and two with contrasting source properties
produced with the semi-numerical simulation \cmfast.
We find the qualitative evolution of the normalised bispectrum during X-ray heating to be
generic, unless the sources of X-rays are, as with QSOs, less numerous and so exhibit more distinct isolated heated profiles.
Assuming mitigation of foreground and instrumental effects are ultimately effective,
we find that we should be sensitive to the normalised bispectrum during the epoch of heating,
so long as the spin temperature has not saturated by $z\approx 19$.
\end{abstract}


\section{Introduction}\label{sec:intro}
One of the priorities of modern astrophysics is to try and understand the
first stars and galaxies,
as well as their subsequent evolution.
The formation of luminous sources drastically changed the properties of the Universe.
For example, radiation from such sources ionized the hydrogen and helium in the Inter-Galactic Medium (IGM),
ultimately causing the Universe to transition from being largely neutral to almost
entirely ionized.
This phase transition is generally referred to as the \textit{Epoch of Reionization (EoR)}.
Remnants of stars, such as black holes and neutron stars, will also
produce X-rays which importantly will heat the neutral IGM.
Simulations suggest that the IGM transitioned from adiabatically
cooling with the background cosmological expansion, to become universally heated.
This transition is often referred to as the \textit{Epoch of Heating (EoH)}
(\citealt{Loeb2013} provide a comprehensive overview of both the EoR and EoH).

The details of sources during the EoH are uncertain,
there is indication that dominant sources of X-rays will be high-mass X-ray binaries
(HMXBs) and
Active-Galactic Nuclei (AGN),
with the hot interstellar-medium contributing to the soft end of the X-ray spectrum
\citep{Mineo2012}.
It is not currently known how much each will ultimately contribute at high-$z$.
AGN are the dominant contributor to the X-ray budget at lower redshift,
but their abundance is seen to rapidly reduce beyond $z=3$ \citealt{Fan2001, Lehmer2016},
although, mini-quasars could still be a major contributor at high redshifts
\citep{Madau2004, Volonteri2009}.
However, it is likely that HMXBs will be the main contributor based on the fact
that in low-redshift galaxies (in the absense of AGN) they dominate the X-ray production \citep{Fabbiano2006},
and that their abundance (in contrast to AGN) is seen to increase with redshift \citep{Gilfanov2004, Mirabel2011, Mineo2012a}.
Simulations also suggest that the very first generation of Population III stars
predominantly formed in binary, or multiple systems \citep{Turk2009, Stacy2010}.

In order to establish which of these scenarios is true
(or indeed if other heating sources might have contributed),
we need observational constraints.
It is the hope that high-$z$ observations of the 21-cm line of neutral hydrogen
will provide a wealth of information about the EoH (as well as the EoR).
The CMB will interact with any neutral hydrogen in its path to us, and
by looking at fluctuations in the CMB at the frequencies associated with the 21-cm
interaction at different redshifts, we can (in principle) make 21-cm maps
and learn about the evolution in the properties of neutral hydrogen with time.

The observable for the 21-cm line is the offset of the brightness temperature\footnote{
Intensity $I_\nu$ is usually described in terms of a
brightness temperature $T_{\rm b}$, defined such that $I_\nu=B(T_{\rm b})$,
where $B(T)$ is the Planck black-body spectrum - well approximated by the
Rayleigh-Jeans formula at the frequencies
relevant to reionization studies.}
($\delta T_{\mathrm{b}}$) relative to that of the CMB ($T_{\mathrm{cmb}}$) \citep{Field1958, Field1959a, Madau1997},

\begin{equation}
\begin{split}
\delta T_{\rm b}=&\frac{T_{\rm s}-T_{\textsc{cmb}}}{1+z}(1-e^{-\tau_{\nu_0}})\,,\\
\approx&\,27\,\frac{T_{\rm s}-T_{\textsc{cmb}}}{T_{\rm s}}\,x_{\textsc{hi}}(1+\delta)\left[\frac{H(z)/(1+z)}{\d v_{\rm r}/\d r}\right]\\
&\times \left(\frac{1+z}{10}\frac{0.15}{\Omega_{\rm m}h^2}\right)^{1/2}\left(\frac{\Omega_{\rm b}h^2}{0.023}\right) \rm mK
\,.\\ \label{eqn:brightTemp}
\end{split}
\end{equation}

\noindent This depends on the cosmological parameters:
the Hubble parameter $H(z)=100\,h$, and
the matter ($\Omega_{\rm m}$)
and baryon ($\Omega_{\rm b}$) density parameters
(where $\Omega_i=\rho_i/\rho_{\rm c}$ and
$\rho_{\rm c}$ is the critical density required for
flat universe). For the analysis performed in this paper we will adopt a
$\Lambda$CDM with $\sigma_8=0.80$,
$h=0.70$, $\Omega_{\rm m}=0.27$, $\Omega_{\Lambda}=0.73$,
$\Omega_{\rm b}=0.044$ and $n_{\rm s}=0.96$.
These values are consistent with the values adopted by 
the simulations of \citealt{Ross2016} analysed in this work and WMAP 7 \citep{Komatsu2010}.
Note that unless otherwise stated the analysis in this paper is done on the mean-subtracted
brightness temperature, i.e. $\delta T_{\rm b} - \langle \delta T_{\rm b} \rangle$.

More important to our discussion here is the dependence of the brightness temperature
on density $\delta$, the neutral fraction $x_{\textsc{hi}}$ (which together measure
the amount of neutral hydrogen gas present and so provide sensitivity to the EoR),
and the spin temperature $T_{\rm s}$ (which measures the relative distributions
of electrons over the two levels associated with the 21cm transition).
Stars produce copious amounts of Lyman-$\alpha$ radiation, which is incredibly
efficient at coupling $T_{\rm s}$ to the thermal temperature of the gas $T_{\rm k}$.
Once Lyman-$\alpha$ coupling is complete, the spin temperature provides a probe of the thermal history of the Universe.
However, the spin temperature will saturate as $T_{\rm s} \gg T_{\textsc{cmb}}$
and so the brightness temperature can lose sensitivity to fluctuations in the gas temperature if it gets very high.

The first generation of 21-cm radio interferometer, such as LOFAR\footnote{The LOw Frequency ARray \url{http://www.lofar.org/}},
MWA\footnote{The Murchison Wide-field Array
\url{http://www.mwatelescope.org/}} and
PAPER\footnote{The Precision Array to Probe Epoch of Reionization
\url{http://eor.berkeley.edu/}},
have been taking data for several years now,
and we are at last starting to see these instruments place some upper-bounds
on the 21-cm power-spectrum, e.g. \citealt{Paciga2011, Dillon2013, Ali2015, Pober2015, Beardsley2016c}
and \citealt{Patil2017b}.
There is also indication from the global experiment EDGES\footnote{The Experiment to Detect the Global EoR Signature \url{http://loco.lab.asu.edu/edges/}} 
(which is a single antenna
experiment observing the mean evolution of the 21-cm signal, rather than attempting
to constrain 21-cm fluctuations across the sky) that some form of coupling followed
by heating is occurring in the redshift range $15<z<21$ \citep{Bowman2018a}.
However, the inferred cosmological signal is far more extreme than expected, and
exhibits an unexpected flat evolution over a large range of redshifts.
If true, new physics beyond our standard models is required to explain this signal \citep{Bowman2018a}.

Given then the challenging nature of the observation (strong foregrounds and ionospheric effects,
both of which are observed with a beam that changes with frequency, must be mitigated),
confirmation from an independent experiment is needed before we can be confident of the result.
\citet{Hills2018} also find that the EDGES fit requires extremely unphysical foreground and ionospheric parameters,
casting doubt on the EDGES result.
It is therefore important that we do not put all our eggs in the exotic-physics basket and continue in parallel,
as we do in this paper, to consider models consistent with our current fiducial astrophysical framework.

The current generation of radio interferometers will not be able to observe the EoH over
the EDGES redshift range (although it is still hoped that one or more may make a statistical
detection of the EoR, and MWA could in principle provide statistical constraints of the EoH at $z<16$).
It is expected that the next generation such as HERA\footnote{The Hydrogen Epoch of Reionization Array
\url{http://reionization.org/}} and the SKA\footnote{The Square Kilometre Array
\url{http://www.skatelescope.org/}} will allow us to observe the EoH.

It has been seen from simulations that the signal will be highly non-Gaussian
during both the EoH and the EoR \citep{Iliev2006, Mellema2006, Watkinson2014, Watkinson2015, Watkinson2015a, Shimabukuro2016a, Majumdar2017}.
As such, it is important that we look to statistics other than the power spectrum,
which can only fully describe a Gaussian field.
This paper studies the bispectrum, which is sensitive to non-Gaussianities in a map,
as measured from the fully numerical EoH simulations of \citet{Ross2016} and \citet{Ross2018}.
We focus on their \textit{X-ray + Stellar} simulation, as low-redshift observations
indicate that HMXBs are most likely to be the dominant X-ray source out of all the observed sources;
we will refer to this simulation as \textit{HMXB} in the remains of the paper.
We will also compare with simulations that include some level of contribution
from X-rays generated by AGN (or QSO); throughout,
we will refer to these as the \textit{HMXB + QSO} and \textit{QSO} simulations \citep{Ross2018}.

In Section \ref{sec:sims} we review the numerical N-body + ray tracing simulations
that we analyse here.
In Section \ref{sec:interp} we discuss the interpretation of the bispectrum.
In Section \ref{sec:scale} we define the \textit{normalised bispectrum}, a version of bispectrum,
which has been normalised so as to remove the amplitude component.
Note that we discuss other common normalisation options in Appendix \ref{app:normdicuss}.
In Section \ref{sec:scale} we also present our findings that the normalised bispectrum
from the \textit{HMXB} simulation exhibits a turn-over at high redshifts,
the scale associated with which corresponds  to the typical
separation of emission regions.
In Section \ref{sec:other_sims}, we will consider how consistent this qualitative
evolution of the normalised bispectrum is across other simulations.
We consider a totally different type of simulation by studying the normalised bispectrum
from the semi-numerical simulation \cmfast as well as the \textit{HMXB + QSO} and \textit{QSO} simulations.
We find that the qualitative evolution is the same for all but the \textit{QSO} simulation.
This simulation differs in that its heated profiles are more distinct, driven
by isolated sources and so imprint a second and stronger turnover corresponding to the typical size of heated regions.
In Section \ref{sec:detect} we show that if foregrounds can be mitigated, the bispectrum should be detectable over the redshift range that the simulations we consider predict the EoH occurred.
Finally, we conclude this work in Section \ref{sec:conc}.

\section{Numerical simulations of X-ray heating}\label{sec:sims} 

\subsection{N-Body simulations}

The underlying cosmic structures are obtained using a high-resolution
$N$-body simulation run with \textsc{\small CubeP$^3$M} code \citep{Harnois2013}.
The simulation follows 4000$^3$ particles in a
(244 Mpc/$h$)$^3$ volume and resolves haloes down to the Jeans mass for \hii (10$^9$ $M_\odot$)
For more details on this $N$-body simulation see \citet{Dixon2016}.

\subsection{Sources}\label{sec:sources}

Our sources always form in dark matter haloes.
Haloes above the Jeans mass for \hii ($\sim 10^9\,M_\odot\le M$ ) are resolved, so
we identify these directly from the N-body simulation.
In addition, haloes with masses below this but greater than the minimum mass
at which atomic line cooling of primordial gas is efficient
(10$^8\,M_\odot$\textless $M$ \textless 10$^9\,M_\odot$) are added using a
subgrid model \citep{Ahn2015a}. Source models are summarized below,

\textbf{Stellar sources}: Stellar sources are assumed to form within dark-matter
haloes with luminosities proportional to their host halo's mass, and have a black body spectra
of 50,000 K, similar to that of O and B stars.
These softer sources do not contribute to heating, so are only important for correctly including ionizations.

\textbf{HMXBs}: As they consist of binaries of stars and stellar remnants,
HMXBs exist in stellar populations. Hence, these sources trace dark-matter distribution,
with their luminosities proportional to the host halo's mass. For more details on the implementation of these sources see \citet{Ross2016}. 

\textbf{QSOs}: We assume that QSOs are much rarer sources that have varying luminosities
uncorrelated with the mass of their host haloes.
We assign QSOs randomly to haloes with $M>10^9\,M_\odot$. The number of QSOs and their luminosities are calculated by using an
extrapolation of the low-redshift luminosity function from \citet{Ueda2014}, but with a shallower co-moving density evolution. In doing so, we assume more QSOs than \citet{Ueda2014}, motivated by the uncertainty surrounding high-redshift QSO populations and for maximal effect. \citep[e.g.][]{Giallongo2015, Parsa2018}
To mimic the variability of observed QSOs these sources are assigned a new
luminosity every 11.5 Myrs.
QSOs live in a given halo for 34.5 Myrs, which is consistent with current estimates \citep[e.g.][]{Borisova2016,Khrykin2017}. 
The simulations analysed here use a spectral index of $-0.8$ and do not include any UV contribution. For more details on these simulations see \citet{Ross2018}.

\subsection{Radiative-Transfer}

The Radiative-Transfer (RT) is calculated using \textsc{\small C$^2$-Ray} code \citep{Mellema2006}
which was updated to accommodate multi-frequency RT in order to correctly model
the effects of hard radiation \citep{Friedrich2012}.
Three such simulations are analysed in this work:
one with both stellar and HMXB sources (\textit{HMXB});
another with stellar, HMXB, and QSO sources (\textit{HMXB+QSO});
and one with stellar and QSO sources (\textit{QSO}).
The stellar component and underlying cosmic structures are identical in all simulations.
The density is smoothed onto an RT grid of size $250^3$.

\hii regions can be unresolved in our simulations, particularly for individual weak sources.
These will appear as partially ionized cells in the simulation,
with a kinetic temperature that is averaged between the hot,
ionized gas phase and the colder, neutral one.
Using the average $T_\mathrm{k}$ of these cells yields a $\delta T_\mathrm{b}$
higher than the true value.
Such cells require special treatment for calculating the correct $\delta T_\mathrm{b}$
as discussed in \citet{Ross2016} and \citet{Ross2018}.

\section{Interpretting the 21-cm Bispectrum}\label{sec:interp}
The bispectrum is defined as,
\begin{equation}
\begin{split}
(2\pi)^3 B(\vect{k}_1, \vect{k}_2, \vect{k}_3) \delta^{\mathrm{D}}(\vect{k}_1 + \vect{k}_2 +  \vect{k}_3  )=
\langle \Delta(\vect{k}_1)\Delta(\vect{k}_2)\Delta(\vect{k}_3)\rangle \,,\\\label{eq:bi_def}
\end{split}
\end{equation}
and is the Fourier pair to the three-point correlation function, which measures
excess probability as a function of three points in real space.

\begin{figure}
\centering
  $\renewcommand{\arraystretch}{-0.75}
  \begin{array}{c}
    \includegraphics[trim=0.9cm 4.2cm 1.75cm 4cm, clip=true, scale=0.4]{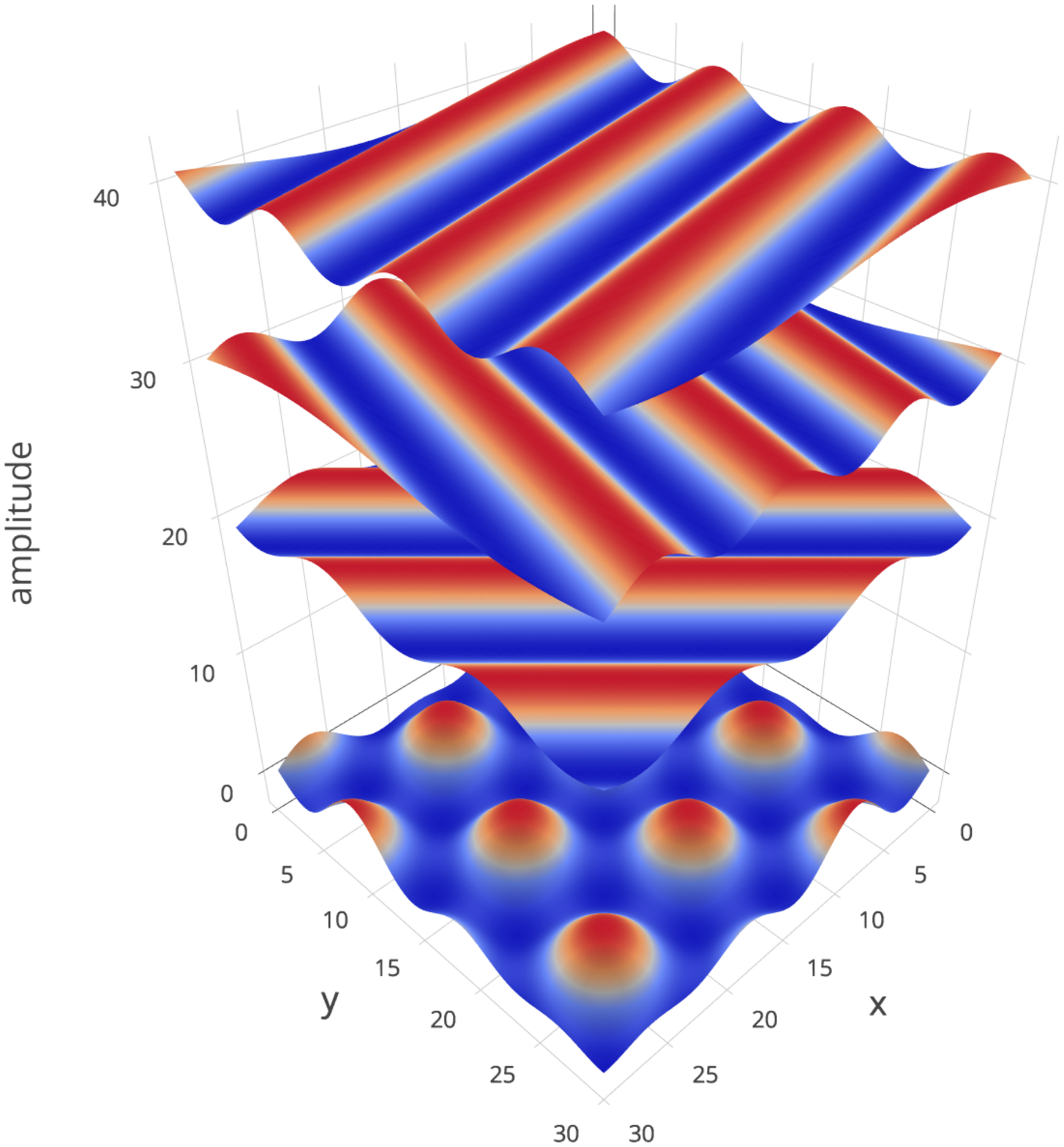}\\
  \end{array}$
  \caption{Visualisation of (from top to bottom) three different 2D waveforms in real space (whose amplitude provides a third dimension).
  These correspond to a set of three $k$-vectors (each with $|\boldsymbol{k}|=0.5$ Mpc$^{-1}$) that form an equilateral triangle.
  Note the amplitude of these three waves is offset by 40, 30, and 20 (from top to bottom).
  The interference pattern of these three waveforms is plotted at the bottom.
  Such a combination of modes produces a regular series of circularly-symmetric
  above-average concentrations of signal separated by less concentrated below-average
  regions of signal.
  For 3D waveforms, these condensed above-average regions of signal will be long filaments with
  a circular cross section.
  }  \label{fig:cartoon}
\end{figure}

\begin{figure}
\centering
  $\renewcommand{\arraystretch}{-0.75}
  \begin{array}{c}
    \includegraphics[trim=0.0cm 0.0cm 0.0cm 0.0cm, clip=true, scale=0.5 ]{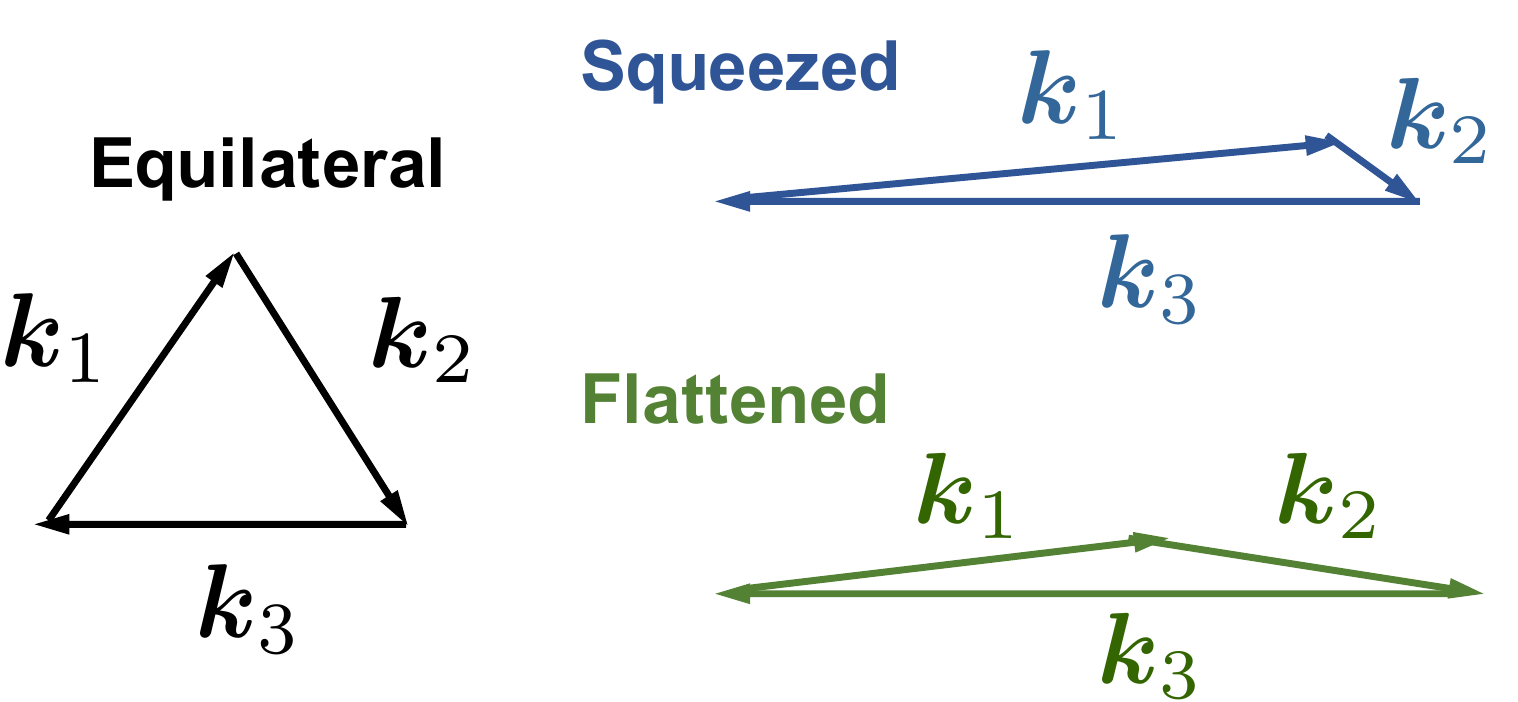}\\
    \includegraphics[trim=0.0cm 0.0cm 0.0cm -0.5cm, clip=true, scale=0.5 ]{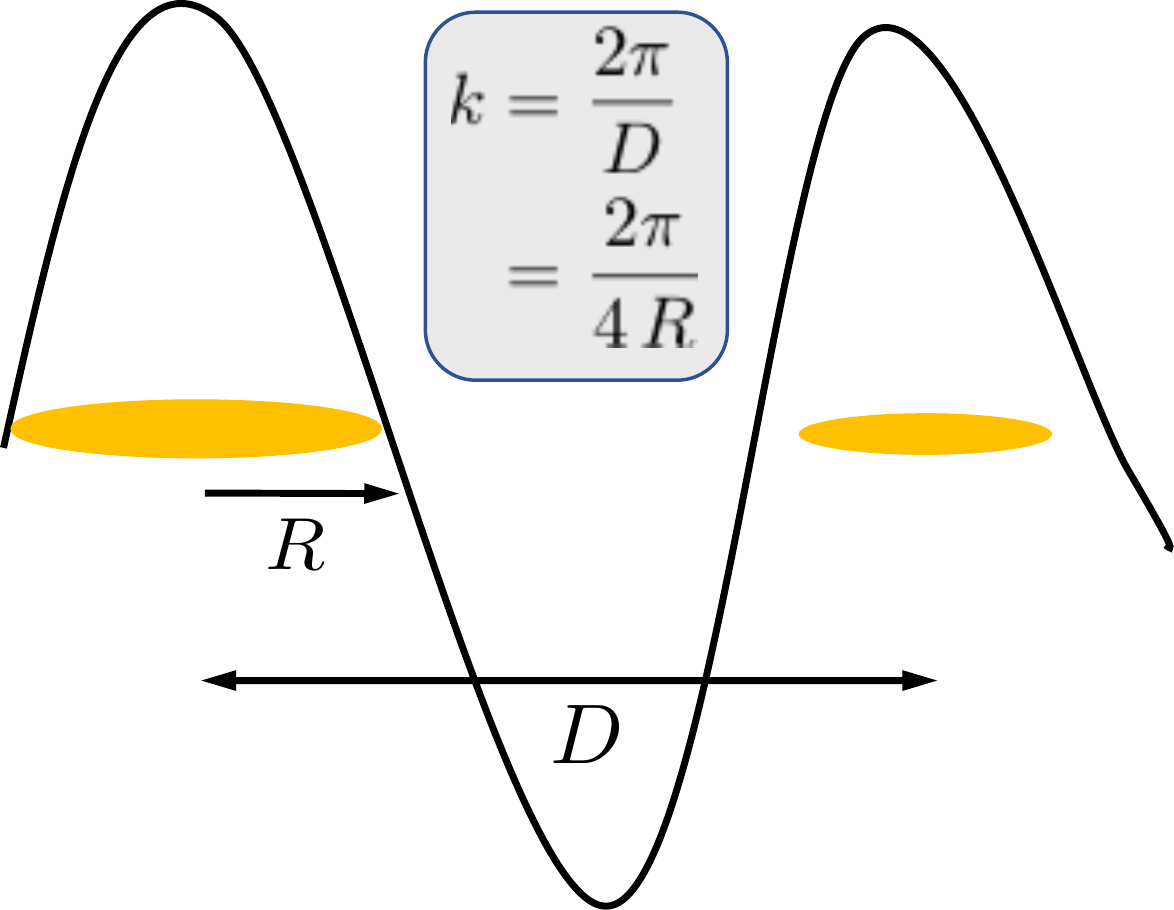}\\
  \end{array}$
  \caption{\textbf{Top} - Visualisation of three extremes of triangle configuration that may be considered
  when measuring the bispectrum.
  \textbf{Bottom} - Illustration of how the radius ($R$) of features and the clustering
  properties (via their separation $D$) correspond to wavenumber $k$.
  }  \label{fig:cartoon2}
\end{figure}

When we calculate the bispectrum, we are probing the degree to which
structure in our real map, is coherent with the three waves
defined by the three $k$-vectors ($\vect{k}_1, \vect{k}_2, \vect{k}_3$) that form a closed triangle in Eqn.~\ref{eq:bi_def}.
Fig.~\ref{fig:cartoon} shows a real-space plot of (from top to bottom)
three 2D waves associated with an equilateral configuration;
i.e. with three different
$\vect{k}_i$ forming a closed triangle, each with $|\vect{k}_i|=0.5$ Mpc$^{-1}$
(see the left black triangle illustrated in the top
panel of Fig.~\ref{fig:cartoon2}).
For the purposes of visual clarity, each waveform's amplitude is offset in the $z$-axis
relative to their true mean of zero.
At the very bottom we show their interference pattern,
i.e. what kind of structure they combine to form in real-space.
In other words, the top three waveforms are the Fourier components of the bottom wave pattern, or dataset.
This equilateral wave combination creates above-average spherically-symmetric concentrations of signal in 2D,
of radius roughly corresponding to $\pi/(2\,|\vect{k}|)$ (see the bottom panel of
Fig.~\ref{fig:cartoon2}).
In 3D these concentrations of signal extend into filaments with a circular cross section.

\citet{Lewis2011} provides a really nice discussion of what certain bispectrum
configurations correspond to in real-space.
As well as the equilateral configuration,
\citet{Lewis2011} consider the flattened and squeezed limits.
Flattened triangles have a large angle between $k_1$ and $k_2$, so that at the
most extreme angles $k_3 \sim k_1 + k_2$, i.e. $k_3$ is much larger than
$k_1$ and $k_2$ (see the bottom-right green triangle illustrated in the top
panel of Fig.~\ref{fig:cartoon2}).
This is somewhat similar to the equilateral in that the combination of such modes form a resultant signal that is concentrated along
filaments in 3D; however, for the flattened configuration these filaments have an ellipsoidal cross section, rather than circular
as for the equilateral configuration.
For very large angles these filaments tend towards planes.
At the other extreme, squeezed triangles have a very small angle between $k_1$ and $k_2$,
so that $k_3$ is very small in comparison (see the top-right blue triangle illustrated in the top
panel of Fig.~\ref{fig:cartoon2}).
This combination results in a modulation of the larger-scale mode on the
smaller scale modes, see \citet{Lewis2011} for an illustration of this type of configuration.

When we calculate the bispectrum we first FFT our dataset,
in doing so we essentially convolve three such waves with our data and average the combination
to produce the three different $\delta(\vect{k}_i)$ corresponding to whatever triangle configuration we are probing.
We then multiply these three $\delta(\vect{k}_i)$ together to get our bispectrum estimate.
The bispectrum is thus sensitive to whether structure in the data is in or out of phase with the three Fourier waves associated with the FFTs. 
The sign of the bispectrum is therefore sensitive to whether the data contains above or below-average concentrations of signal.
A positive bispectrum tells us there are concentrations of above-average signal surrounded by below-average regions.
A negative bispectrum tells us that there are concentrations of below-average
regions of signal surrounded by above-average regions of signal \citep{Lewis2011}.

A real 21-cm map is unlikely to exhibit such distinct structures as discussed above,
instead the topology of the map will result in a non-zero bispectrum for a range of triangle shapes,
with its sign depending on whether the bispectrum is driven by above or below-average concentrations of signal.
It will be the relative amplitudes of the bispectrum between different triangle configurations
that will provide some information as to the nature of structure within the dataset.
For example, the bispectrum will have greatest amplitude for the equilateral
configuration on a given scale if,
\begin{enumerate}
\item the signal is concentrated in clumps that follow the filaments of the
equilateral interference pattern to some degree, and/or;
\item the distribution is such that the signal filaments are also separated
by a similar scale to the filaments in the equilateral interference pattern.
Like the separation $D$ of the two yellow ellipses in the bottom plot of Fig.~\ref{fig:cartoon2} for which $k=2\pi/D$;
\item the bispectrum for the equilateral configuration at a given scale will be further boosted if
signal is concentrated in clumps of similar shape and size to the circular cross-section of the filaments
corresponding to the equilateral interference pattern.
Like the yellow ellipse in the bottom plot of Fig.~\ref{fig:cartoon2}
for which $k=2\pi/(4\,R)$.
\end{enumerate}

The bispectrum will be a more noisy statistic to measure
than the power spectrum (as we will see later for Gaussian noise the bispectrum
covariance is connected to the triple product of the noise power spectrum, see also \citealt{Yoshiura2015}),
and is also challenging to visualise (given that it is
a function of two $\vect{k}$ vectors rather than just one).
We therefore restrict our analysis to the spherically-averaged bispectrum in the discussion that follows.
Whenever the bispectrum is measured from gridded data,
a binwidth of at least one pixel must be allowed on each triangle side.
Therefore we never probe the bispectrum of a perfectly defined triangle; we instead measure the average
bispectrum for a selection of different (but very similar) triangles.
We choose to further bin the bispectrum in order to reduce sampling noise in the statistic.
For all equilateral configurations we bin over cos$(\theta)\pm 0.05$, where
$\theta$ is the angle in radians between $k_1$ and $k_2$.
As well as the equilateral configuration, we consider configurations where $k_2 = N\,k_1$
(for which we restrict ourselves to integer factors of $N$).
The bispectra for these configurations are presented as a function of $\theta/\pi$ radians and are binned
over $\theta\pm 0.1$ radians.
For both binning choices we have checked that this binning choice produces a
bispectrum consistent with the unbinned calculations.

We use the FFT bispectrum algorithm described in \citet{Watkinson2017} to
measure the bispectrum, this provides a very fast way to measure the bispectrum
(for 250 pixels per side our code takes $<2$s per binned triangle configuration
on a Macbook pro with 2.9 GHz Intel Core i5 using 16 threads).
We refer the reader to \citet{Watkinson2017} and references therein for
details of this algorithm.

\section{The bispectrum due to X-ray heating as driven by HMXBs}\label{sec:scale}

\begin{figure}
\centering
  \begin{tikzpicture}
    \node[anchor=south west,inner sep=0] at (0,0) {$\renewcommand{\arraystretch}{-0.75}
  \begin{array}{c}
    \includegraphics[trim=1.35cm 3.1cm 0.7cm 2.0cm, clip=true, scale=0.245]{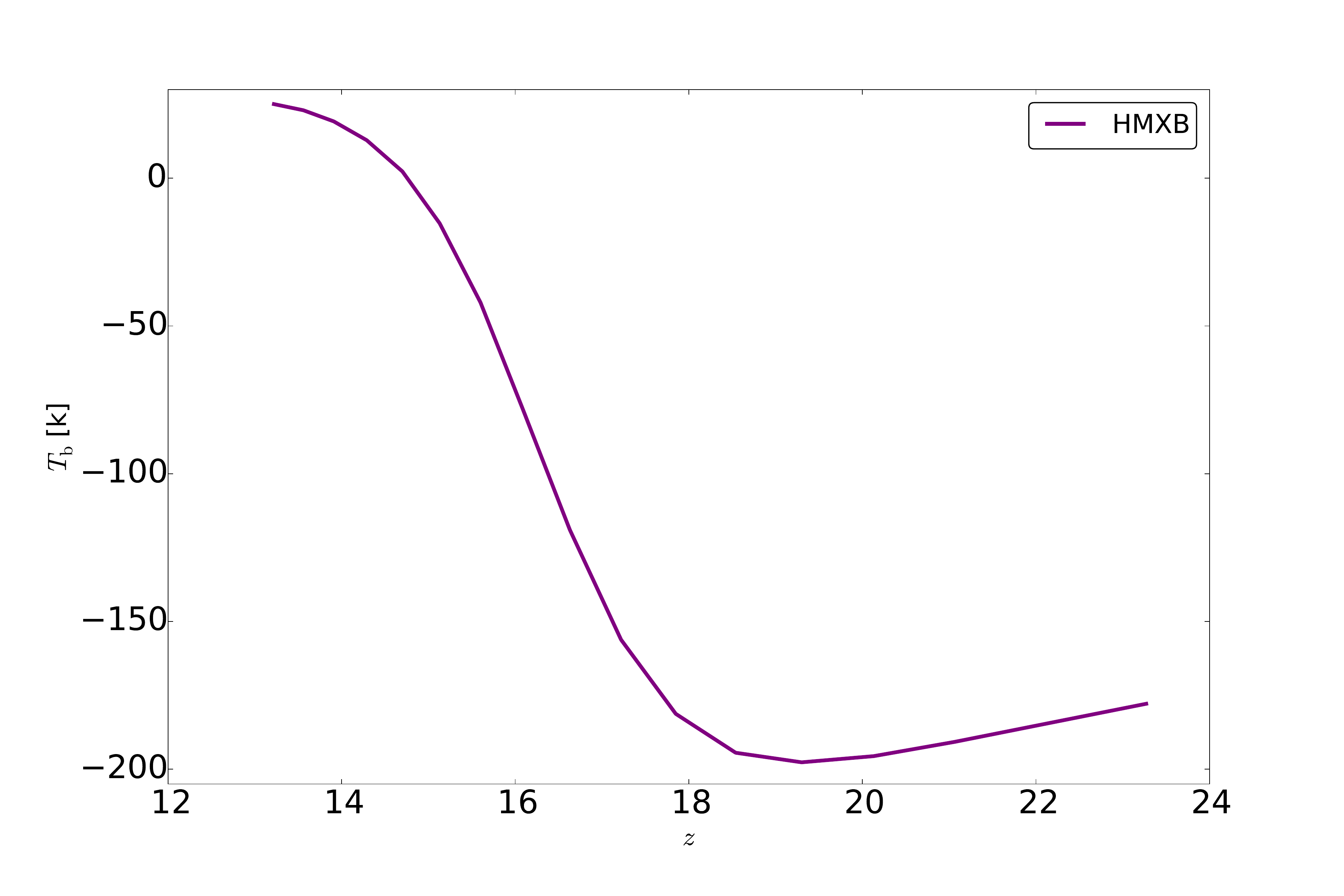}\\
    \includegraphics[trim=1.35cm 3.1cm 0.7cm 2.0cm, clip=true, scale=0.245]{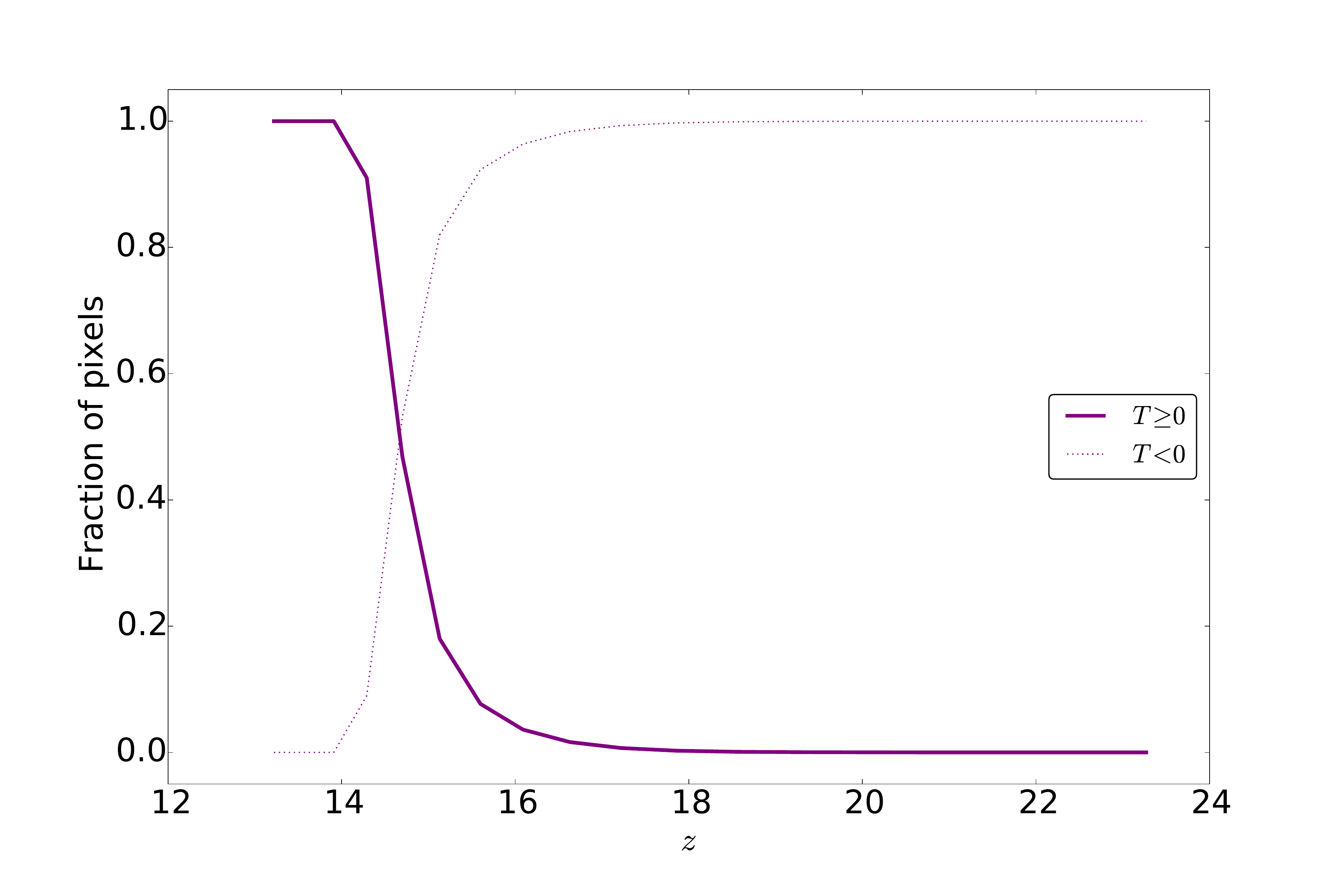}\\
    \includegraphics[trim=1.35cm 0.5cm 0.7cm 2.0cm, clip=true, scale=0.245]{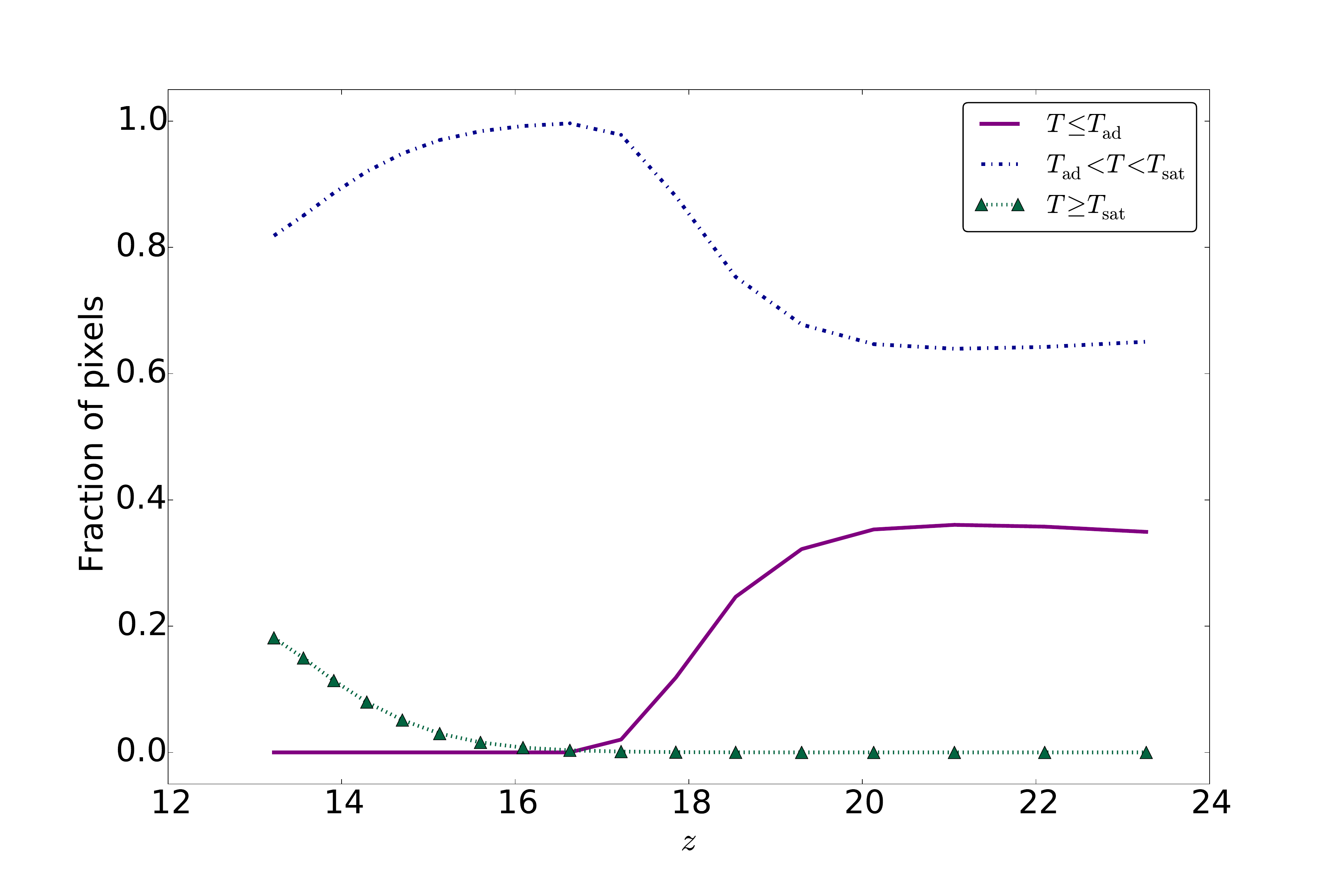}
  \end{array}$};
  \draw[orange, dashed, thick] (2.65,0.61) -- (2.65,15.55);
  \draw[green, dashed, thick] (5.85,0.60) -- (5.85,15.55);
  \end{tikzpicture}\\
  \caption{
  \textbf{Top} -
  Average brightness temperature as a function of redshift for the \textit{HMXB} simulation.
  \textbf{Middle} - Fraction of pixels in emission (fraction in absorption is shown in the thin dashed line)
  \textbf{Bottom} - Fraction of pixels that are at (1) at the saturated limit ($T_{\mathrm{s}}\gg T_{\mathrm{cmb}}$),
  (2) still cooling adiabatically, and (3) heated but not yet saturated.
  The green line marks the redshift at which heating is commencing in the simulation and the orange the point at which the map passes into emission (on average)}
  \label{fig:aveT_fracs}
\end{figure}
In this section, we will discuss the bispectrum as measured from the \textit{HMXB} simulation during X-ray heating.
Throughout this discussion we will make reference to several plots that summarise the progress of heating in the \textit{HMXB} simulation.
In Fig.~\ref{fig:aveT_fracs} we show the brightness temperature evolution (top);
the fraction of pixels in emission and absorption (middle);
and the fraction of saturated (i.e. with $T_{\mathrm{s}}\gg T_{\mathrm{cmb}}$), unheated, as well as those that are heated but
not yet saturated (bottom) for the \textit{HMXB} simulation.
We do not show the ionized fraction as it never reaches more than a few percent throughout
the simulation, and any ionization is concentrated in the very hottest regions,
and therefore has minimal impact on our discussion \citep{Ross2016}.
We have marked on these plots when heating commences in the simulation with the
green dotted line and when the simulation transitions into emission (on average) with an orange dotted line.

\begin{figure}
\centering
  $\renewcommand{\arraystretch}{-0.75}
  \begin{array}{c}
    \includegraphics[trim=3.9cm 1.0cm 5cm 1.0cm, clip=true, scale=0.32]{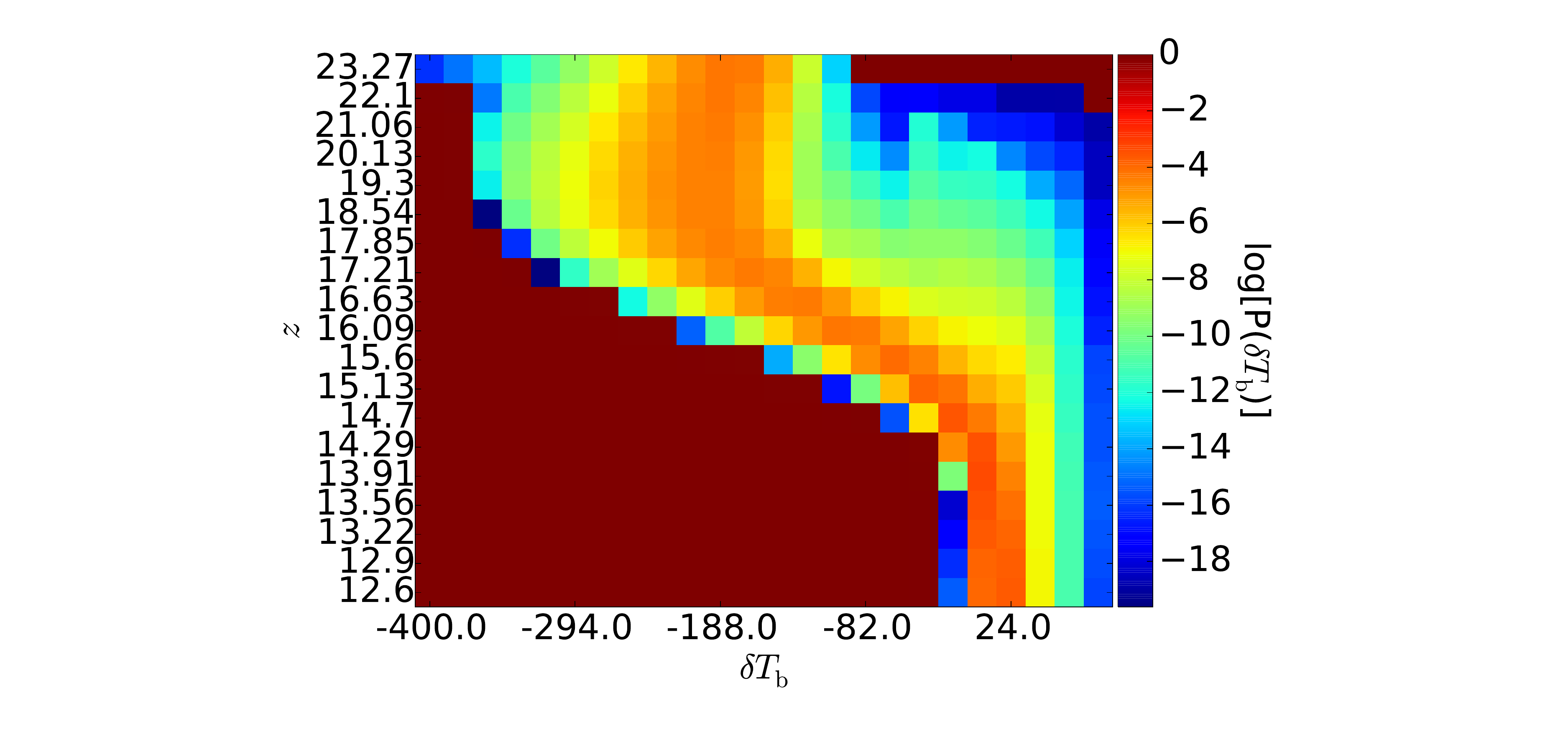}\\
    \includegraphics[trim=1.2cm 3.1cm 21.0cm 5.0cm, clip=true, scale=0.32]{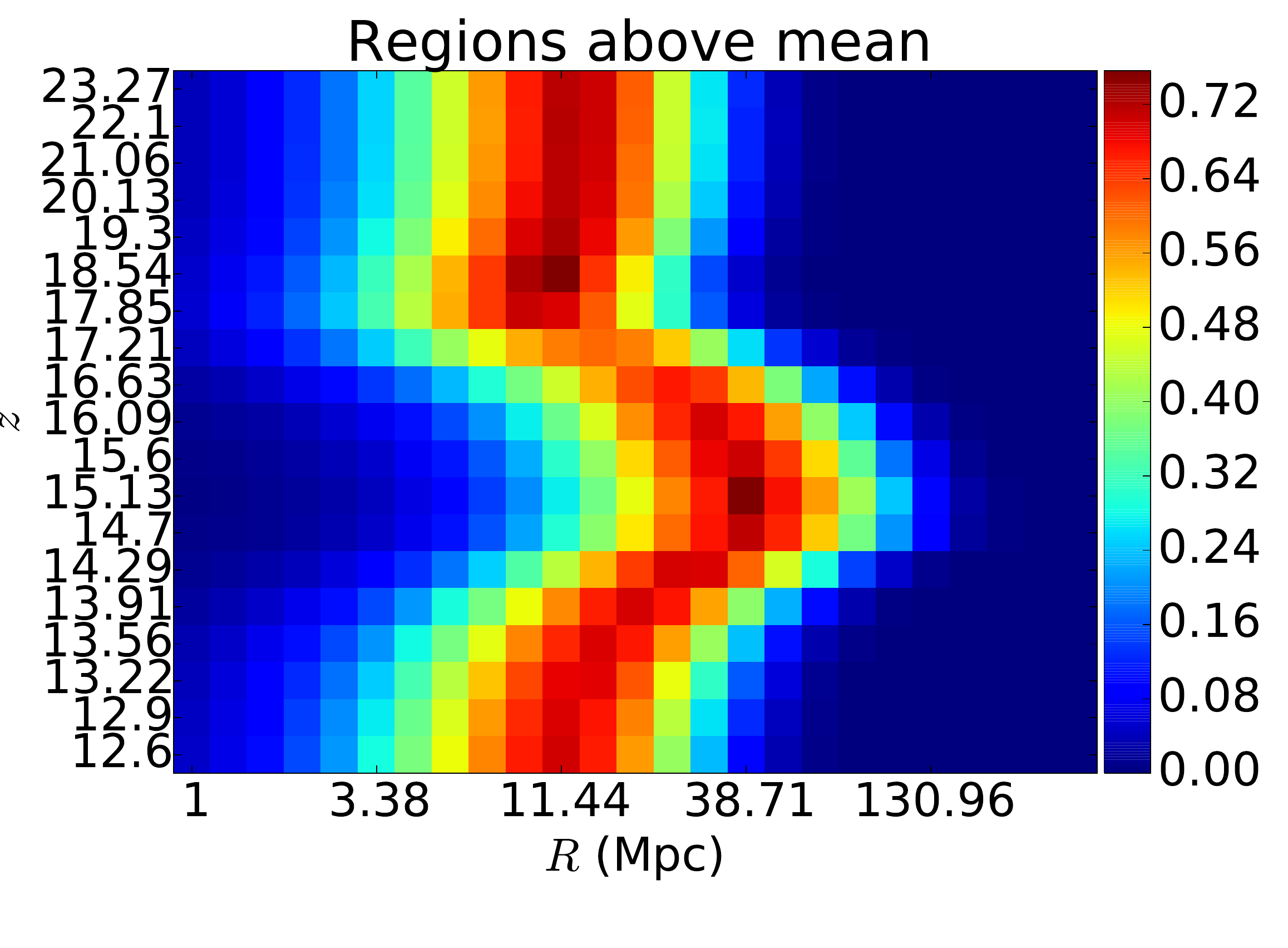}\\
  \end{array}$
  \caption{\textbf{Top} - The \textit{HMXB}'s 1D histogram for every redshift
  (color bar represents the log of the probability to highlight the PDF tails).
  It is at $17.85<z<20.00$ that the most cold pixels are wiped out by the formation of
  heating sources, at this point the bispectrum starts to gain amplitude with a shape
  close to that seen at $17.85$.
  \textbf{Bottom} - PDFs of characteristic size of emission regions for all redshifts, the colour bar describes d$P$/d$R$.
  }
  \label{fig:hist_vs_z}
\end{figure}

Also useful for tracking the progress of heating is the probability-density distribution (PDF) of the
brightness temperature at different redshifts.\footnote{The PDF of the brightness temperature is
generated from the unsmoothed datacubes} We therefore plot the log of the brightness-temperature PDF in Fig.~\ref{fig:hist_vs_z}.
Since the bispectrum is measuring the coherence between the above and below-average $\delta T_{\mathrm{b}}$ regions
and the waves associated with the three modes under consideration,
we also plot the probability distribution of the characteristic radius of above-average $\delta T_{\mathrm{b}}$ regions
in the bottom plot of Fig.~\ref{fig:hist_vs_z}
(measured by binarising the maps by above and below-average $\delta T_{\mathrm{b}}$ regions and using the
mean-free-path method of \citealt{Mesinger2007}, in which randomly
seeded trajectories are traced through the datacube until a phase transition is
met using monte-carlo methods).
Again, each row corresponds to the PDF at a different redshift.
We see that there is an evolution from small to large above-average $\delta T_{\mathrm{b}}$ regions over the range
$17.21\le z\le 14.70$ and then a reduction over the range $14.70\le z\le 13.22$.\footnote{The characteristic
sizes of the below-average $\delta T_{\mathrm{b}}$ regions evolve with redshift in a very similar way to the
above-average $\delta T_{\mathrm{b}}$ regions.}

\subsection{The normalised bispectrum for equilateral configurations}\label{sec:equiresults}

We have studied several common normalisations for the bispectrum (see the appendix for details),
and find that both the raw bispectrum $B(k_1, k_2, k_3)$  (with units of mK$^3$ Mpc$^{6}$)
and the dimensionless bispectrum $(k_1, k_2, k_3)^2/(2\,\pi^2)\,B(k_1, k_2, k_3)$
(which despite its name retains units of mK$^3$) exhibit regimes in which the amplitude
flips from strongly positive to strongly negative (and vice versa).
This occurs as the contribution to the statistic from non-Gaussianity gets very small
so that it fluctuates about zero and combines with a strong non-zero bispectrum
amplitude due to contributions from the power in the map.
It is common in large-scale structure studies to normalise out the contribution of
the power spectrum to the bispectrum by instead plotting
$Q(k_1, k_2, k_3) = B(k_1, k_2, k_3)/[P(k_1)\,P(k_2) + P(k_1)\,P(k_2) + P(k_1)\,P(k_3)]$,
which does suppress the sign fluctuations in the bispectrum.
However, if data is not without units (as is the case for 21-cm data which has units of mK,
so that $Q(k_1, k_2, k_3)$ has units of mK$^{-1}$), then $Q(k_1, k_2, k_3)$ retains a contribution
from the power spectrum, the degree of which is scale dependent.
The $Q(k_1, k_2, k_3)$ statistic is therefore not appropriate for use outside of large-scale structure studies.
We have detailed our findings in Appendix \ref{app:normdicuss} for the curious
reader and to support comparison with other studies of the 21cm bispectrum made in the main text.

It is more common in signal processing and time-series analysis to use the following normalisation first defined by \citet{BrillingerD.R.Rosenblatt1967},
\begin{equation}
\mathcal{B}(k_1, k_2, k_3) = \frac{B(k_1, k_2, k_3)}{\sqrt{P(k_1)\,P(k_2)\,P(k_3)}}\,,\\
\end{equation}
which isolates the contribution from the non-Gaussianity to the bispectrum, by normalising out the amplitude part of the statistic \citep{Hinich1968, Kim1978, Hinich1995, Hinich2005}.
\citet{BrillingerD.R.Rosenblatt1967} argue that $\mathcal{B}(k_1, k_2, k_3)$ is the correct normalisation choice for the bispectrum.
$\mathcal{B}$ has units of $\sqrt{V}$, we therefore instead consider the dimensionless quantity,
\begin{equation}
b(k_1, k_2, k_3) = \frac{B(k_1, k_2, k_3)}{\sqrt{(k_1\,k_2\,k_3)^{-1}\,P(k_1)\,P(k_2)\,P(k_3)}}\,.\\
\end{equation}
This statistic is directly proportional to the ensemble average of the three phases associated with $k_1$, $k_2$ and $k_3$; see \citet{Eggemeier2017}.
We will concentrate on this normalisation for the rest of this paper and refer to it as the \textit{normalised bispectrum} throughout.

\begin{figure}
\centering
  $\renewcommand{\arraystretch}{-0.75}
  \begin{array}{c}
    \includegraphics[trim=1.5cm 3.1cm 0.0cm 1.5cm, clip=true, scale=0.25]{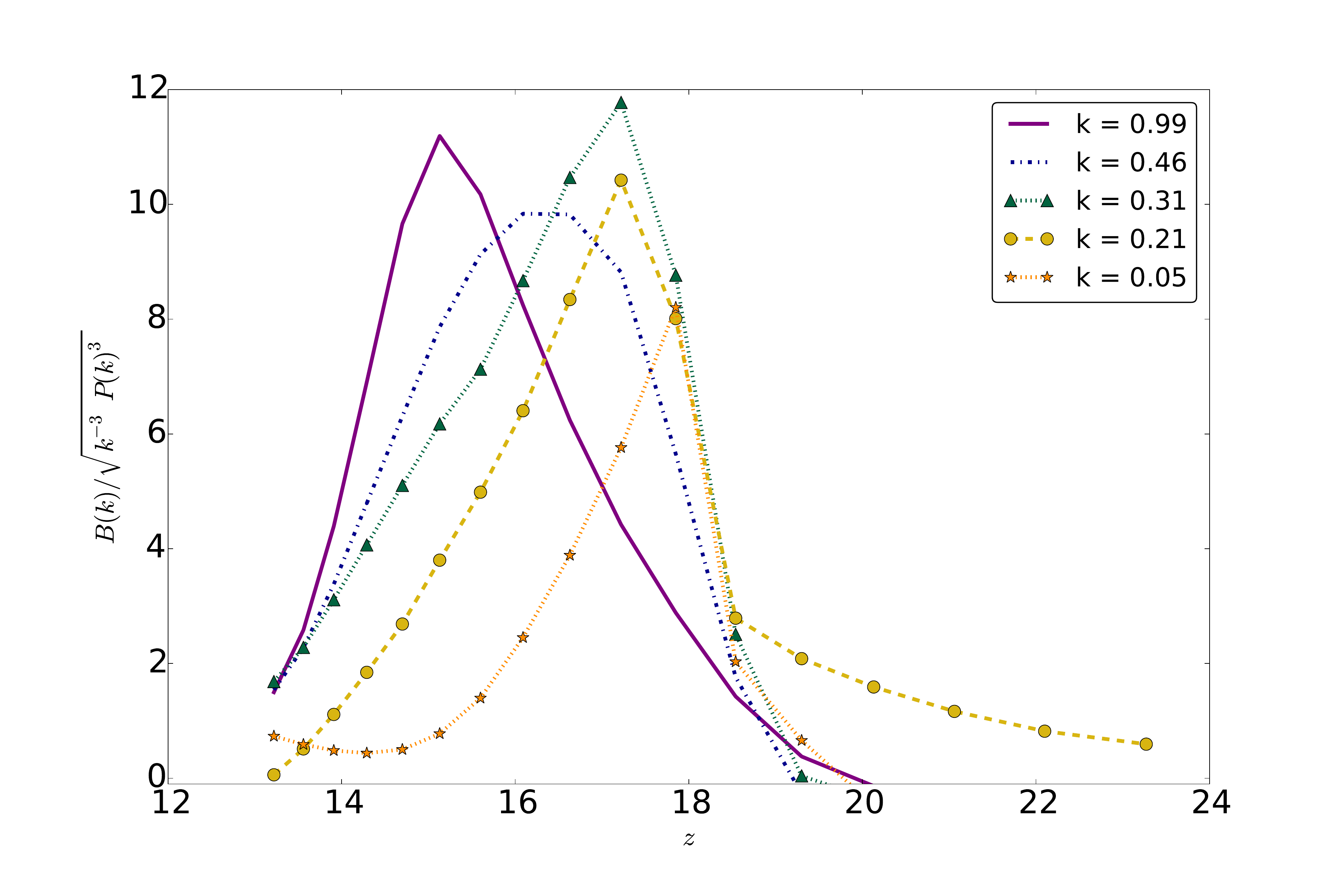}\\
    \includegraphics[trim=1.5cm 0.0cm 0.0cm 1.5cm, clip=true, scale=0.25]{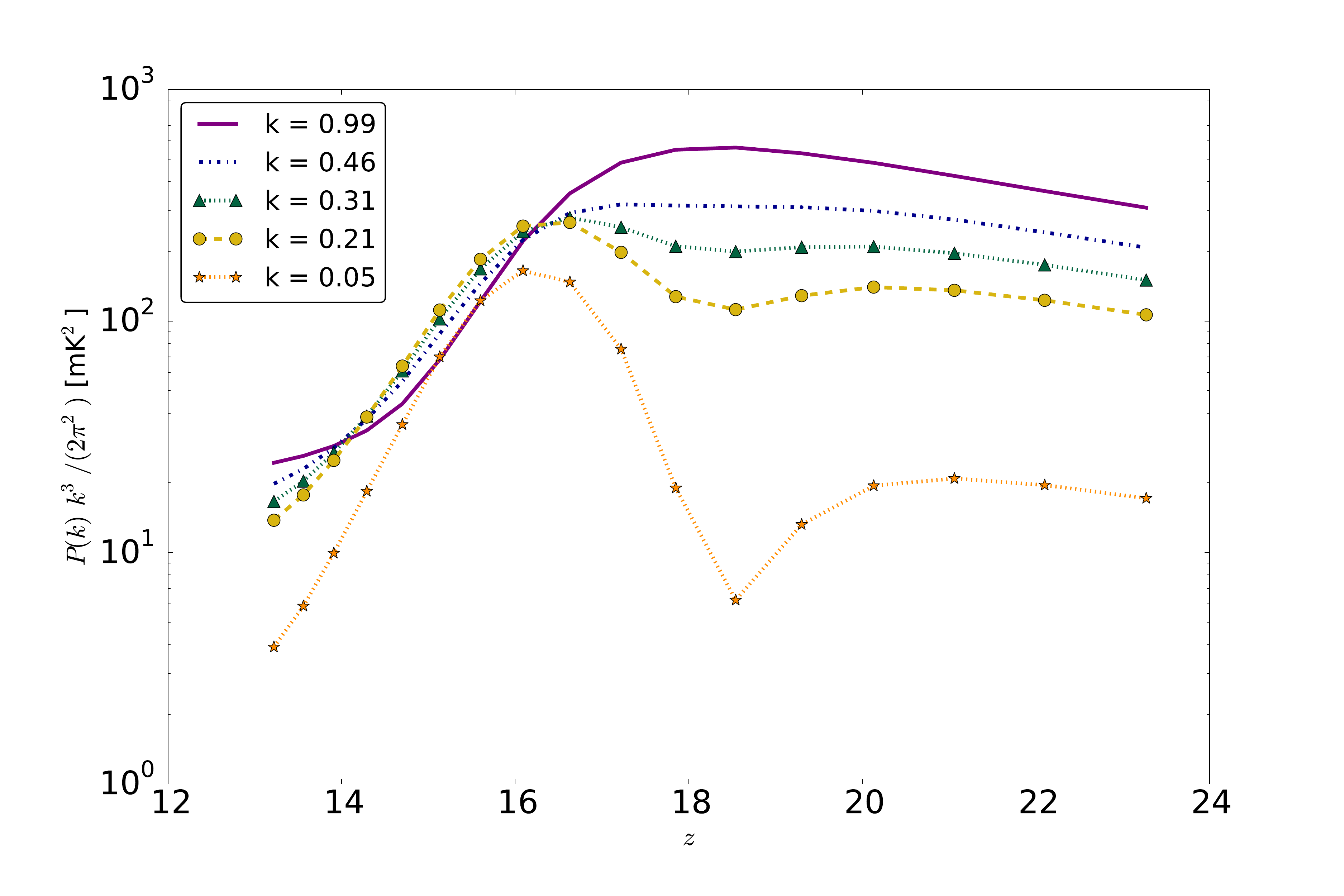}\\
  \end{array}$
  \caption{\textbf{Top} - Equilateral spherically-averaged normalised bispectrum
  measured from the mean-subtracted \textit{HMXB} simulation
  as a function of $z$ for various $k$ scales.
  \textbf{Bottom} - Corresponding spherically-averaged power spectrum.
  Each scale peaks at a different redshift and for most scales
  the normalised bispectrum starts to grow from $z=20$.
  The scales associated with the strongest non-Gaussianity (seen
  at $z=17.22$) starts increasing from the beginning of the simulation.}
  \label{fig:QvsZ_dimensional}
\end{figure}

\begin{figure}
\centering
  $\renewcommand{\arraystretch}{-0.75}
  \begin{array}{c}
  \includegraphics[trim=1.5cm 0.5cm 0.0cm 2.5cm, clip=true, scale=0.25]{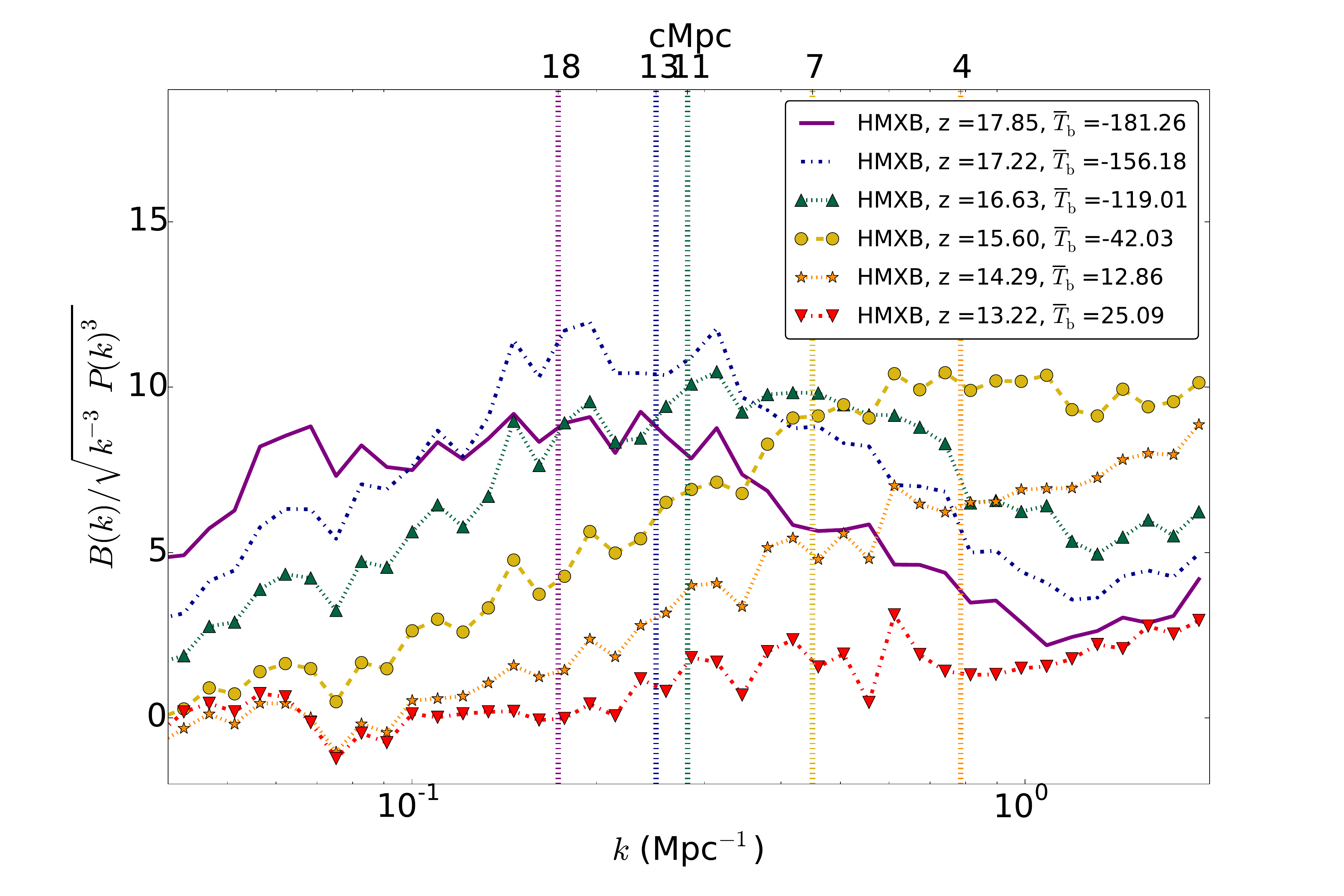}\\
  \end{array}$
  \caption{Evolution of the spherically-averaged normalised bispectrum
  with $k$ for equilateral configurations of $k$ vectors for the \textit{HMXB} simulation.
  Vertical dotted line correspond to the characteristic separation of emission regions
  as measured by granulometry.
  These lines correspond to from left to right $z = {17.85, 17.22, 16.63, 15.60, 14.29}$
  with colours following the legend's redshift relation.
  We see that there is clearly a correlation between this scale and the position of the
  peak of the turnover in the bispectrum.
  Note that rise in small-scale non Gaussianity as driven by the density field
  wipes out the turnover feature (see the orange dotted line with stars and the
  red dot-dashed line with triangles)
  }
  \label{fig:bvsk}
\end{figure}

We first plot the equilateral normalised bispectrum as a function of redshift
for a selection of $k$ scales; see the top panel of Fig.~\ref{fig:QvsZ_dimensional}.
It is very clear from comparing this to the corresponding power spectra in
the bottom plot of Fig.~\ref{fig:QvsZ_dimensional},
that the normalised bispectrum is providing us with new information
that is not possible to infer from the power spectrum alone.
The bispectrum peaks at increasingly high redshifts with increasing scale (decreasing $k$),
compare in the top plot the purple line ($k=0.99$ Mpc$^{-1}$  - small scale - which
peaks at $z\sim15$) with the orange dotted line with stars
($k=0.05$ Mpc$^{-1}$  - large scale - which peaks at $z\sim18$).
The power spectrum on the other hand has more of a turnover feature on small
scales (e.g. the purple line $k=0.99$ Mpc$^{-1}$) that rapidly drops off to smaller scales
at $z< 18$ and then exhibits a peak for larger scales at $z\sim16$;
this is particularly evident in the orange dotted line with stars ($k=0.05$ Mpc$^{-1}$).

In Fig.~\ref{fig:bvsk} we plot the spherically-averaged normalised bispectrum
for the equilateral configuration as a function of $k$ for a selection of redshift
(for the equivalent plot of the power spectrum, see the bottom plot of Fig.~\ref{fig:Qvsk_contrast} in the appendix).
We see that the normalised bispectrum starts to grow around the time that heating kicks in, and is maximised at
$z = 17.22$ (see the blue dot-dashed line in Fig.~\ref{fig:bvsk}) when the fraction of unheated pixels is approaching 0\%,
(refer to the bottom plot of Fig.~\ref{fig:aveT_fracs}),
i.e. $z=17.22$ coincides with the point at which most of the simulated volume has
experienced some level of heating.\footnote{Note that Fig.~\ref{fig:aveT_fracs} only
provides an estimate of the unheated pixels based on the brightness temperature
corresponding to the theoretical adiabatic kinetic temperature.}
After $z = 17.22$ the normalised bispectrum drops in amplitude with redshift.
The normalised bispectrum exhibits a turnover whose peak shifts to smaller scales with reducing redshift until $z=15.60$
(see the purple solid line, blue dot-dashed line, green dotted line with triangles,
and the yellow dashed line with circles in the Fig.~\ref{fig:bvsk}).

Between $17.85<z<20$ the shape of the bispectrum is very similar in shape
to that at $z=17.85$, but with a smaller amplitude,
which decreases with increasing redshift.
This is evident from at the yellow dashed line with circles in Fig.~\ref{fig:QvsZ_dimensional},
which shows the evolution of $k=0.21$~Mpc$^{-1}$ with $z$.
As can be seen from the top panel of Fig.~\ref{fig:hist_vs_z},
which shows the brightness-temperature log PDF for each redshift,
it is around $z=17.85$ that the most cold regions (overdense regions in which sources are yet to form)
are starting to be wiped out by the formation of the first stars in these regions.
\footnote{In these simulations
overdense regions in which stars have yet to form are the coldest regions
as the signal is in absorption and a large overdensity will make the signal more
extremely negative as $\delta \overline{T}_{\rm b}(\vect{x}) = (1+\delta)\, \langle (1 - \Tcmb/\Ts) \rangle$.
Of course, in reality such regions would likely be shock heated and this is
an effect that should be studied in the future.}
As more heated regions switch on, the level of coherence in the map
(and so to the degree of non-Gaussianity) will increase
on the scales associated with the typical separation of sources
(which at early times will coincide with the separation of saturated regions).

As well as a turnover that shifts to smaller scales (larger $k$),
there is also an increase in the small-scale bispectrum with decreasing redshift.
By $z = 14.29$, the normalised bispectrum exhibits a monotonic increase
(from roughly zero on large scales/ small $k$) linearly
towards smaller scales (see the orange dotted line with stars in Fig.~\ref{fig:bvsk}).

The growth in small scale non-Gaussianity with decreasing redshift,
is most easily seen in the plots of the equilateral bispectrum as a function of redshift
(top panel of Fig.~\ref{fig:QvsZ_dimensional}).
We see that the small-scale (large-$k$) bispectrum starts increasing from $z=20$
(see the purple solid line in Fig.~\ref{fig:QvsZ_dimensional}) which
coincides with the point at which heating is becoming notable (see the
green dashed line on Fig.~\ref{fig:aveT_fracs}).
The small-scale (large-$k$) normalised bispectrum then peaks at $z\sim 15$
(as the map passes into emission; see the orange dashed line in Fig.~\ref{fig:aveT_fracs}),
before starting to drop in amplitude.
By the end of the simulation the heating has saturated the spin temperature,
and as we will see in Section~\ref{sec:cross},
the non-Gaussianity is driven by fluctuations in the density field.

\begin{figure*}
\begin{minipage}{176mm}
\begin{tabular}{c}
  \includegraphics[trim=1.8cm 1.2cm 0.75cm 0.0cm, clip=true, scale=0.5]{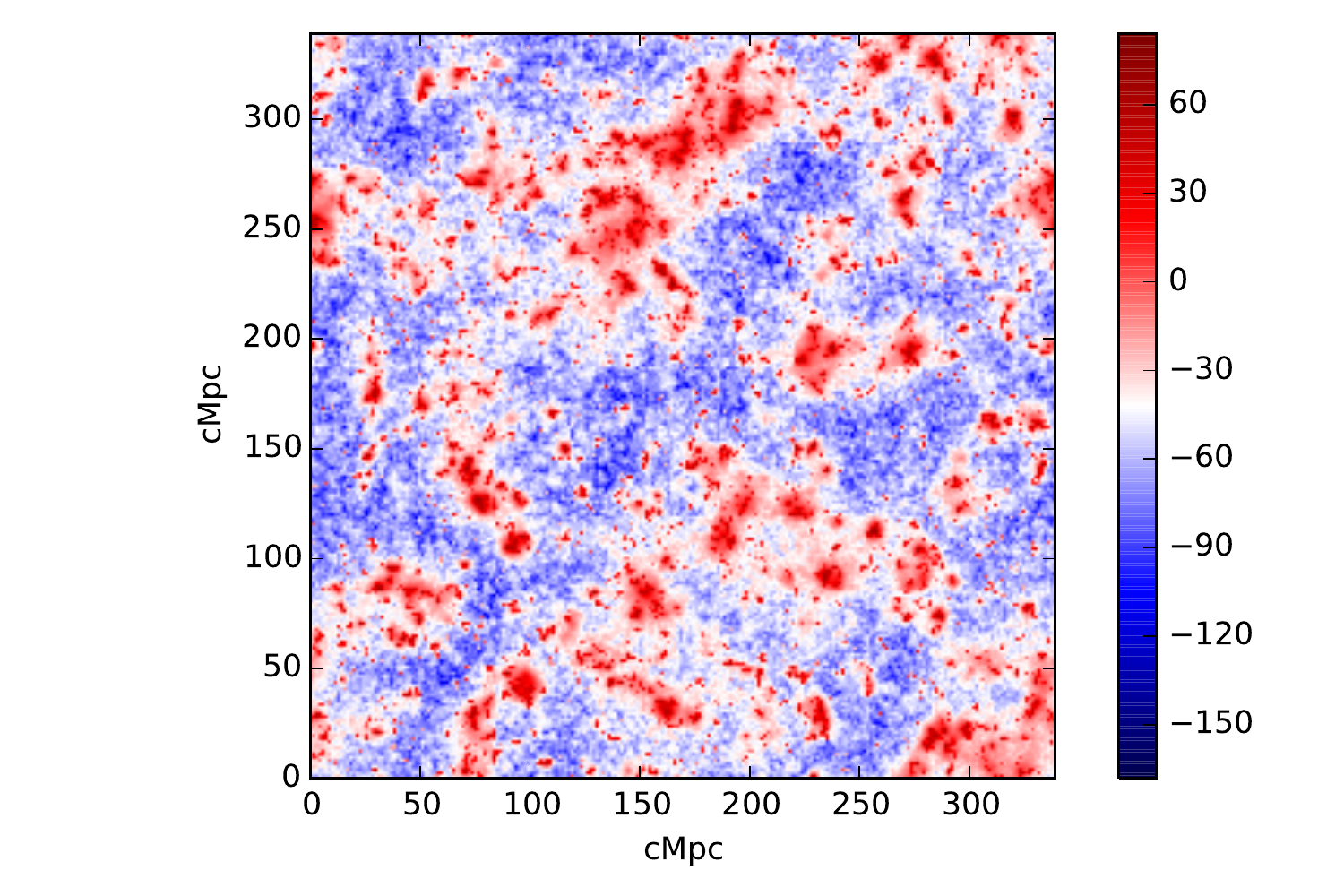} \includegraphics[trim=3.45cm 1.2cm 0.75cm 0.0cm, clip=true, scale=0.5]{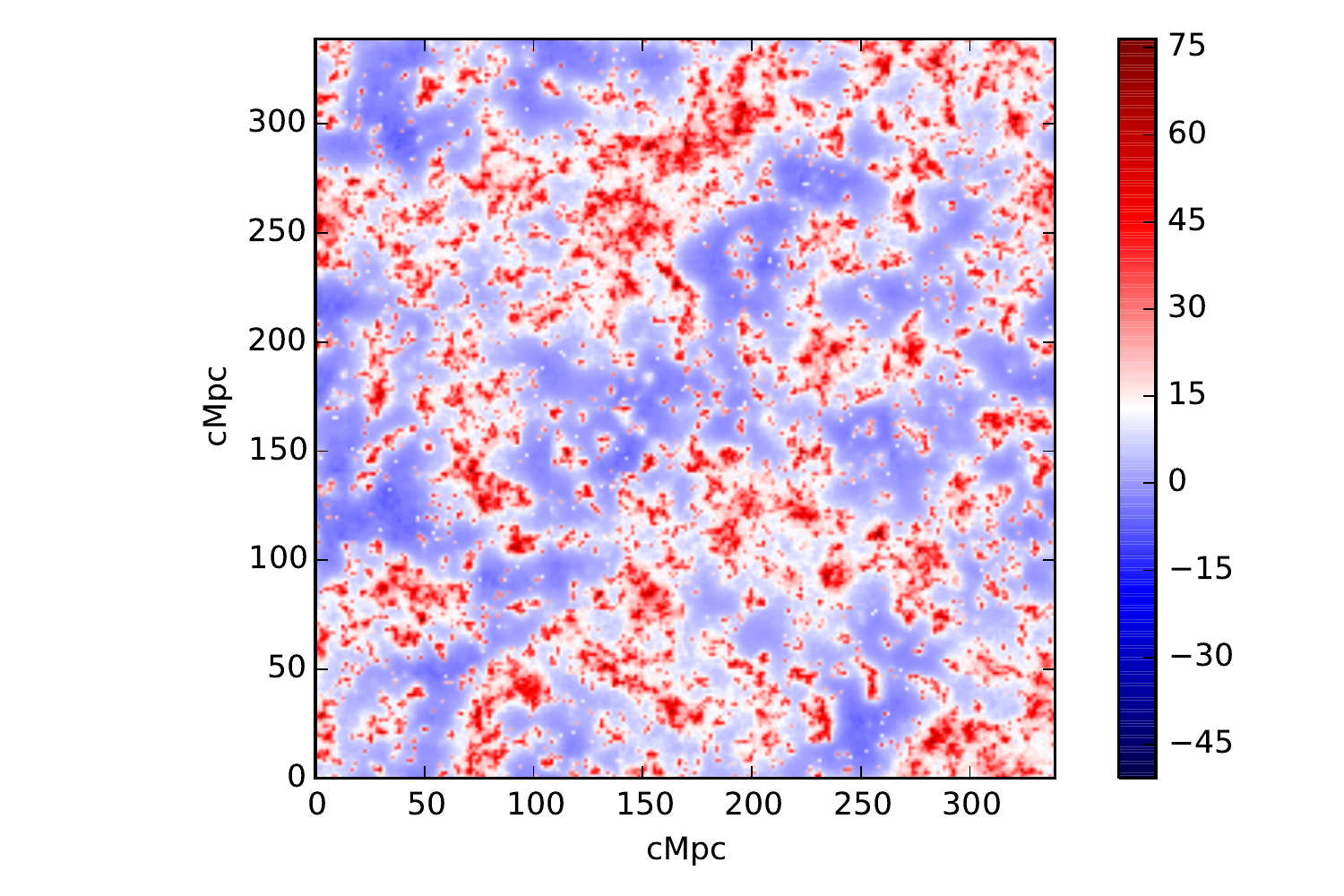} \includegraphics[trim=3.45cm 1.2cm 0.75cm 0.0cm, clip=true, scale=0.5]{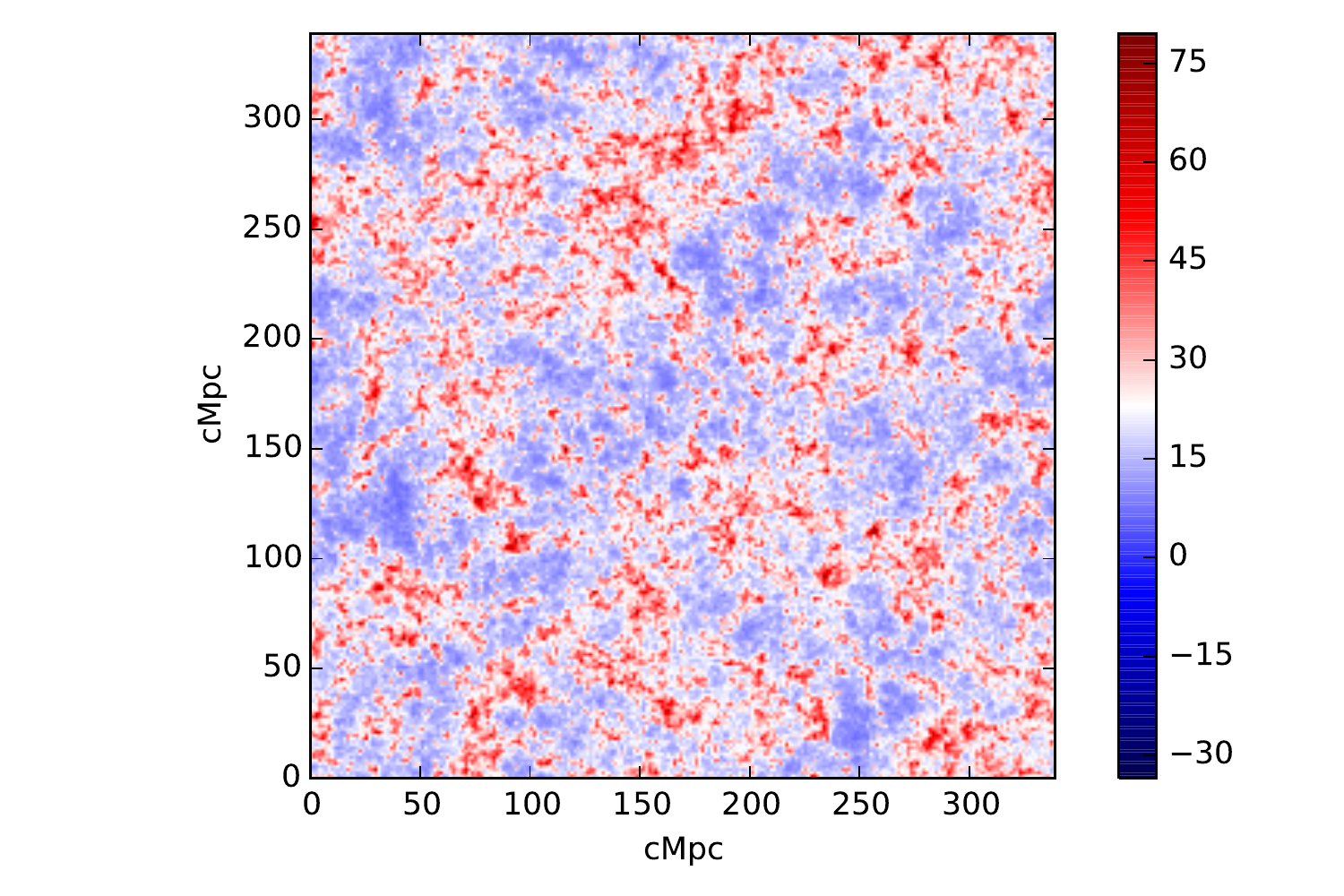}\\
\end{tabular}
\caption{Slices taken from the \textit{HMXB} simulation at $z=15.60, 14.29, 13.56$ from left
to right when the respective average brightness temperatures in the maps are
$\overline{T}_{\mathrm{b}} = -42.03, 12.86, 23$ mK.
White depicts regions with mean brightness temperature, blue highlights below-average $\delta T_{\mathrm{b}}$ regions,
and red the above-average $\delta T_{\mathrm{b}}$ regions; the colour bar is the brightness temperature in mK.
As the background brightness temperature increases the above-average $\delta T_{\mathrm{b}}$ regions become less
spherically symmetric as they fragment into smaller less spherical regions,
therefore the equilateral bispectrum will become less strong relative to the
other configurations with decreasing redshift.
}\label{fig:HMXBmaps}
\end{minipage}
\end{figure*}

This increase in small scale-structure power occurs as a background of X-rays
heat regions with below-average kinetic temperature, located away from the centre of heated regions.
This reduces the contrast between the hottest and coldest regions.
Early on, non-Gaussianities will therefore be driven by the larger-scale features in the map,
e.g. the distribution of the extremely hot regions relative to the cold.
As the contrast between such features is reduced, the small-scale fluctuations (modulating the large-scale brightness-temperature fluctuations) will have increasing influence.
This can be seen in the maps of the \hmxb simulation shown in Fig.~\ref{fig:HMXBmaps} for (from left to right)
$z = 15.60, 14.29, 13.56$.
We also see this, at some level, in the PDF plots of Fig.~\ref{fig:normZoo},
this shows how the typical size of above average $\delta T_{\mathrm{b}}$ regions shrink beyond $z=15.13$
and ultimately returns to the same scale as it was prior to heating
(in this simulation Lyman-$\alpha$ coupling is assumed to be complete and
so the density field drives the non-Gaussianity in the maps prior to heating).

\subsection{Synthetic datacubes to relate the normalised bispectrum to physical properties of heated regions}

In Section \ref{sec:equiresults} we have shown that there is an
evolving feature in the normalised bispectrum that must
connect with some physical features in the \hmxb simulation.
There are two main contributions to the bispectrum in such simulations,
one comes from the clustering of hot regions
and the other comes from the profile shape of features,
as per our discussion of what features various triangle configurations correspond to in Section~\ref{sec:interp}
and Fig.~\ref{fig:cartoon2}.
This concept is similar to the Halo model (see \citealt{Cooray2002} for a review) where
the power spectrum, bispectrum and other higher-order polyspectra
can be analytically calculated by considering the contribution of halo clustering and halo
profile to the non-Gaussianity as independent.
This assumption of independence is less appropriate to the EoH as heated profiles
are not as isolated from one another as they are for dark-matter haloes,
they instead overlap and combine to form a complex topology of heated regions.

We can attempt to better understand what drives the bispectrum of \hmxb
by creating synthetic datacubes which isolate certain physical features in the original
simulations.
First, we make binary maps from the \hmxb simulation that are 1 in regions
that are in absorption, and 0 in regions that are in emission.
The motivation for such a cut is to isolate the most heated regions in our datacubes.
We then use the granulometry method (see \citealt{Kakiichi2017}
for details on this method) to get a measure of the typical separation $D$ of
the emission regions at different redshifts.\footnote{We do not use the mean-free
path method of \citet{Mesinger2007} to measure the typical separation of saturated regions,
as emission regions are quite small and at many redshifts quite isolated,
therefore the mean-free-path method would return a size distribution biased towards
scales larger than those in which we are interested.}

In Fig.~\ref{fig:bvsk}, we overplot $k=2\,\pi/D$ with dotted vertical lines, using the
redshift-colour relation defined by the legend.
These lines correspond to the wavenumber one would expect to be associated with a
wave that would be coherent with the distribution of such hot regions.
There is clearly a positive correlation between the separation of emission regions
and the turnover in the normalised bispectrum.
This implies that the bispectrum is boosted on the scales due, at least in part,
to the clustering of the most hot regions in the \hmxb datacubes.

\begin{figure}
\centering
$\renewcommand{\arraystretch}{-0.75}
  \begin{array}{c}
  \includegraphics[trim=1.1cm 3.5cm 0.0cm 0.50cm, clip=true, scale=0.23]{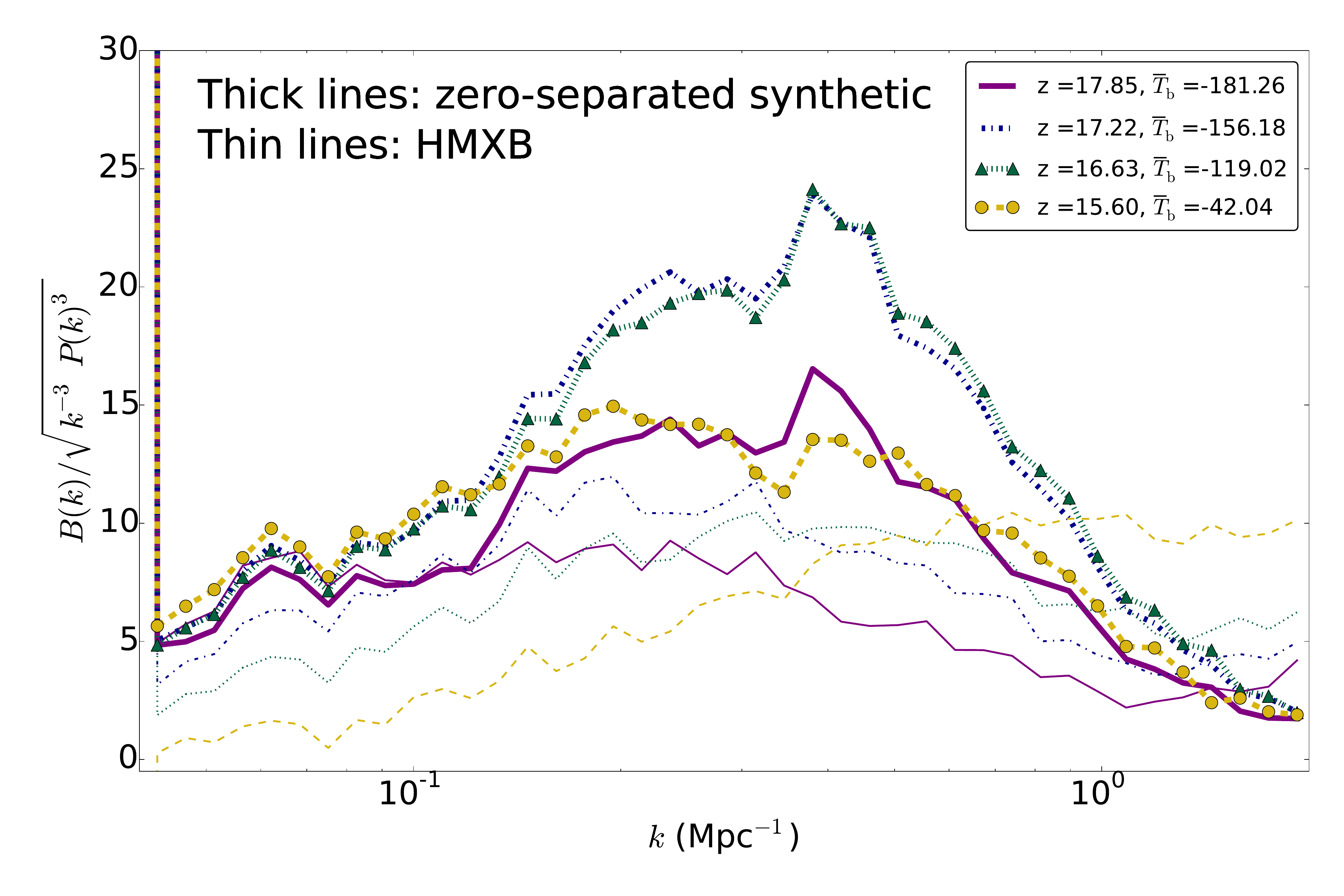}\\
  \includegraphics[trim=1.1cm 0.5cm 0.0cm 0.50cm, clip=true, scale=0.23]{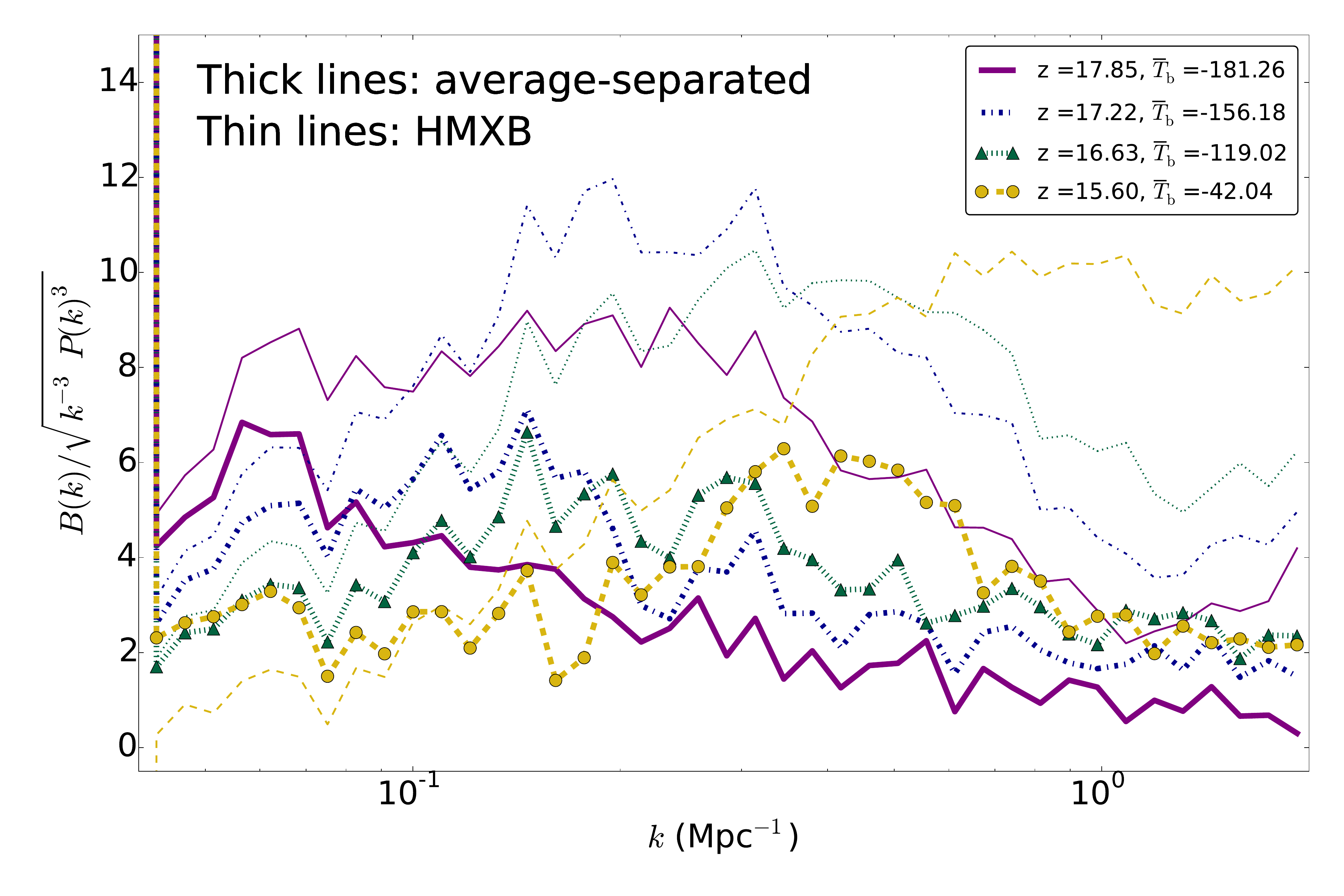}\\
  \end{array}$
  \caption{Comparison of the equilateral spherically-averaged $b(z)$ of the \hmxb
  (thin lines) to that of a synthetic repainted datacube (thick lines),
  whose pixels are divided into two independent phases based on the pixel values
  in the original \hmxb datacubes.
  \textbf{Top} - Phase 1, where all pixels that correspond to $\Tb \le 0$ in the \hmxb simulation
  are randomly assigned brightness temperatures from a Gaussian distribution (with same mean and variance as the equivalent
  subset of pixels in the \hmxb simulation),
  and phase 2 where pixels that correspond to $\Tb \ge 0$ in \hmxb are also randomly assigned brightness
  temperatures using the mean and variance as the equivalent
  subset of pixels from \hmxb.
  The turnover is seen at a roughly fixed scales at all $z$ in the zero-separated synthetic map,
  therefore the turnover that evolves to smaller scales in the \textit{HMXB}
  simulation can not solely be driven by clustering.
  \textbf{Bottom} - Phase 1 where pixels that are below the average in the \hmxb simulation
  are randomly assigned a temperature by sampling a Gaussian distribution according
  to the statistics of below-average pixels in the \hmxb simulation;
  and phase 2 as in phase 1 but for below-average pixels.
  We see that the scales over which $b(z)$ rises moving from smallest $k$
  (largest scales) to larger $k$ (smaller scales) roughly correlate with the scale
  of a similar turnover in the bispectrum of the mean-separated synthetic maps.
  }
  \label{fig:Qvsk_synthetic}
\end{figure}

Such a measure of separation cannot tell us about the coherence in the distribution
of emission regions.
We therefore make zero-separated synthetic datacubes in which pixels belong to one of two phases.
Phase 1, in which pixels associated with emission regions in the \hmxb
datacubes are randomly assigned brightness temperatures by sampling from a Gaussian distribution
(whose mean and variance is measured from the corresponding subset of pixels in the
\hmxb datacubes). Similarly for phase 2, all other pixels are randomly assigned
a $\delta T_{\mathrm{b}}$ by sampling from a Gaussian distribution
(whose mean and variance is set by the distribution of the
subset of pixels that are in absorption in the corresponding \hmxb datacube).
In such synthetic maps, the only sources of non-Gaussianity are the size of
regions in each phase and the relative distribution of such regions.
We refer to this synthetic dataset as zero-separated.

The thick lines in the top plot of Fig.~\ref{fig:Qvsk_synthetic} shows the normalised
equilateral bispectrum measured from zero-separated synthetic datacubes
for a reflective range of $z$.
We have also included the normalised bispectrum from \hmxb for reference
(thin lines, the colour of which correspond to the redshifts in the legends).
We see that the bispectrum from the synthesised datacube also exhibits a turnover,
but over a fixed range of scales.
For all $z$ plotted, the normalised bispectrum exhibits a broad peak over $0.2<k<0.7$~Mpc$^{-1}$ with a
narrow spike at $k=0.4$~Mpc$^{-1}$;
i.e. we see no evolution of the turnover to smaller scales with redshift.
Therefore, whilst there will be some contribution to the
amplitude of the bispectrum, primarily between around $0.2<k<0.7$~Mpc$^{-1}$,
from the distribution and size of the most hot regions,
this cannot be the only driver of the evolution we are seeing in the bispectrum.
Clearly, the details of the heating profiles surrounding the most hot regions
in the map must play a major part in driving the correlation we see between the
typical separation of the most heated regions and the scales of maximal
non-Gaussianity.\footnote{Note we also have considered the separation of
saturated regions.
The bispectrum from such fields looks very similar to that of the
zero-split synthetic datacube.}

We have already considered the typical size of above-average $\delta T_{\mathrm{b}}$ regions in
Fig.~\ref{fig:hist_vs_z},
and we see a characteristic size that is constant (of order 10 Mpc),
then increases from $z=17.22$ to $z=14.7$ and decreases again until $z=13.22$ it
settles back to the same characteristic scale as at $z>17.22$,
at which point it becomes roughly constant (again of order 10 Mpc) with decreasing redshift.
So it is not immediately obvious that we can make a connection with the characteristic
size of above-average $\delta T_{\mathrm{b}}$ regions and the scales that exhibit maximal non-Gaussianity in $b(z)$.
But as discussed with such measures of characteristic size, we ignore the level of coherence in the distribution of the regions of interest.
So we again use synthetic datacubes to try and probe the coherence in the size and distribution
of above and below-average $\delta T_{\mathrm{b}}$ regions.
To do so, we again split a separate set of synthetic datacubes into two phases,
one consisting of pixels that are above-average $\delta T_{\mathrm{b}}$ in the \textit{HMXB} simulation, and another consisting of pixels that are below average in the \textit{HMXB} simulation.
Pixels that belong to each phase are again randomly assigned brightness temperatures from a Gaussian
so that the mean and variance of each phase is the same as that of the two corresponding
subsets of pixels in the \hmxb simulation.
We refer to this synthetic dataset as average-separated.

We plot the equilateral spherically-averaged normalised bispectrum as measured from such average-separated
synthetic datacubes with thick lines in the bottom panel of Fig.~\ref{fig:Qvsk_synthetic}.
Again we plot the corresponding \hmxb bispectra (thin lines using the same redshift-colour
relation as in the legend).
There is a turnover in the bispectra from these average-separated synthetic datacubes
that corresponds to the scale of the large-scale edge of the turnovers
seen in the \hmxb's normalised bispectrum;
i.e. the scale at which the normalised bispectrum is seen to start increasing
(as we move from small $k$ to large $k$).
We therefore conclude that both the size and distribution of the above-average $\delta T_{\mathrm{b}}$ regions
define the large-scale edge of the turnover we see in the normalised
$\delta T_{\mathrm{b}}$ bispectrum.

\subsection{Contribution of the density, spin-temperature and their cross-terms}\label{sec:cross}

To try and gain further intuition as to what is driving the evolution of
the 21-cm bispectrum during X-ray heating,
we can break the bispectrum down into contributions from bispectra of the
two fields that drive the brightness temperature during the epoch of heating,
namely the density field and spin-temperature
field (we assume the neutral fraction is 1 throughout).
Because we can expand
$\delta T_{\mathrm{b}} = T_0\,(1 - \Tcmb/\Ts + \delta - \delta\,\Tcmb/\Ts)$,
we can write,

\begin{equation}
\delta^T\,\overline{T} = \delta T_{\mathrm{b}} - \overline{\delta T_{\mathrm{b}}} =
T_0\,\left(
\delta
- \delta_{\mathrm{x}}\,\overline{\psi\,\delta}
- \delta_\psi\,\overline{\psi}\right)\,,
\end{equation}
where $\delta$ is the matter overdensity, $\delta_\psi$ is the field contrast of
$\psi= T_{\mathrm{cmb}}/T_{\mathrm{s}}$,
the cross-field contrast is given by $\delta^{\mathrm{x}} = (\delta\,\psi/\overline{\delta\,\psi} -1)$,
and $T_0 = 27\,[(\Omega_{\rm b}\,h^2)/0.023] \,\sqrt{ [0.15/(\Omega_{\rm m} \,h^2)]\,[(1+z)/10.0] }$~mK.
These variables are summarised in Table~\ref{tbl:xterms} for ease of reference.
With this breakdown of $\delta^T\,\overline{T}$ in hand,
we can expand the 21-cm equilateral bispectrum as (dropping explicit mention of $k$ dependence for clarity),

\begin{equation}
\begin{split}
\overline{T}^3\,\langle \delta^T\,\delta^T\,\delta^T\rangle &= T_0^3\Big\{
\langle \delta\,\delta\,\delta \rangle \\
& - 3\,(\overline{\psi\,\delta})\, \langle \delta\,\delta\,\delta_{\mathrm{x}} \rangle
- 3\,\overline{\psi}\, \langle \delta\,\delta\,\delta_\psi \rangle \\
& + 3\,(\overline{\psi\,\delta})^2\, \langle \delta\,\delta_{\mathrm{x}}\,\delta_{\mathrm{x}} \rangle
+ 6\,(\overline{\psi\,\delta})\,\overline{\psi}\, \langle \delta\,\delta_{\mathrm{x}}\,\delta_\psi \rangle \\
& + 3\,(\overline{\psi})^2\, \langle \delta\,\delta_\psi\,\delta_\psi \rangle
- (\overline{\psi\,\delta})^3\,\langle \delta_{\mathrm{x}}\,\delta_{\mathrm{x}}\,\delta_{\mathrm{x}} \rangle \\
& - 3\,(\overline{\psi\,\delta})^2\,\overline{\psi}\, \langle \delta_{\mathrm{x}}\,\delta_{\mathrm{x}}\,\delta_\psi \rangle
- 3\,(\overline{\psi\,\delta})\,(\overline{\psi})^2\, \langle \delta_{\mathrm{x}}\,\delta_\psi\,\delta_\psi \rangle \\
& - (\overline{\psi})^3\,\langle \delta_\psi\,\delta_\psi\,\delta_\psi \rangle
\Big\}\,.
\end{split}\label{eqn:xprods}
\end{equation}

\begin{table}
    \caption{Terms used to understand the contribution of the underlying fields, i.e. $\delta$ and $T_{\rm s}$, to the evolution
    of the 21-cm bispectrum during the epoch of heating.}
    \begin{center}
      \begin{tabular}{ll}
        \hline
         $\delta = (\rho - \overline{\rho})/\overline{\rho}$ & \parbox{3.5cm}{Density}\\
         \\
         $\delta_\psi = (\psi - \overline{\psi})/\overline{\psi}$ & \parbox{3.5cm}{Inverse of spin temperature, where $\psi = \frac{T_{\textsc{cmb}} }{T_{\rm s}}$}\\
         \\
         $\delta_{\rm x} = (\delta\,\psi - \overline{\delta\,\psi})/\overline{\delta\,\psi}$ & \parbox{3.5cm}{Cross-product of $\psi$ and $\delta$}\\
        \hline
      \end{tabular}
    \end{center}
    \label{tbl:xterms}
\end{table}

\begin{figure}
\centering
  $\renewcommand{\arraystretch}{-0.75}
  \begin{array}{c}
    \includegraphics[trim=0.9cm 3.5cm 0.0cm 2.7cm, clip=true, scale=0.228]{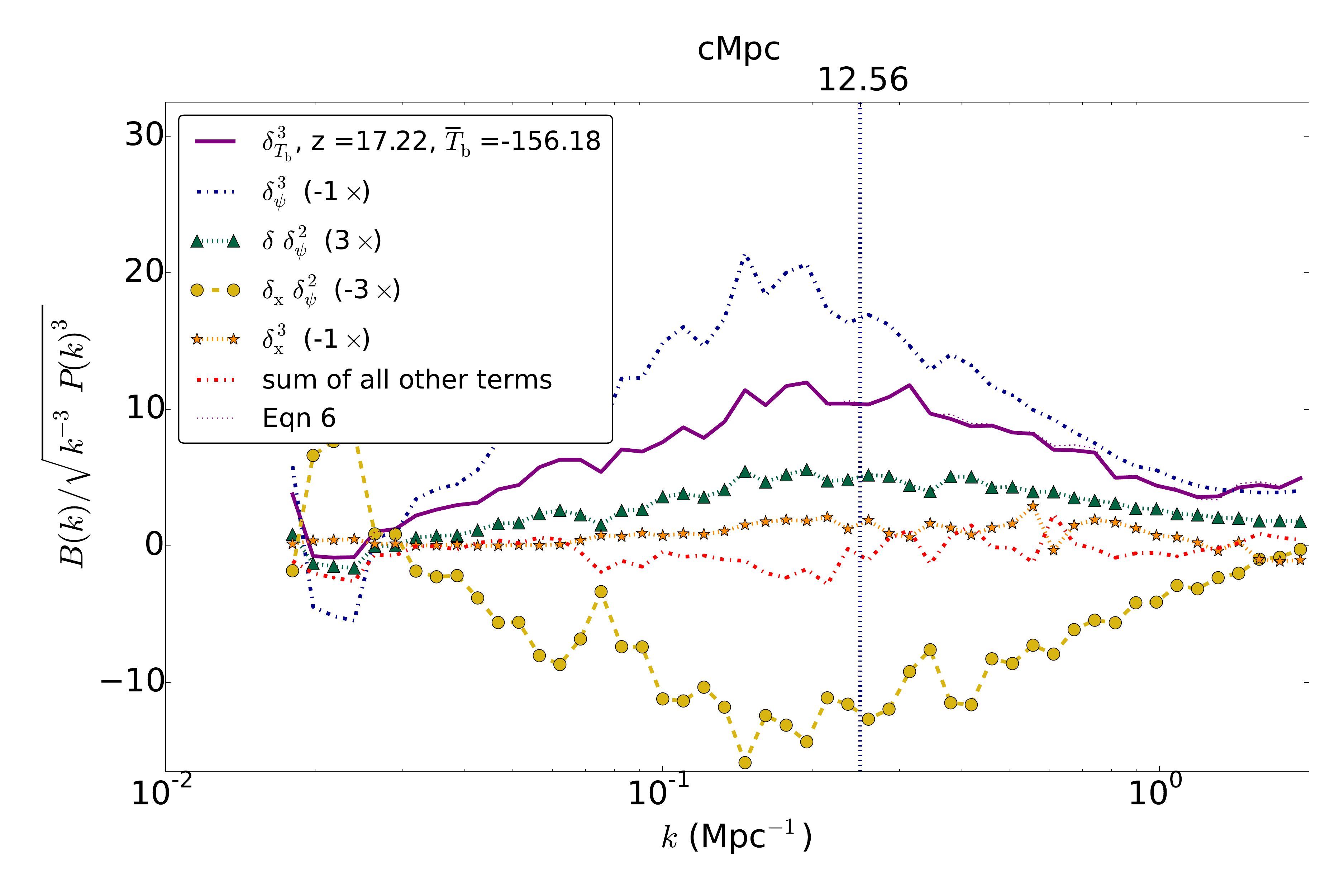}\\
    \includegraphics[trim=0.9cm 3.5cm 0.0cm 2.7cm, clip=true, scale=0.23]{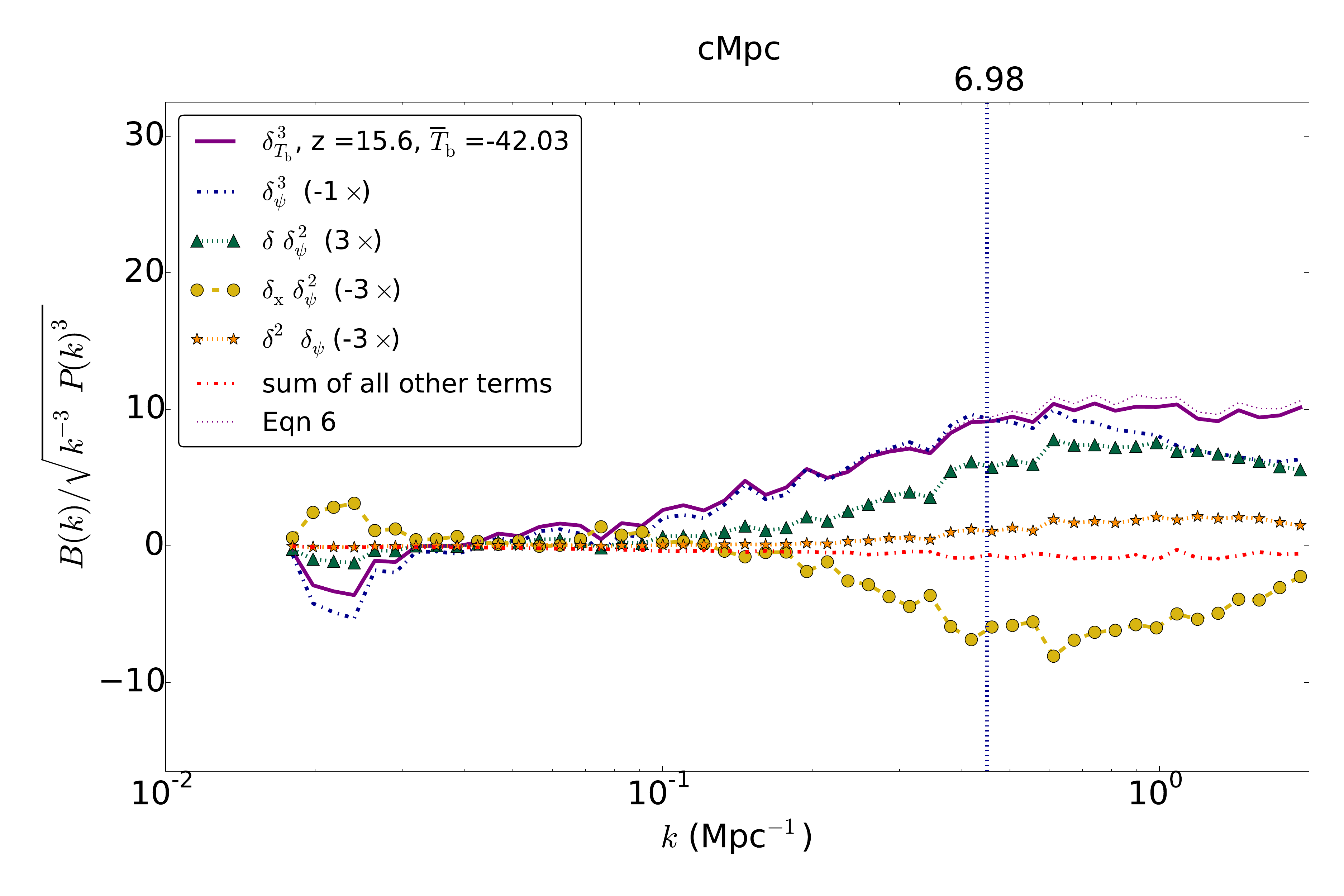}\\
    \includegraphics[trim=0.9cm 0.5cm 0.0cm 2.7cm, clip=true, scale=0.23]{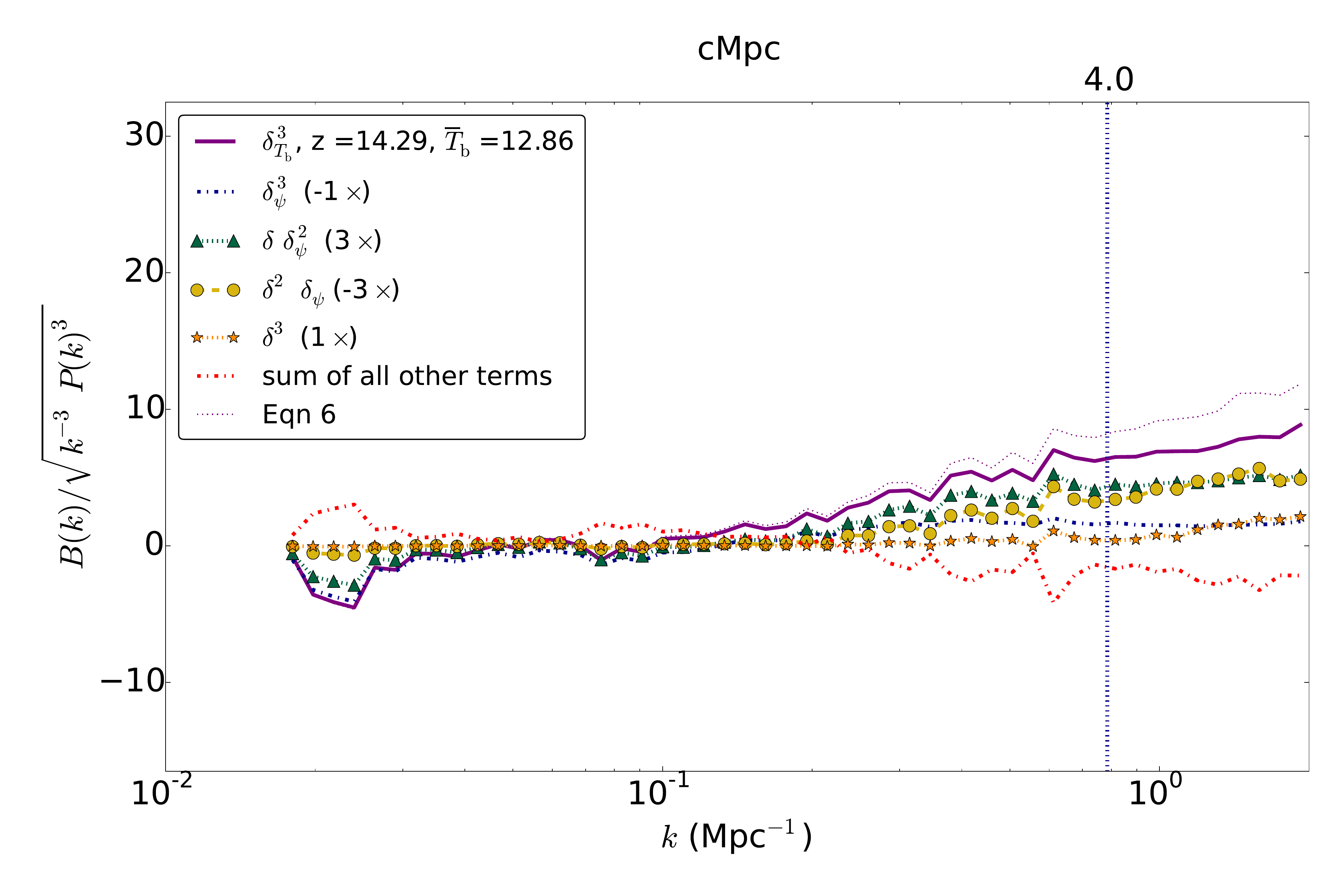}\\
  \end{array}$
  \caption{Spherically-averaged normalised bispectrum as a function of $k$ with
  the density, spin temperature and their cross-product contributions
  for the equilateral configuration, for from top to bottom $z = 17.22, 15.60, 14.29$
  and for the \hmxb simulation.
  The brightness-temperature bispectrum is shown with the solid purple line.
  During the early stages of heating the cross-product field has a lot of influence on the
  equilateral bispectrum mainly through $\langle \delta_{\mathrm{x}}\, \delta_\psi^2\,\rangle$
  (yellow dashed line w/ circles in the top and middle panels) and
  $\langle \delta_{\mathrm{x}}^2\, \delta_\psi\rangle$
  (orange dotted line w/ stars in the top panel).
  At later times the density field comes to dominate over the cross terms through
  $\langle \delta^3\rangle$ (orange dotted line w/ stars in the bottom panel)
  and $\langle \delta^2\, \delta_\psi \rangle$ (yellow dashed line w/ circles in the bottom panel).
  }
  \label{fig:Qvsk_CrossProds1}
\end{figure}

In Fig. \ref{fig:Qvsk_CrossProds1} we plot the
spherically-averaged normalised bispectrum for the brightness
temperature along with the contributions from $\delta$, $\Ts$ and
their cross-product field as described in Table \ref{tbl:xterms}
and Eqn.~\ref{eqn:xprods}.
We only explicitly plot a selection of the most dominant of these at
any given redshift and plot the the collective contribution from the rest of the terms
together.
Fig. \ref{fig:Qvsk_CrossProds1} shows from top to bottom $z = 17.22, 15.6, 14.29$.

For most of the simulation, the normalised bispectrum is dominated by fluctuations
in the spin temperature
$\langle \delta_\psi^3\rangle$ (blue dot-dashed lines)
and to a lesser extent by the cross-bispectra of the density and spin temperature
$\langle \delta\, \delta_\psi^2\rangle$ (green dotted line w/triangles).
In the \hmxb simulation it is sources in the more dense regions that produce heating,
and so the spin temperature and the density field will be positively correlated.
As a result the $\psi$ and $\delta$ will be anti-correlated and above-average
heated regions will correspond to a below-average $\psi$.
As we see, the $\langle \delta_\psi^3\rangle$ (blue dot-dashed lines; note we plot $-\langle \delta_\psi^3\rangle$)
is indeed negative as in $\psi$ the non-Gaussianity is coming from concentrations
of below-average $\psi$ regions in a more diffuse above-average $\psi$ background.
In contrast, the contribution from $\langle\delta_{\mathrm{x}}\,\delta_\psi^2\rangle$
(which is dominant at early times) is positive;
see the yellow dashed line w/circles in the top and middle panels,
(noting that we plot $-3\langle\delta_{\mathrm{x}}\,\delta_\psi^2\rangle$).
This means that at the point when the contribution from the spin temperature is most strong
(which occurs at $z=18.54$ in the \hmxb simulation when the contrast between the
most hot and the most cold regions is at its most extreme - see Fig.~\ref{fig:hist_vs_z}),
the brightness-temperature bispectrum is suppressed by the contribution
of $\langle\delta_{\mathrm{x}}\,\delta_\psi^2\rangle$
opposing that from the spin temperature.
The bispectrum therefore peaks slightly later than one might naively expect
from an argument based on the contrast between extreme cold regions
and extreme hot regions being maximal and so boosting the degree of non-Gaussianity.

As the background of X-rays heats up the cooler areas, and more
and more regions become saturated (at which point they basically follow the
fluctuations in the density field), the influence of the cross-products reduces.
This is most clearly seen in the reduction in the contribution of $\langle \delta_{\mathrm{x}}\, \delta_\psi^2\rangle$
(yellow dashed line w/ circles) relative to the other contributing terms
between the top panel and middle panels of Fig.~\ref{fig:Qvsk_CrossProds1}.
The $\langle \delta_{\mathrm{x}}^3\rangle$ (orange dashed line w/stars in the top panel)
is also one of the more influential terms at early times.

As the influence of the cross products decreases,
the influence of fluctuations in the density field on the normalised bispectrum increases (mostly on smaller scales).
This can be seen in the relative increase in the influence of $\langle \delta\, \delta_\psi^2\rangle$,
seen by tracking the green dotted line with triangles from the top to bottom panels
of Fig.~\ref{fig:Qvsk_CrossProds1}.
The influence of $\langle \delta^2\, \delta_\psi\rangle$ also starts to have influence during
the mid phases of heating (orange dashed line w/triangles in the middle panel).
Towards the end of heating
the density field starts to dominate, see $\langle \delta^2\, \delta_\psi\rangle$
(yellow dashed line with circles in the bottom panel) and $\langle \delta^3\rangle$ (orange dashed line w/triangles in the bottom panel).
By the end of the simulation (not shown) the density field drives the normalised bispectrum
through $\langle \delta^2\, \delta_\psi\rangle$ and $\langle \delta^3\rangle$.

We have marked the typical separation of emission regions with the vertical blue dotted line.
The turnover is more prominent in the normalised bispectrum of $\langle \delta_\psi^3\rangle$ (blue dot-dashed lines)
than it is in the brightness-temperature bispectrum (purple solid lines),
and the correlation between the typical separation and this turnover is also more
clear.
The middle panel of Fig.~\ref{fig:Qvsk_CrossProds1} shows nicely how this turnover
is ultimately suppressed by the increasing domination of small-scale structure in the density field.

Note that at later times Eqn.~\ref{eqn:xprods} overestimates the true bispectrum.
This is because the influence of the neutral fraction (which we have assumed to be
totally negligible in deriving Eqn.~\ref{eqn:xprods}) can no longer be ignored.
However, as is clear from comparing the true brightness-temperature normalised
bispectrum (purple solid line) with Eqn.~\ref{eqn:xprods}
(thin purple dot-dashed line), ionizations simply damp the amplitude of the bispectrum,
rather than qualitatively alter it.

\subsection{The normalised bispectrum for isosceles configurations}

\begin{figure}
\centering
  $\renewcommand{\arraystretch}{-0.75}
  \begin{array}{c}
    \includegraphics[trim=0.9cm 0.0cm 0.0cm 1.0cm, clip=true, scale=0.23]{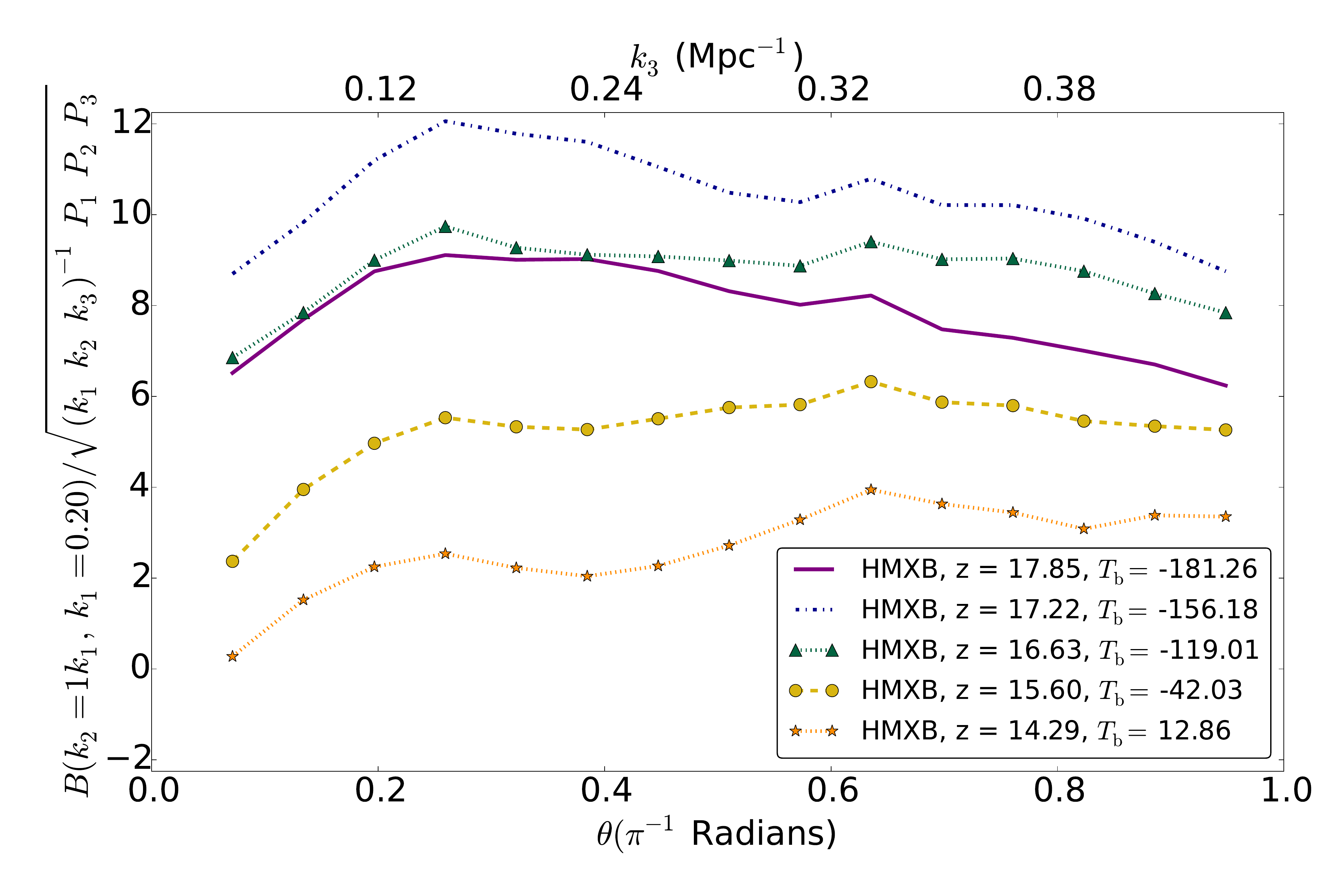}\\
    \includegraphics[trim=0.9cm 0.0cm 0.0cm 1.0cm, clip=true, scale=0.23]{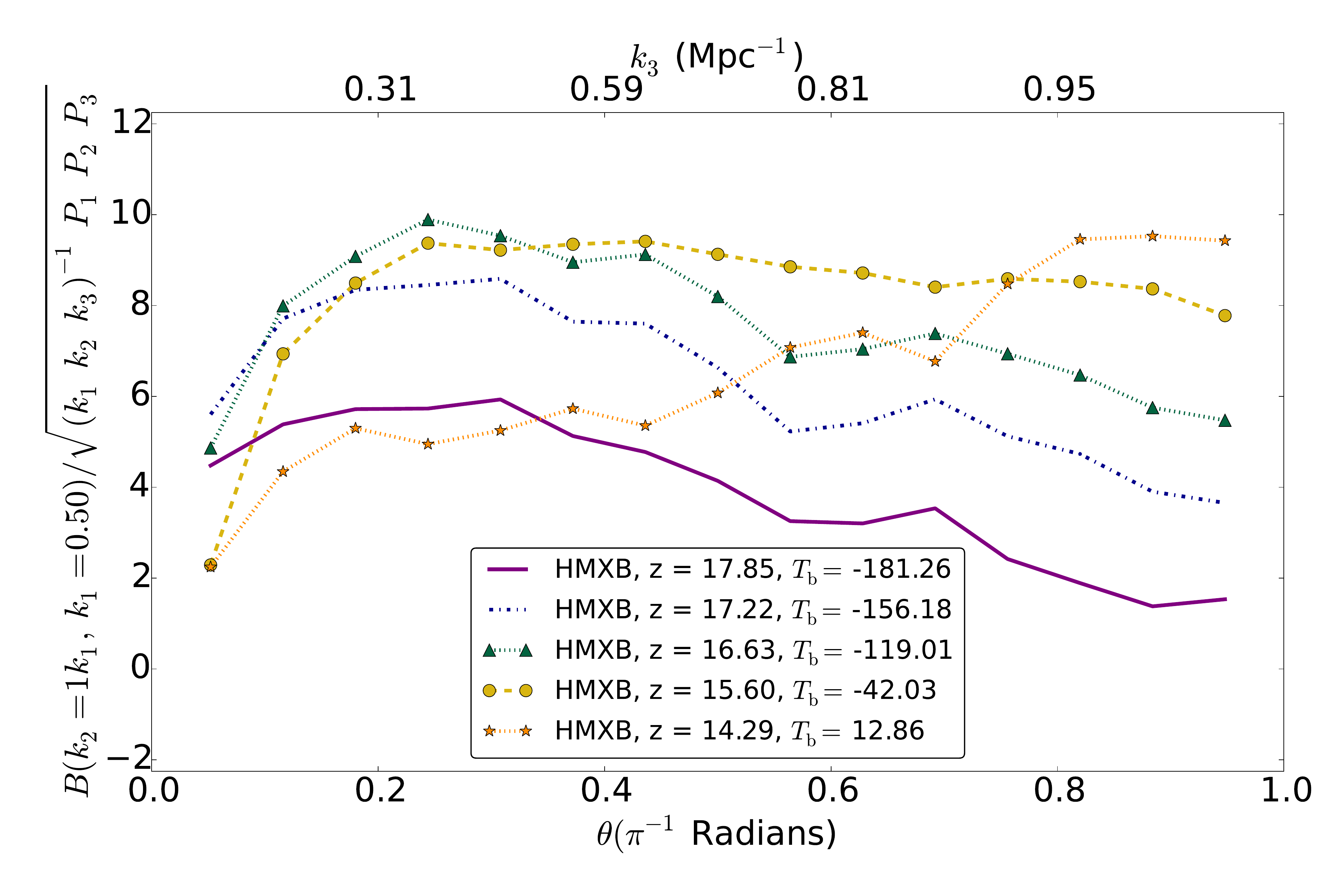}\\
    \includegraphics[trim=0.9cm 0.0cm 0.0cm 1.0cm, clip=true, scale=0.23]{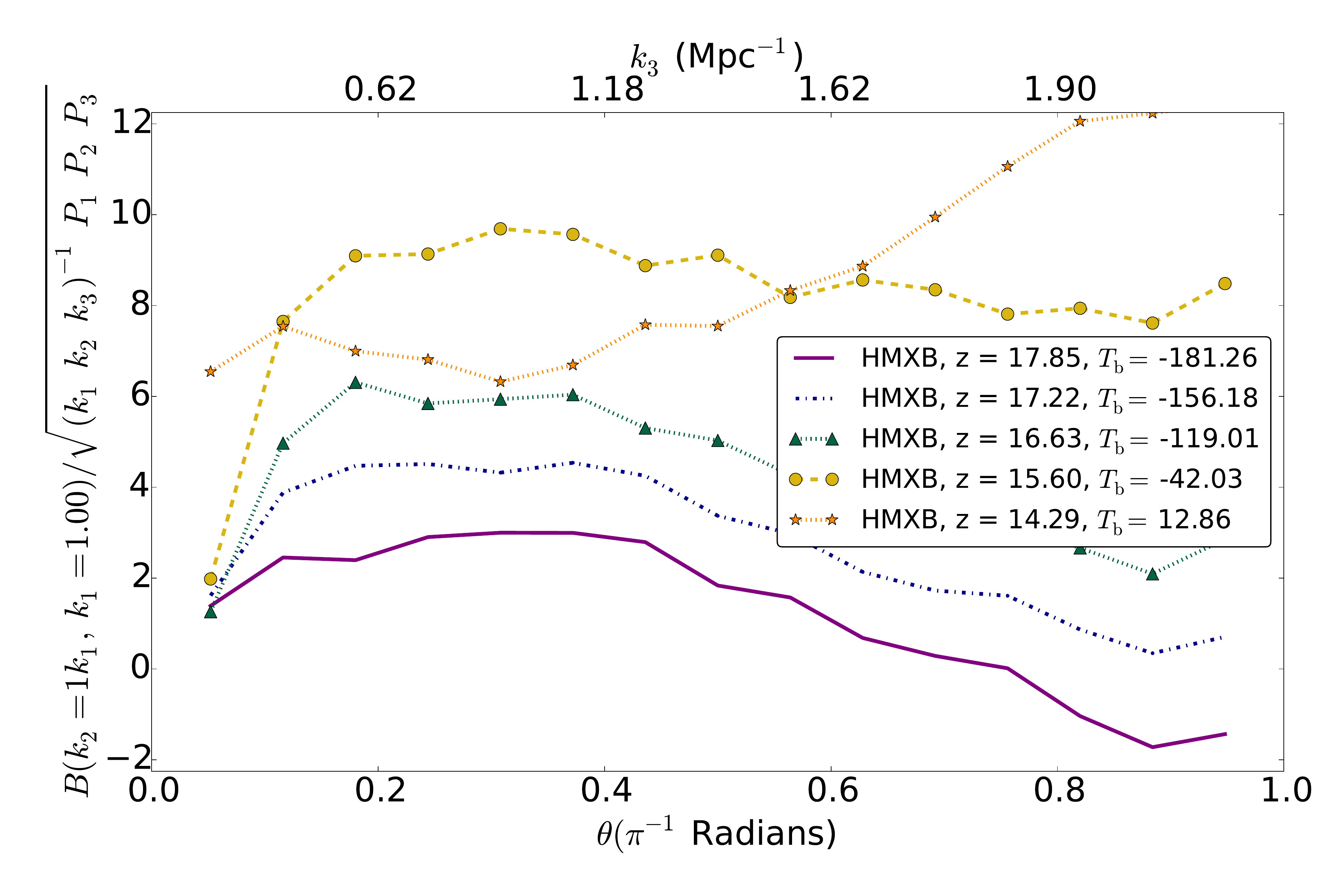}
  \end{array}$
  \caption{Normalised bispectrum as a function of angle between $k_1$ and $k_2$ for a range
  of redshifts for the isoceles configuration,
  i.e. where $k_1 = k_2$.
  \textbf{Top} - $k_1 = k_2 = 0.2$ Mpc$^{-1}$;
  \textbf{middle} - when $k_1 = k_2 = 0.5$ Mpc$^{-1}$;
  and \textbf{bottom} when $k_1 = k_2 = 1.0$ Mpc$^{-1}$.
  Here we show results for the mean-subtracted \hmxb simulation.
  The general trends of the normalised bispectrum as a function of redshift we see in
  the equilateral configuration are the same on a variety of scales.
  Whilst the large-scale details of the distribution and shape of heated profiles dominates the signal,
  we see a peak around the equilateral configuration.
  This is due to heating sources following the filamentary structure of the underlying
  dark matter field and heating profiles being roughly symmetrical around sources.
  }
  \label{fig:isoceles}
\end{figure}

Until this point we have focused on the equilateral configuration,
but of course, this is just one of many possible configurations of triangle
that may be formed by three $\vect{k}$ vectors.
We therefore consider the isosceles configuration in this section.
We focus on the isosceles as other configurations we looked at during our studies
for this paper were qualitatively quite similar.

Early in the heating process the most heated regions are concentrated
around sources and are quite symmetric in their profiles shapes due to the long mean-free path of X-rays. The most extreme hot regions will therefore follow the underlying filamentary structure of the cosmic web, whilst exhibiting a level of spherical symmetry. We therefore expect the normalised bispectrum to be maximal for configurations close to equilateral during the early phases of the heating process.
This can be seen at $z=17.85$ (purple solid line), $z=17.22$ (blue dot-dashed line)
and $z=16.63$ (green dotted line with triangles) in the three panels of
Fig.~\ref{fig:isoceles}. These plots show the spherically-averaged normalised
isoceles bispectra for a range of $k_3$ (defined by the angle $\theta$ between
$k_1$ and $k_2$) for $k_2= k_1=0.2$~Mpc$^{-1}$ (top), $k_2= k_1=0.5$~Mpc$^{-1}$ (middle)
and $k_2= k_1=1.0$~Mpc$^{-1}$ (bottom).
While the map is in absorption, configurations that are close to equilateral,
i.e. $\theta\approx \pi/3$ radians, have the largest bispectrum.
Note also from this plot, that similar to the normalised equilateral bispectrum,
the normalised isosceles bispectrum has maximum amplitude at $z=15.60$ on small scales (large $k$)
(see the yellow dotted line with circles in the bottom plot of Fig. \ref{fig:isoceles}),
and at $z=17.22$ on large scales (see the blue dot-dashed line in the top plot Fig. \ref{fig:isoceles}).
Of course, as multiple HMXB sources drive a given heated region, there will be
deviation from spherical symmetry in the heated features of the map,
and so we would also expect a strong bispectrum from flattened triangle configurations.
This is seen in Fig.~\ref{fig:isoceles} in which a positive normalised bispectrum persists
as the \vect{k} triangle is flattened by an increasing angle between $\vect{k}_1$
and $\vect{k}_2$.

As the background brightness temperature rises, and the small-scale structure starts driving
the bispectrum, the spherical symmetry of heated profiles
becomes less of a dominant feature and there will be more non-Gaussianity
coming from ellipsoidal profiles, even plane-like features.
This is, for example, seen at $z=15.6$ (yellow dashed line with circles) in Fig.~\ref{fig:isoceles},
where the normalised bispectrum becomes roughly flat for most angles (it even increases to larger angles at $k=0.2$~Mpc$^{-1}$),
but still drops off at the smaller angles, $\theta<0.2\pi$~radians.
After the map moves into emission the normalised bispectrum exhibits a U
shape on small scales (reminiscent of what is seen in the reduced bispectrum of the density field).
This can be seen at $z=14.29$ (orange dotted line with stars) in the bottom plot of Fig.~\ref{fig:isoceles}.

\section{Consistency of qualitative evolution of the normalised bispectrum across various simulations of X-ray heating}\label{sec:other_sims}

\hmxb is but one simulation, and as with all simulations, it makes certain assumptions
regarding the nature of the dominant sources of heating and their spectra.
As the parameter space for the EoH remains wide open,
we will now briefly consider how generic the features we see in the \hmxb bispectrum
are to other simulations.

\begin{figure}
\centering
  $\renewcommand{\arraystretch}{-0.75}
  \begin{array}{c}
    \includegraphics[trim=1.2cm 3.15cm 0.0cm 2.3cm, clip=true, scale=0.245]{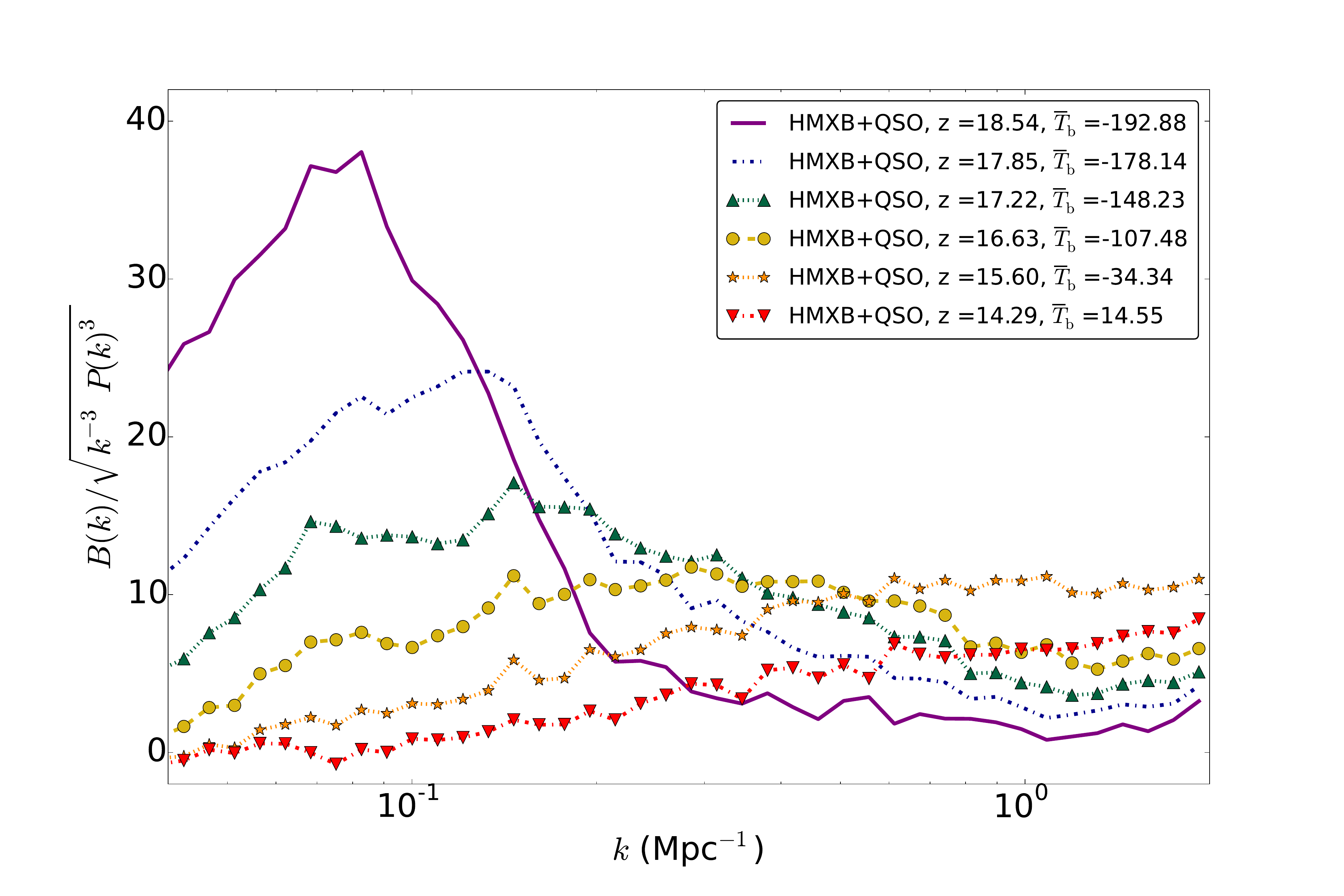}\\
    \includegraphics[trim=1.2cm 3.15cm 0.0cm 2.3cm, clip=true, scale=0.245]{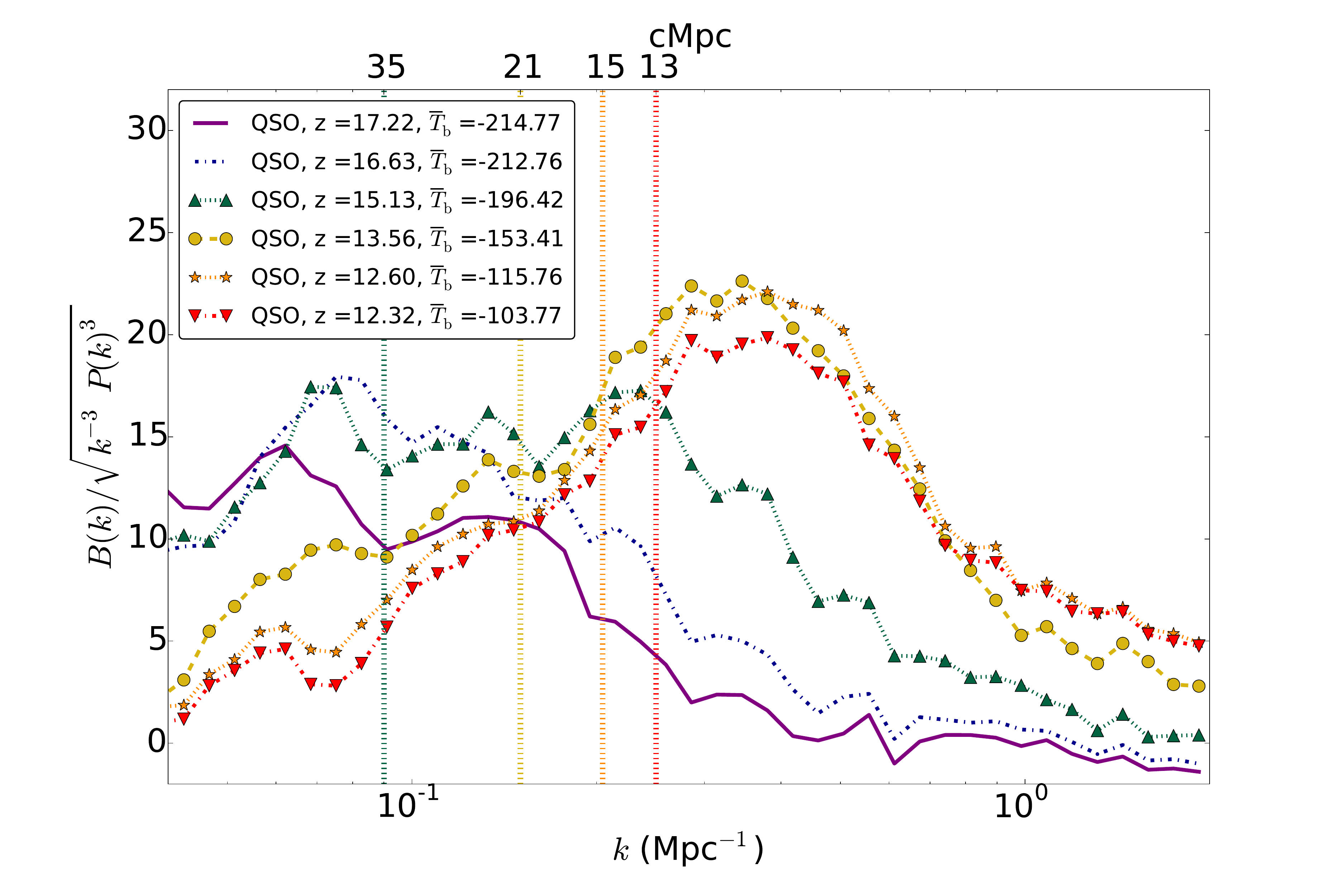}\\
\includegraphics[trim=1.2cm 0.0cm -2.7cm 0.9cm, clip=true, scale=0.222]{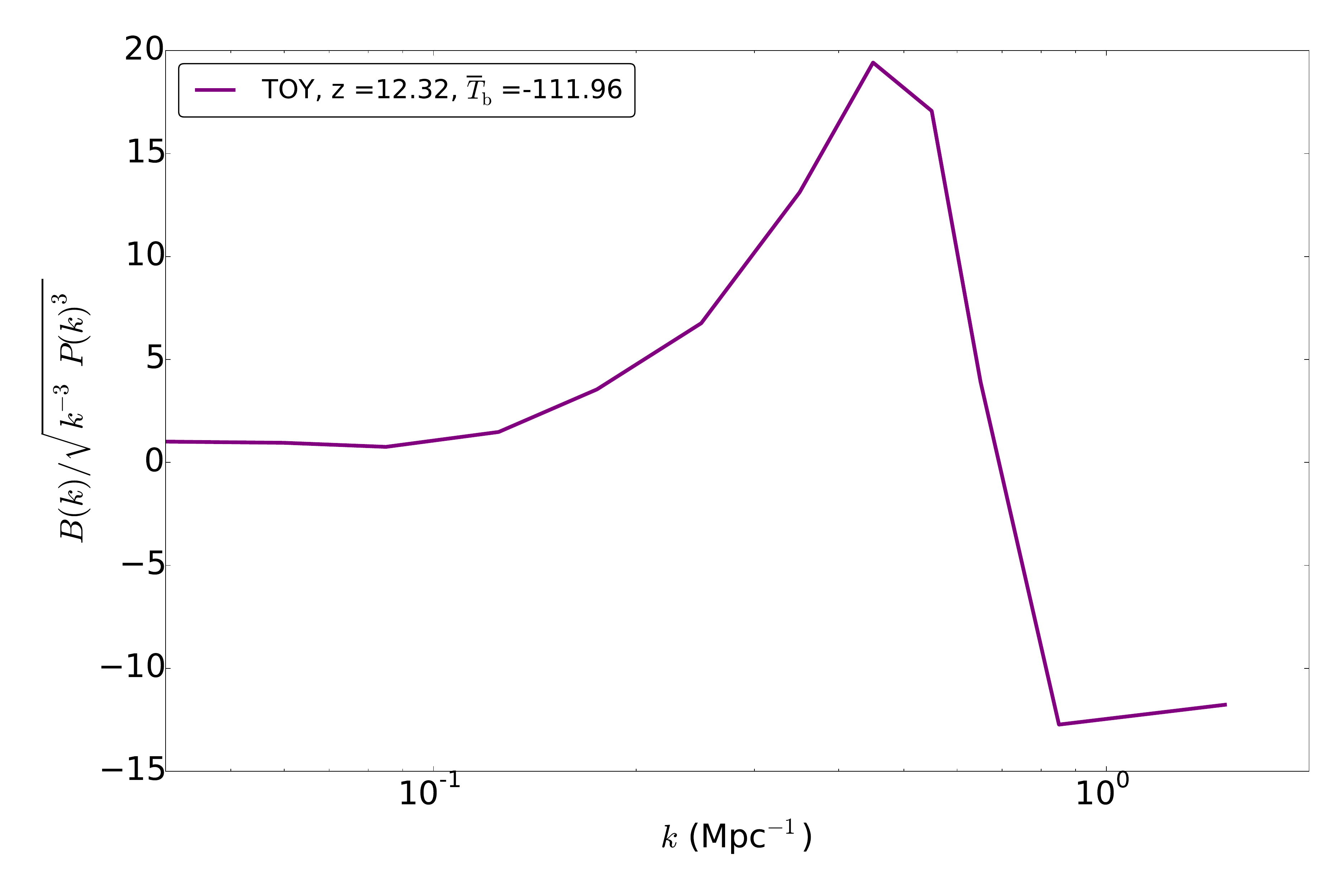}\\
  \end{array}$
  \caption{Normalised bispectrum for the equilateral configuration from the \textit{HMXB+QSO} simulation (\textbf{top}),
  and the \textit{QSO} simulation (\textbf{middle}) and \textit{toy} simulation (\textbf{bottom}).
  In the \textit{HMXB+QSO} simulation we see a similar feature of a turnover moving from large to small scales as in \hmxb.
  The amplitude is much bigger at early times than in \hmxb as the heated profiles from QSO's are very spherically symmetric.
  The \textit{QSO} simulation (\textbf{middle}) does not exhibit a turnover that correlates
  with the typical separation of emission regions (marked with the dotted vertical lines for,
  from left to right, $z =$ {15.13, 13.56, 12.60, 12.32}).
  The \textit{toy} simulation has randomly scattered Gaussian heated profiles
  designed to roughly reproduce the properties of \qso.
  The turnover we see in the \textit{toy} simulation's bispectrum (\textbf{bottom})
  is very similar to that seen in \qso (\textbf{middle}),
  which indicates that the shape and size of the heating profiles around QSOs is
  a major driver of this feature, with clustering playing a subdominant role.
  }
  \label{fig:QSOonlyQ}
\end{figure}

\begin{figure*}
\begin{minipage}{176mm}
\begin{tabular}{c}
  \includegraphics[trim=1.8cm 0.1cm 0.75cm 0.38cm, clip=true, scale=0.5]{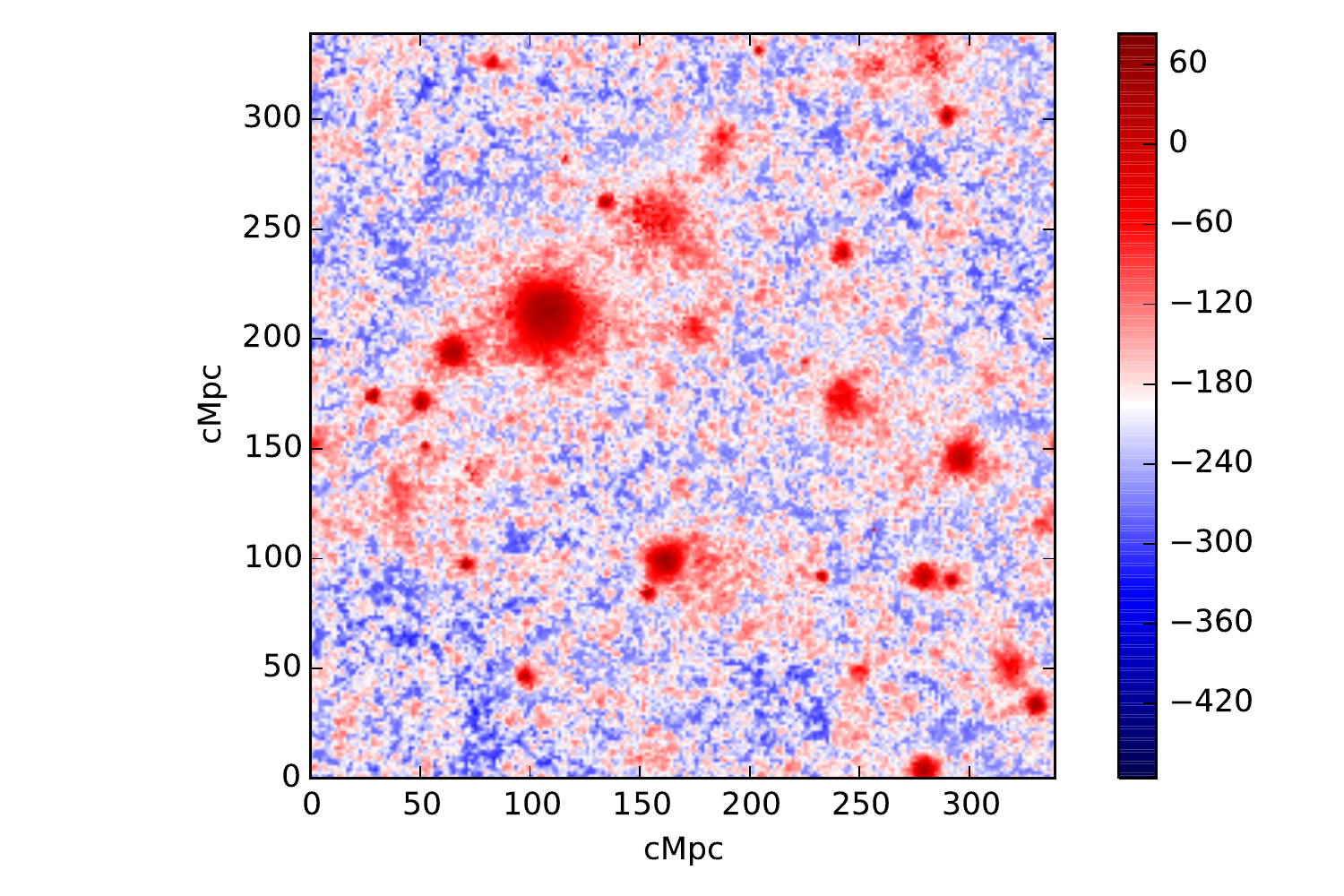} \includegraphics[trim=3.45cm 0.1cm 0.75cm 0.38cm, clip=true, scale=0.5]{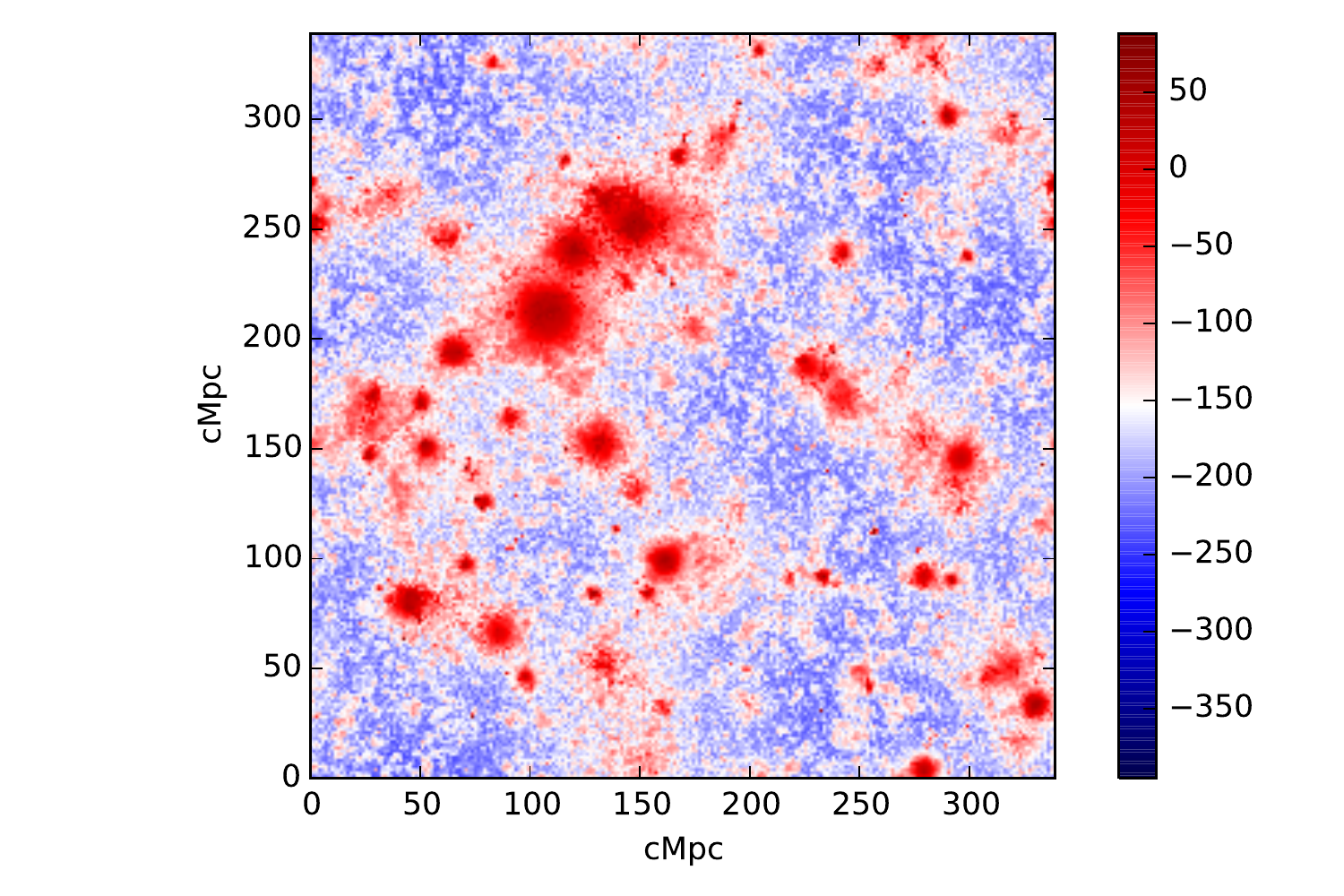} \includegraphics[trim=3.45cm 0.1cm 0.75cm 0.38cm, clip=true, scale=0.5]{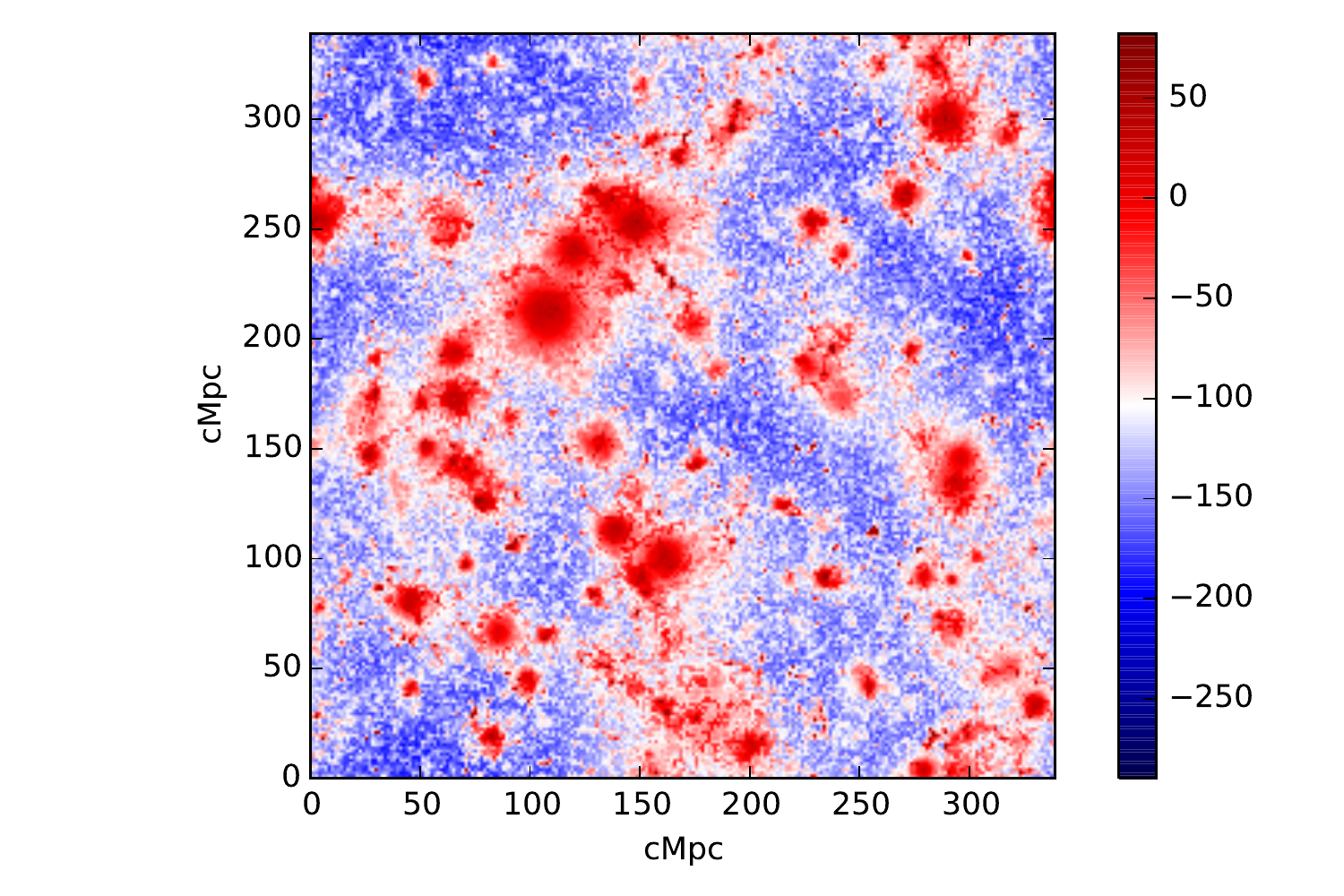} \\
\end{tabular}
\caption{Slices taken from the QSO simulation during a similar regime
where the background brightness temperature is increasing.
Corresponding from left to right to $z=15.13, 13.56, 12.32$ and $\overline{T}_{\mathrm{b}} = -196.42, -153.41, -103.77$ mK.
The heated profiles around quasars are more well defined in comparison with those of the \textit{HMXB} simulation.
QSO sources are also fewer, more isolated and generate more spherically symmetric heating profiles.
}\label{fig:QSOmaps}
\end{minipage}
\end{figure*}

First we consider what happens when AGN (aka QSOs) are allowed to contribute
to the X-ray heating budget, as per the \textit{HMXB + QSO} and \textit{QSO}
simulations described in Section~\ref{sec:sources}.
We show the equilateral normalised bispectrum for these two simulations in the top and middle
panels of Fig.~\ref{fig:QSOonlyQ}.
The top panel is from \textit{HMXB + QSO} simulation;
the normalised bispectrum of this simulation exhibits a very similar turnover
feature shifting to smaller scales with decreasing redshift.
However, there are differences, the heating process kicks in earlier and produces a large bispectrum at $k\sim 0.07$~Mpc$^{-1}$,
which maximises at $z=18.54$
(when the contrast between the hottest and coldest pixels is maximised in \hmxb).
The amplitude is greater than it is in the \hmxb simulation,
which is to be expected as QSO's produce a more spherically symmetric heated profile
(note that because of this, the equilateral configuration also has a much larger
normalised bispectrum relative to that of other configuations).
Another difference is the boost in non-Gaussianity at $k\sim 0.07$~Mpc$^{-1}$ seems to persist to lower redshifts,
which must be driven by the distribution of QSO heating profiles.
The contribution of the QSO distribution and profile shape quickly gets washed out by the HMXB heating profiles.
The bispectrum therefore drops in amplitude from $z=18.54$ and then by $z=17.22$,
looks very similar to the \hmxb bispectrum.

The normalised bispectrum of the \textit{QSO} simulation (see the middle panel of Fig.~\ref{fig:QSOonlyQ})
exhibits more isolated heated regions with very spherically symmetric profiles around each QSO;
see the maps in Fig.~\ref{fig:QSOmaps}.
We therefore would not necessarily expect that it would exhibit the same qualitative bispectrum evolution
as the \hmxb simulation.
Indeed the spherically-averaged equilateral normalised bispectrum of the \textit{QSO} simulation
is quite different and it is therefore useful to compare it with that of the \textit{HMXB} simulation.
Instead of a single turnover, there is a multimodality to the bispectrum,
dominated by an early turnover at $k\sim 0.07$~Mpc$^{-1}$
(similar to that seen in the \textit{HMXB + QSO} simulation, but with a lower amplitude).
Later the bispectrum becomes dominated by a turnover at smaller scales,
peaking around $k\sim 0.4$~Mpc$^{-1}$.
There does not seem to be a clear correlation between the typical separation of emission
regions (shown with the vertical dashed lines in Fig. \ref{fig:QSOonlyQ}) and
the features we see in the \qso simulation,
as was the case for the \hmxb simulation.

It is not possible to say how much of the non-Gaussianity we see in the \qso simulation
comes from the distribution of heated profiles and how much from the profile shapes.
We therefore look at the bispectrum for randomly distributed QSO-like heating profiles,
by constructing a toy model in which spin-temperature profiles around
randomly-distributed sources are modelled as a Gaussian.
This produces a brightness-temperature profile that is qualitatively similar to model B for mini-QSO in \citet{Ghara2015}.
Before populating a datacube with source profiles,
every pixel is assigned a fixed background spin temperature in line with
the lowest brightness temperatures seen in the \qso simulation
(assuming a mean density and fully neutral IGM).
We then randomly distribute Gaussian spin-temperature profiles,
sampling the $\sigma$ of the profile from a triangular function
(whose mode and maximum are chosen to reproduce the most common-sized and maximal $\Tb$ profiles
we observe in the \qso simulation).
We set the minimum of our triangular selection function to $\sigma=0$,
and choose a mode and maximum $\sigma$ to produce an above-average $\Tb$ profiles
with a mode of $R=7$ Mpc and maximum $R=12.5$~Mpc.
Note that we did not tweak these values at all to tune the resulting bispectrum.
The number of sources was fixed so that at $z=13.55$ the average brightness temperature
in the toy datacube matched the original \qso simulation.
We find that despite merely rising the background brightness temperature to produce a
toy datacube at $z=12.32$,
the mean brightness temperature of the toy ($\Tb = -111.96$~mK) matches well with \qso ($\Tb = -103.77$ mK).

We show the normalised bispectrum for $z=12.32$ from such a toy model
in the bottom panel of Fig. \ref{fig:QSOonlyQ}.
Note we do not plot other redshifts, because the $\Ts$ profiles are quite narrow and so,
for the range of redshifts we consider in this section,
raising the background brightness temperature does little in changing the size of the
resulting $\Ts$ profile size, i.e. the bispectrum is unchanging with redshift.
This turnover in the bispectrum over a fixed scale range with increasing background brightness
temperature is consistent with what we see in the \qso bispectrum,
once the source number has reached a point at which there are QSOs in most halos
(see the yellow dashed line with circles, the orange dotted line with stars,
and the red dot-dashed line with upturned triangles in the middle panel of Fig. \ref{fig:QSOonlyQ}).
We can see that this regime (where the turnover becomes fixed in scale) is associated
with the source number becoming roughly constant by simply comparing the middle
map of Fig. \ref{fig:QSOmaps} ($z=13.56$; yellow dashed-line with circles)
with the left map ($z=15.13$; green dotted line with triangles)
and right map ($z=12.32$; red dot-dashed line with upturned triangles).
The left map has fewer heated regions than the other two
(which look very similar despite being $\sim 50$~mK apart in their mean brightness temperatures),
and these are on average bigger than those seen at the lower redshifts.

Whilst the turnover in the toy model peaks at the same scales as that in \qso,
it is much sharper and on small scales falls off to negative amplitude on larger $k$
(i.e. on small scales under-densities are driving the bispectrum).
Therefore we conclude that the bispectrum we see in the \qso simulation
must be sensitive to both the profile size and the distribution of profiles,
with the scale at which the late-time bispectrum peaks corresponding to the bubble
profile size.
The reason we do not see such sensitivity to a profile size so clearly
in the \hmxb simulation is because the profiles are less well defined
and do not exhibit a strong characteristic scale
(compare the \hmxb maps in Fig.~\ref{fig:HMXBmaps} with the \qso maps in~\ref{fig:QSOmaps}).

\begin{figure}
\centering
  $\renewcommand{\arraystretch}{-0.75}
  \begin{array}{c}
    \includegraphics[trim=1.3cm 0.05cm 0.0cm 2.3cm, clip=true, scale=0.255]{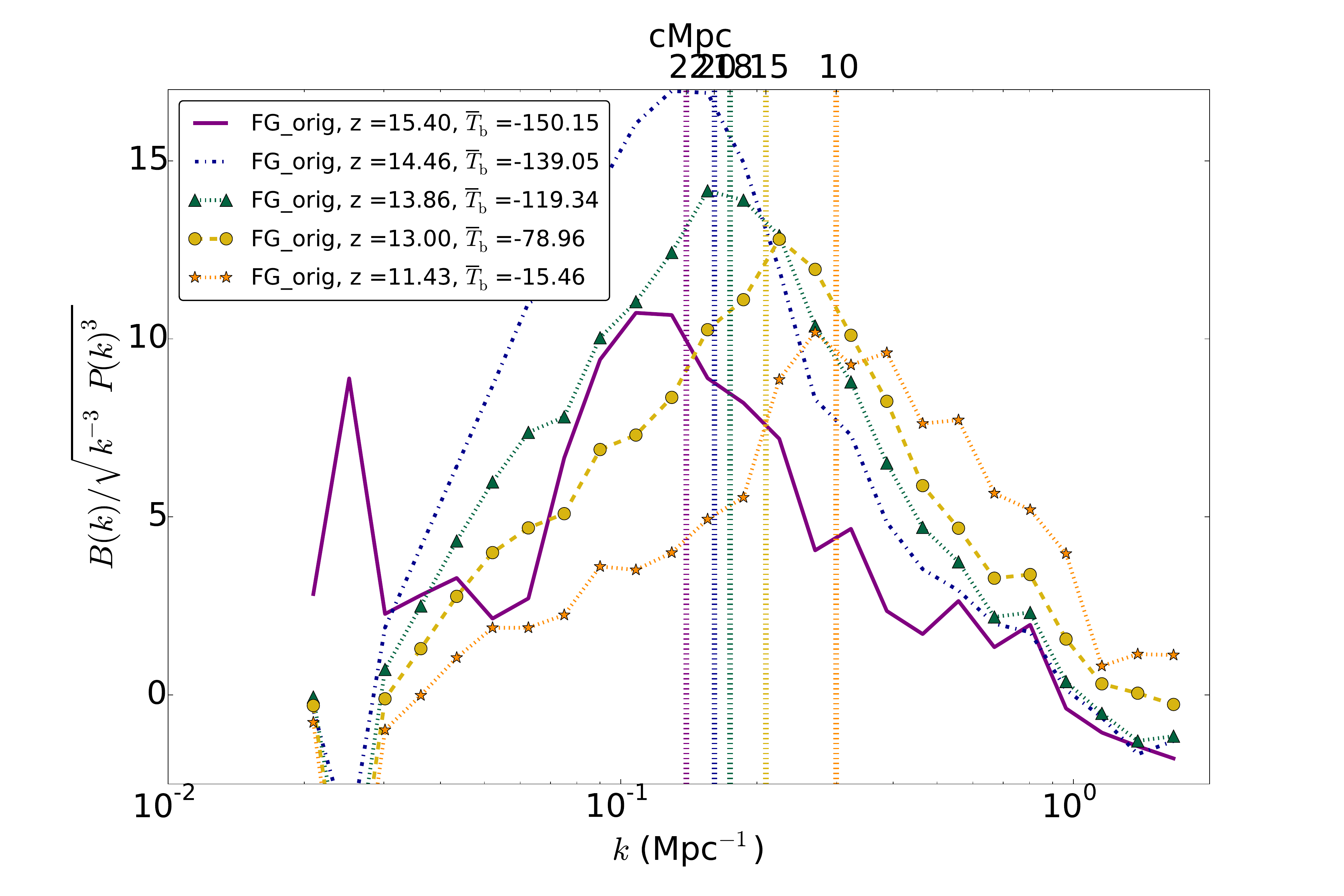}\\
  \end{array}$
  \caption{Normalised bispectrum for the equilateral configuration, for
  $z = {15.40, 14.46, 13.86, 13.00, 11.43}$ when $\xh = {1.00, 0.99, 0.99, 0.98, 0.95}$ in the Faint Galaxies 21cmFAST simulation.
  We see a qualitatively very similar evolution during the epoch of heating as seen in the \textit{HMXB} with a turnover associated with the typical separation of region $\delta T_{\mathrm{b}} >0$.
  Reionization commences before this simulation reaches a stage at which it the bispectrum is driven solely by the density field.}
  \label{fig:tocmfastFG}
\end{figure}

Next we check whether the features we see in the normalised bispectrum
from \hmxb are seen in semi-numerical simulations.
To do so we utilise one of the most popular semi-numerical simulations of the epoch of
heating and reionization - \cmfast
(we refer readers to \citet{Signal2010} for details on this code).
We have measured the equilateral bispectrum from two contrasting \cmfast simulations -
namely the \textit{faint galaxies} and \textit{bright galaxies} simulations from \citet{Greig2017a}.
The simulations we consider were generated for another project and so have similar,
but not identical, resolution (200 pixels per 300 Mpc side).
Fig.~\ref{fig:tocmfastFG}, shows the equilateral normalised bispectrum as a function of
$k$, we only show the \textit{faint galaxies} simulation for the sake of brevity.\footnote{The \cmfast normalised bispectrum is much smoother with $k$ than
that of the \textit{HMXB} simulation, despite the fact we used the same binning for both bispectra analysis.
This likely stems from fundamental differences in the way semi-numerical and
numerical codes operate.
Semi-numerical codes average over the density field on varying scales in order to
perform the integrals associated with coupling and heating,
as well as to numerically apply the \citet{Furlanetto2004a} excursion-set model
for reionization.
It is easy to see how statistics from such an approach would be less "noisy"
than a fully numerical simulation.}
In both simulations, we see qualitatively similar evolution of the normalised
bispectrum seen in Fig.~\ref{fig:bvsk};
i.e. a positive turnover forming on large scales (small $k$) during the early stages of heating,
which then drops in magnitude as it shifts to smaller scales with reducing redshift.
This turnover again correlates well with the typical separation of emission regions during this phase
(which are again overplotted with dotted lines whose colour-redshift relation agrees with
that of the legend).
The \textit{faint galaxies} and \textit{bright galaxies} models were chosen by
\citet{Greig2017a} to create contrasting simulated
datasets for 21CMMC paramater studies, which suggests that such features should be qualitatively generic so long as
X-rays sources are hosted by most star-forming haloes,
as the case with HMXBs.

Neither \cmfast simulations reach a stage in which we see the monotonic
increase in the normalised bispectrum with $k$ associated with the late phases of the heating
process when the influence of the density field on the bispectrum is becoming substantial.
It is very likely that this is because in both \cmfast models,
reionization has started before the stage at which this feature in the
\hmxb simulation sets in.
As seen by \citet{Majumdar2017}, the bispectrum becomes negative over a range of scales once reionization
commences,
we also see similar behaviour in the normalised bispectrum from the \cmfast simulations we have considered
when the ionized fraction becomes substantial.
We defer analysis of the normalised bispectrum during the EoR to future work
as the focus of this work is the EoH.

\citet{Shimabukuro2016a} have also studied the bispectrum during the epoch of heating
and reionization as predicted by \cmfast.
However, it is hard to compare their results with our \hmxb analysis as
they do not provide the brightness-temperature evolution of the semi-numerical \cmfast
simulation that they analyse.
The statistic they use is also different from ours.
In our paper, we use only the real part of the FFTed signal in calculating our bispectrum.
This is because we use the FFT estimator of \citet{Watkinson2017} with real FFTs,
and so our bispectrum is forced to be real.
This is reasonable as the imaginary term will cancel out in any binned calculation
of the bispectrum.
On the other hand, \citet{Shimabukuro2016a} measure
$\mathrm{abs}[B(k)] = \sqrt{\mathrm{Re}[B(k)]^2 +  \mathrm{Im}[B(k)]^2 }$,
which is not technically speaking the bispectrum, even if you were to include the
imaginary contribution.
Their Fig. 1, which plots the equilateral $k^6/(2\,\pi)\,\mathrm{abs}[B(k)]$
with $k$, looks quite different to $k^6/(2\,\pi)^2\,B(k)$ from the \hmxb simulation
(provided in the bottom panel of our Fig.~\ref{fig:normZoo} in Appendix \ref{app:normdicuss}).
The amplitude of their statistic varies with redshift, but does not vary much with
scale at a given redshift (i.e. the bispectrum is flat) except for one redshift
at which it exhibits a monotonic increase with $k$.

\section{Detectability of the Bispectrum}\label{sec:detect}
We have shown that the bispectrum should contain valuable information unavailable
from the power spectrum, however it is also more difficult to detect.
Therefore, for the remainder of this paper we will examine the detectability of the features
discussed in preceding sections of this paper.

There will likely be residuals in 21-cm datasets after calibration and foreground removal,
and we will consider the impact of these on the bispectrum in future works.
But in the absence of consensus on the best methods for mitigating foregrounds and instrumental effects,
we feel it is reasonable, for the purposes of this work,
to consider a best-case scenario where the noise on the bispectrum is due
solely to instrumental noise and sample variance.

Instrumental noise is Gaussian and so has a bispectrum of zero.
The covariance of the noise bispectrum is however not zero and therefore contributes to the error $\Delta_{\mathrm{N}}B(\vect{k}_1, \vect{k}_2, \vect{k}_3)$ on our measurement of the bispectrum.
For a Gaussian field, it is possible to write the covariance of its bispectrum $B_{\mathrm{N}}$ as,
\begin{equation}
\begin{split}
\mathrm{Cov}\left[ B_{\mathrm{N}}(\vect{k}_1, \vect{k}_2, \vect{k}_3)\,B_{\mathrm{N}}(\vect{k}_1, \vect{k}_2, \vect{k}_3) \right] &= \left[\Delta_{\mathrm{N}}B(\vect{k}_1, \vect{k}_2, \vect{k}_3)\right]^2 = \\
k_f^3\,\frac{s_{123}}{V_{123}}\,P(k_1)\,P(k_2)\,P(k_3)\,,\\
\end{split}\label{eqn:err}
\end{equation}
where $k_{\rm f}=2\,\pi/L$ is the fundamental $k$ scale, $V_{123} \approx 8.0\pi^2\,k_1\,k_2\,k_3\,(s\,k_{\rm f})^3$
is the the number of fundamental triangles in units of $k_{\rm f}^3$, $s\,k_{\rm f}$ is
the binwidth, and $s_{123} = 1, 2, 6$ for general, isoceles and equilateral triangle configurations respectively; see \citet{Scoccimarro1998a, Scoccimarro2004, Liguori2010}.
This is a convenient way to measure the noise on the bispectrum as it allows us to us
to utilise existing power-spectrum error-estimation pipelines.

We can also use Eqn.~\ref{eqn:err} to estimate the error contribution from sample variance $\Delta B_{\mathrm{sv}}$ a statistical error deriving from the limited sample volume of any observation.
This sample variance is generally assumed to be Gaussian when estimating power spectrum errors and the error due to this is taken to be proportional to the 21-cm power spectrum.

\citet{Mondal2016a} show that the non-Gaussianity of the signal must be taken into account when calculating the sample variance error on the 21-cm power spectrum during reionization (see also \citealt{Mondal2015} and \citealt{Mondal2015a}).
There is therefore strong motivation to perform similar studies into the sample-variance
error on the bispectrum.
However, for the purposes of this work,
where we are simply after an order of magnitude approximation,
the Gaussian approximation to the sample-variance covariance will suffice.

We use \toolscm to generate noise cubes in Fourier space, sample using the $uv$ footprint of SKA-LOW and natural weighting,
and then measure the power spectrum.\footnote{tools21cm maybe downloaded from
here https://github.com/sambit-giri/tools21cm} \toolscm uses the noise
and telescope models of \citet{Giri2018a} who assume SKA-LOW will be composed of
a total of 512 antenna with a diameter of $D_{\mathrm{stat}} = 35$~m, with 224 randomly distributed in a core of radius 500~m.
The rest of the antenna are arranged in 48 clusters (each with 6 randomly placed stations)
lying on a three-arm spiral with a total radial extent of 35~km from the core centre.
We refer the reader to \cite{Ghara2017a} and \citet{Giri2018a} for details on this.
We assume a total integration time of 1000~hours and a bandwidth of 8~MHz.
We calculate the box length $L$ that would correspond to the survey volume
(which we calculate using \ccalc\footnote{\url{http://cxc.harvard.edu/contrib/cosmocalc/} })
assuming that the FoV of SKA is $\Omega_{\mathrm{FoV}}=\lambda^2/D_{\mathrm{stat}}$.

We calculate the error on the bispectrum due to sample variance according to
Eqn.~30 of \citet{Mondal2015a} (which is equivalent to Eqn.~9 in \citealt{Mellema2013b}), i.e.
\begin{equation}
\begin{split}
P_{\mathrm{sv}}(k) = \frac{(2\pi)^2\,P(k)^2}{ L^3\,k^2\,s\,k_{\mathrm{f}} }\,.
\end{split}\label{eqn:SVerr}
\end{equation}

Our total error on the bispectrum is then given by $\Delta B = \Delta B_{\mathrm{N}} + \Delta B_{\mathrm{sv}}$.
However, we need the error on $b(k_1,k_2,k_3) = B(k_1,k_2,k_3)/\sqrt{(k_1\,k_2\,k_3)^{-1}P(k_1)\,P(k_2)\,P(k_3)}$.
In principle, there are correlated errors on the power spectrum that we should worry about,
but as long as the error is dominated by the bispectrum,
then we can approximate $\Delta b(k_1,k_2,k_3) = \Delta B / \sqrt{(k_1\,k_2\,k_3)^{-1}P(k_1)\,P(k_2)\,P(k_3)}$ \citep{Scoccimarro2004}.

Note that we have checked that error calculation on the bispectrum as calculated using
Eqn.~\ref{eqn:err} is consistent with the noise bispectrum sensitivity
calculations of \citet{Yoshiura2015}.
However, our errors are slightly larger on smaller scales which is to be expected as the number
of core antenna we assume is roughly half that used by \citet{Yoshiura2015}
to be in keeping with the latest SKA design specifications.

\begin{figure}
\centering
  $\renewcommand{\arraystretch}{-0.75}
  \begin{array}{c}
    \includegraphics[trim=1.0cm 0.05cm 0.0cm 0.15cm, clip=true, scale=0.2275]{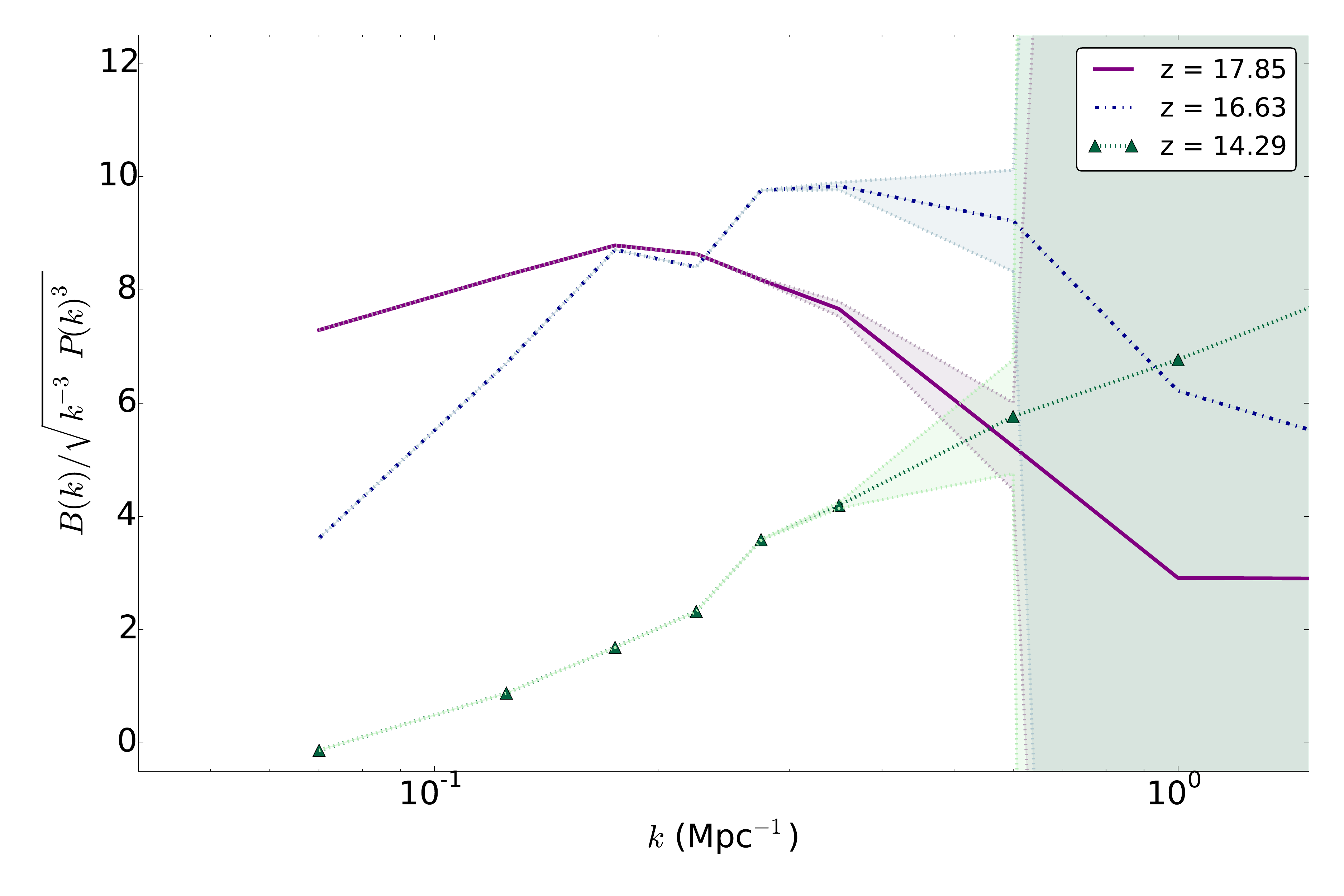}\\
  \end{array}$
  \caption{Normalised bispectrum for the equilateral configuration, for
  $z = {17.85, 16.63, 13.22}$. This has been binned using bin edges
  in $k$/Mpc$^{-1}$ of bin edges [0.04, 0.1, 0.15, 0.2, 0.25, 0.3, 0.4, 0.8, 1.2, 1.8, 2.5].
  The noise error at each redshift is depicted by the shaded regions.
  If foregrounds and instrumental effects can be mitigated,
  we should have good sensitivity to the bispectrum at $k<0.6$ Mpc$^{-1}$.
  Note the normalisation masks the appearance of the noise getting stronger
  with redshift, instead highlighting the detectability of the signal.}
  \label{fig:noise}
\end{figure}

Fig.~\ref{fig:noise} shows the normalised bispectrum from the early ($z=17.85$), mid ($z=16.63$) and late ($z=13.22$) stages of the heat process.
We have overplotted shaded regions that correspond to the normalised-bispectrum noise error for each redshift.
On top of the binning over cos$(\theta)\pm 0.05$ as done in the rest of this paper we
bin the statistic further in $k$ (see the figure caption for binning details).
Note that the normalisation step masks the usual trend of error magnitude getting stronger with
redshift, instead highlighting the detectability which connects
to the amplitude of the signal as much as to the noise level itself.
We find that, if foreground and instrumental effects are successfully mitigated,
we should have sensitivity to the bispectrum at $k<0.6$~Mpc$^{-1}$,
the gradient and amplitude evolution in this $k$ range would provide us with
valuable information about the timing and nature of heating.
Importantly, it is in these early stages and at these scales that the impact on the bispectrum
of including QSOs on top of HMXBs is most predominant.
We therefore conclude that the 21-cm bispectrum should provide a valuable tool for understanding
the properties of stars and galaxies, even during the epoch of heating.
As shown in \cite{Watkinson2015a}, the skewness should also provide a useful probe of the epoch of heating.
Given that reality will likely make detecting the bispectrum harder than we find here,
it is likely that the skewness will have a role to play in combination with the power spectrum and bispectrum.

\section{Conclusions}\label{sec:conc}

In this paper we have presented analysis of the 21-cm normalised bispectrum from fully-numerical
simulations of the epoch of heating, assuming that the only source of X-rays is HMXBs.
In the associated appendix we have also shown that our choice of bispectrum
normalisation is the best option for analysing 21-cm data.
We have found that if HMXB-like X-ray sources drive heating, then the equilateral
bispectrum will be strongest in amplitude compared to other configurations and
will exhibit a turnover that shifts from large to small scales with reducing redshift.
We find that the scale at which this turnover peaks is correlated with
the typical separation of emission regions.
It is clear from our analysis that the bispectrum is driven by a complex interplay
between the shape and size of heated profiles and their distribution.
Cross-terms between the density field and spin temperature dominate at early times
reflecting this complex interplay.
As X-rays heat the cooler regions of the maps, small-scale sub-structure in the
heated regions start to dominate the 21-cm bispectrum, introducing more power on smaller scales
than on large.
Ultimately, by the end of the simulation, fluctuations in the density field totally
dominate the 21-cm bispectrum.

We consider how generic the qualitative evolution of the bispectrum is by analysing
two contrasting semi-numerical simulations.
We observe very similar qualitative behaviour as in the numerical simulation in which
HMXBs dominate the evolution.
We also consider how the bispectrum is changed if QSOs are included into the
numerical simulation, providing a second source of X-rays.
At early times the presence of QSOs produces a stronger equilateral bispectrum,
but still exhibits a turnover that shifts to smaller scales with decreasing redshift.
By the mid phases of the heating process its normalised bispectrum is
indistinguishable from that of the HMXB simulation.
By analysing a third numerical simulation in which only QSOs provide X-ray radiation,
we show that the bispectrum will look quite different than it would if HMXBs
(or a similarly wide-spread source of X-rays) drive heating.
At early times clustering of sources introduces a large-scale turnover feature.
This drops in amplitude as the contrast between the most hot and the most cold
regions decrease and is replaced by a turnover that is driven by the typical size
of the heated profiles surrounding the heating sources.

We consider the observability of the bispecrum with phase-1 of SKA-LOW and find
that, assuming foregrounds and instrumental effects are effectively mitigated,
we should be able to detect the bispectrum during the Epoch of Heating at $k<0.6$ Mpc$^{-1}$.
Measuring the bispectrum should therefore provide a major boost to the information
available from the power spectrum alone.
Further work is required to get a better handle on the effect of sample variance
and other complications to observing statistics such as the bispectrum;
for example calibration and foreground removal residuals, and beam effects.

\section*{Acknowledgements}
CAW would like to thank Suman Majumdar, Claude Schmit and Cathryn Trott
for useful discussions surrounding the interpretation of the bispectrum.
CAW and JRP acknowledge financial support from the European Research Council
under ERC grant number 638743-FIRSTDAWN.
II \&HR would like to acknowledge support by the Science and Technology Facilities Council
(grant numbers ST/P000525/1 and ST/I000976/1) and the Southeast Physics Network (SEPNet).
GM \& SKG acknowledges support by the Swedish Research Council grant
2016-03581. We acknowledge that the results in this paper have been achieved using the PRACE Research Infrastructure resources Curie based at the Tre\'s Grand Centre de Calcul (TGCC) operated by CEA near Paris, France and Marenostrum based in the Barcelona Supercomputing Center, Spain.
Time on these resources was awarded by PRACE under PRACE4LOFAR grants 2012061089 and 2014102339 as well as under the Multi-scale Reionization grants 2014102281 and 2015122822. Some of the numerical computations were done on the Apollo cluster at The University of Sussex.


\bibliographystyle{mn2e}


\appendix
\section{Other Bispectrum normalisations}\label{app:normdicuss}
Throughout this paper we have focussed on what we call the "normalised bispectrum",
but there are several other normalisation choices for the bispectrum.
We will use this appendix to illustrate why we find the normalised bispectrum
to be the best choice for 21cm analysis, mainly because it suppresses random flips
in sign when the data is close to non-Gaussianity by removing the contribution of
the power spectrum to the bispectrum amplitude.

In cosmology it is common to consider either the raw bispectrum $B(k_1, k_2, k_3)$, the reduced bispectrum defined as,
\begin{equation}
Q(k_1, k_2, k_3) = \frac{B(k_1, k_2, k_3)}{[P(k_1)\,P(k_2) + P(k_1)\,P(k_2) + P(k_1)\,P(k_3)]}\,,\label{eqn:Qk}
\end{equation}
or the dimensionless bispectrum $(k_1, k_2, k_3)^2/(2\,\pi^2)\,B(k_1, k_2, k_3)$ e.g. \citet{Scoccimarro1999, Shimabukuro2016a, Majumdar2017}.

\begin{figure}
\centering
  $\renewcommand{\arraystretch}{-0.75}
  \begin{array}{c}
  	\includegraphics[trim=1.2cm 3.1cm 0.0cm 2.5cm, clip=true, scale=0.25]{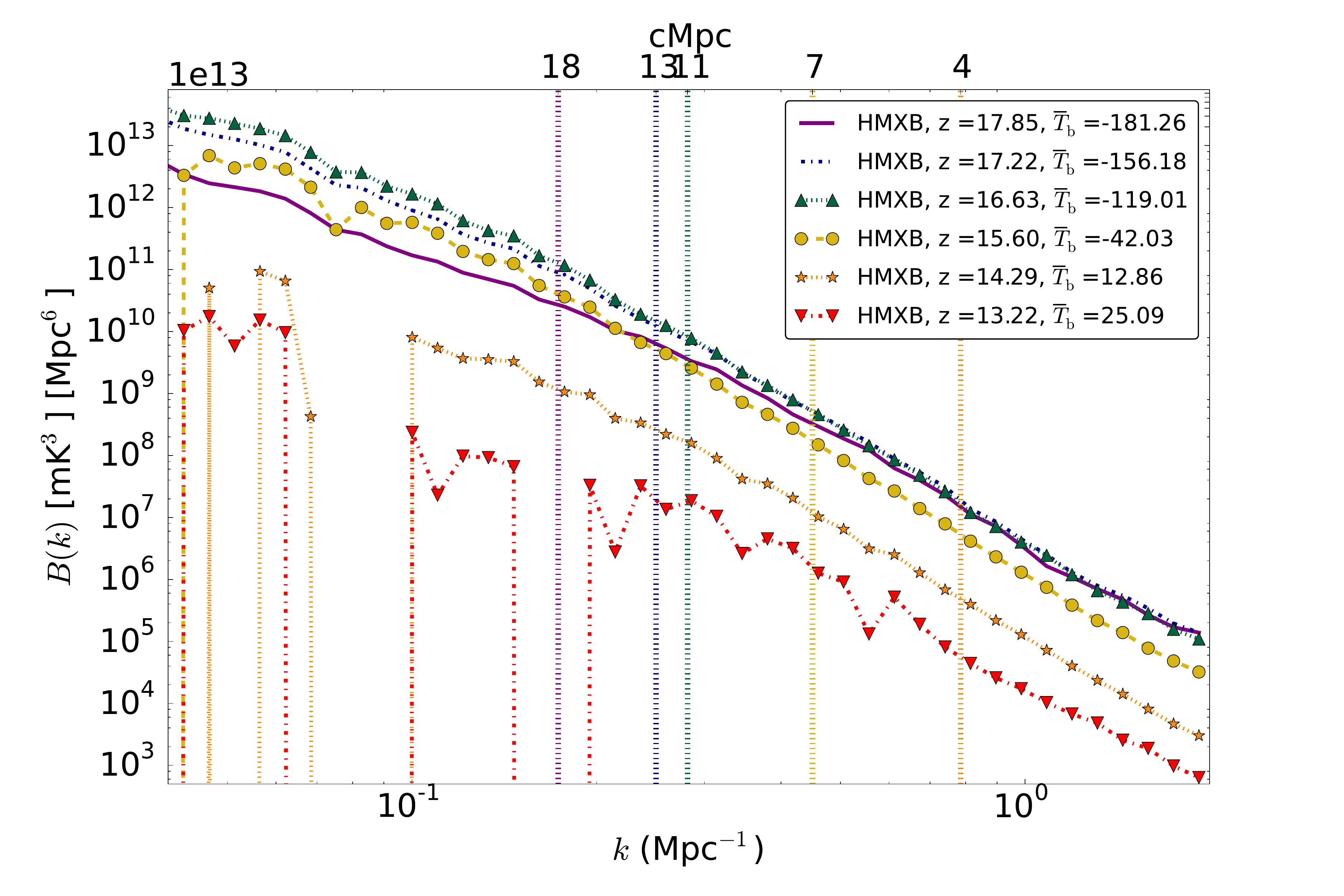}\\
    \includegraphics[trim=1.2cm 0.5cm 0.0cm 2.3cm, clip=true, scale=0.25]{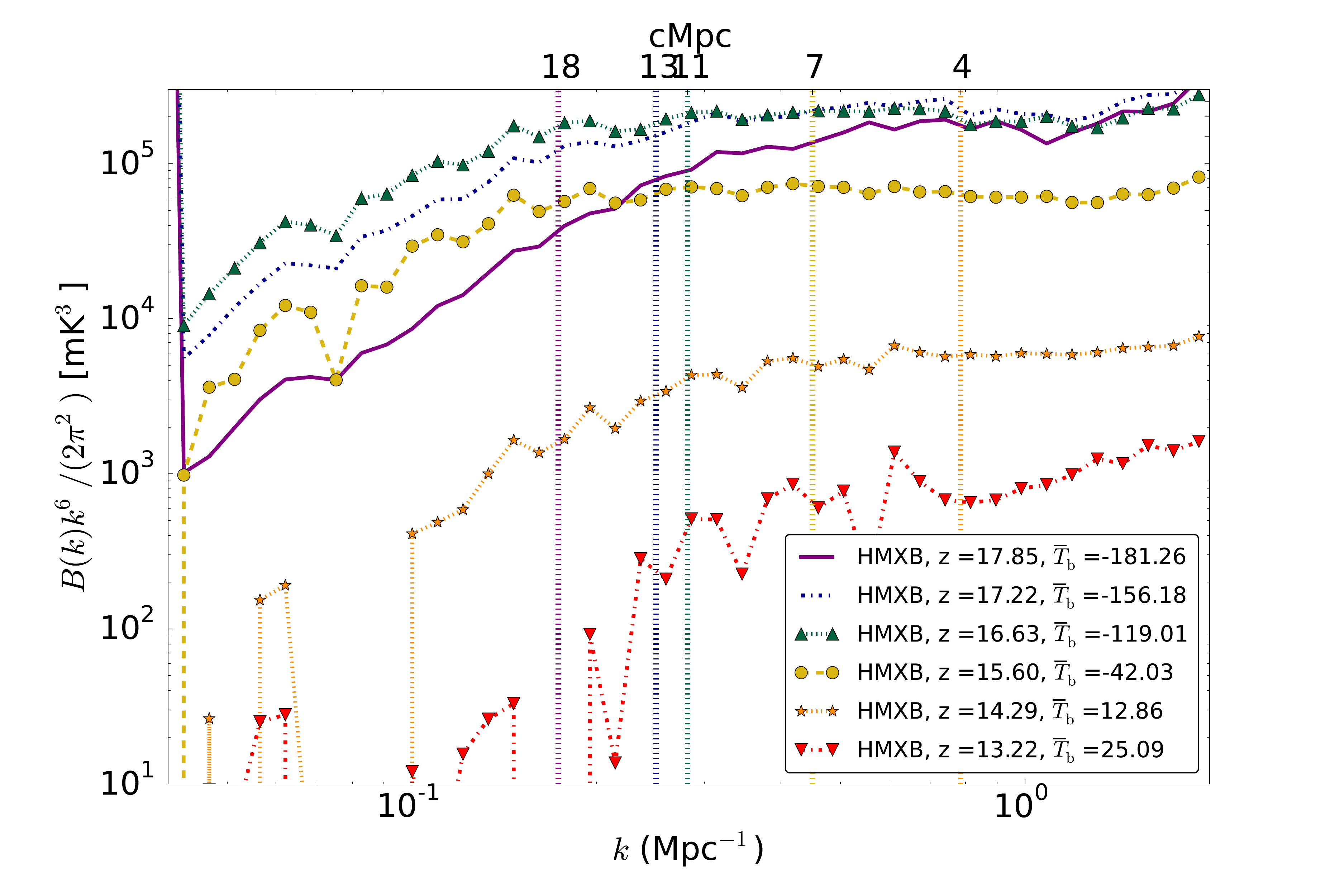}\\
  \end{array}$
  \caption{\textbf{Top} - Spherically-averaged $B(k)$
  with $k$ for the equilateral configuration of $k$ vectors for the \textit{HMXB} simulation.
  \textbf{Bottom} - As top, but including a normalisation factor of $k^6/(2\,\pi^2)$ to $B(k)$.
  Vertical dotted lines correspond to the scales associated with characteristic
  radius of above-average $\delta T_{\mathrm{b}}$ regions as measured by the mean-free-path method.
  }\label{fig:normZoo}
\end{figure}

We plot the spherically-averaged raw bispectrum $B(k)$ of the \textit{HMXB} simulation for the equilateral configuration
is shown in the top plot of Fig. \ref{fig:normZoo}.
The bispectrum of the brightness-temperature field has units of mK$^3$ Mpc$^{6}$.
In the bottom plot of Fig. \ref{fig:normZoo} we have normalised out the volume dimension
of the statistic by instead plotting the dimensionless bispectrum
$(k_1, k_2, k_3)^2/(2\,\pi^2)\,B(k_1, k_2, k_3)$ (with units of mK$^3$).
The first thing to note about both these statistics is that they exhibit
wild fluctuations from positive to negative amplitude at certain redshifts and scales;
see the red dot-dashed line with inverted triangles for $k<0.2$ Mpc$^{-1}$
and the orange dotted line with stars at $k<0.15$ Mpc$^{-1}$ in both plots of Fig. \ref{fig:normZoo}.
This occurs as the contribution to the statistic from non-Gaussianity is oscillating around zero.
There is then a strong non-zero amplitude coming from the power in the maps.
If more excessive binning is used, these flips in sign can produce spurious features in the statistic.
It is for this reason that we strongly advocate the use of the normalised bispectrum described
in the main part of this paper as it isolates the contribution due to non-Gaussianity
in the bispectrum and therefore does not suffer from such artificial features.

Comparing the raw bispectrum with the dimensionless bispectrum, we see that the monotonic drop from large (small $k$) scales to small (large $k$) is ultimately due to dimension rather than anything physical in the map.\footnote{The prefactor in the
$(k_1, k_2, k_3)^2/(2\,\pi^2)\,B(k_1, k_2, k_3)$ normalisation derives from the spherically averaging the bispectrum.
The area under this function is connected to the skew as a function of d ln$k$.}
A similar evolution from high amplitude at large $k$ to low amplitude at small $k$ in the raw $B(k)$ is seen in the plots of \citet{Majumdar2017},
who studies the spherically-averaged raw bispectrum during reionization.
During reionization it is ionized regions (therefore below-average $\delta T_{\mathrm{b}}$ regions)
that introduce non-Gaussianities beyond that from the density field,
and so the EoR bispectrum is negative on many scales.
During the EoH, we find that the bispectrum is positive;
this tells us that it is the heated regions, i.e. the regions that are above-average $\delta T_{\mathrm{b}}$,
that are introducing non-Gaussianity to the maps.

There is some sense from the evolution in the large-scale power of $k^6/(2\,\pi^2)\,B(k)$
(see the turnover evolving in the bottom plot of Fig. \ref{fig:normZoo}) that there is some characteristic scale in
the \hmxb simulation that gets bigger and then smaller with decreasing redshift.
We saw a similar evolution in scale is seen in Fig. \ref{fig:hist_vs_z}
where we plot the PDF of the characteristic radius of above-average $\delta T_{\mathrm{b}}$ regions.
However if we translate the mean of these PDFs to $k$-scales $k=2\pi/(4\,R)$ and mark these onto the bottom plot of Fig. \ref{fig:normZoo} (vertical dotted lines whose colour defines the redshift), we see that there is not a clear cut connection with the features seen in this statistic, even qualitatively.

$Q(k_1, k_2, k_3)$ is motivated by large-scale structure studies as work on non-linear perturbation theory predicted that the
density-field bispectrum should exhibit non-Gaussianities such that
$B_{ \mathrm{tree} }(k_1, k_2, k_3) = 2\,F_2\,(k_1, k_2)\,P(k_1)\,P(k_2) + \mathrm{cyc.}$,
where the kernel $F_2\,(k_1, k_2)$ is derived from the equations of motion
for gravitational instabilites (to second order, or tree level - see \citet{Scoccimarro2000} for the full expression).
As such, $Q_{ \mathrm{tree} }(k_1, k_2, k_3)$ is time and (approximately) scale
independent \citep{Fry1984, Scoccimarro2000} for the density field.
We refer the curious reader to \citet{Bernardeau2001} (and references therein)
for more details of perturbation theory and its predictions.

\begin{figure}
\centering
  $\renewcommand{\arraystretch}{-0.75}
  \begin{array}{c}
    \includegraphics[trim=1.5cm 3.1cm 0.0cm 2.3cm, clip=true, scale=0.25]{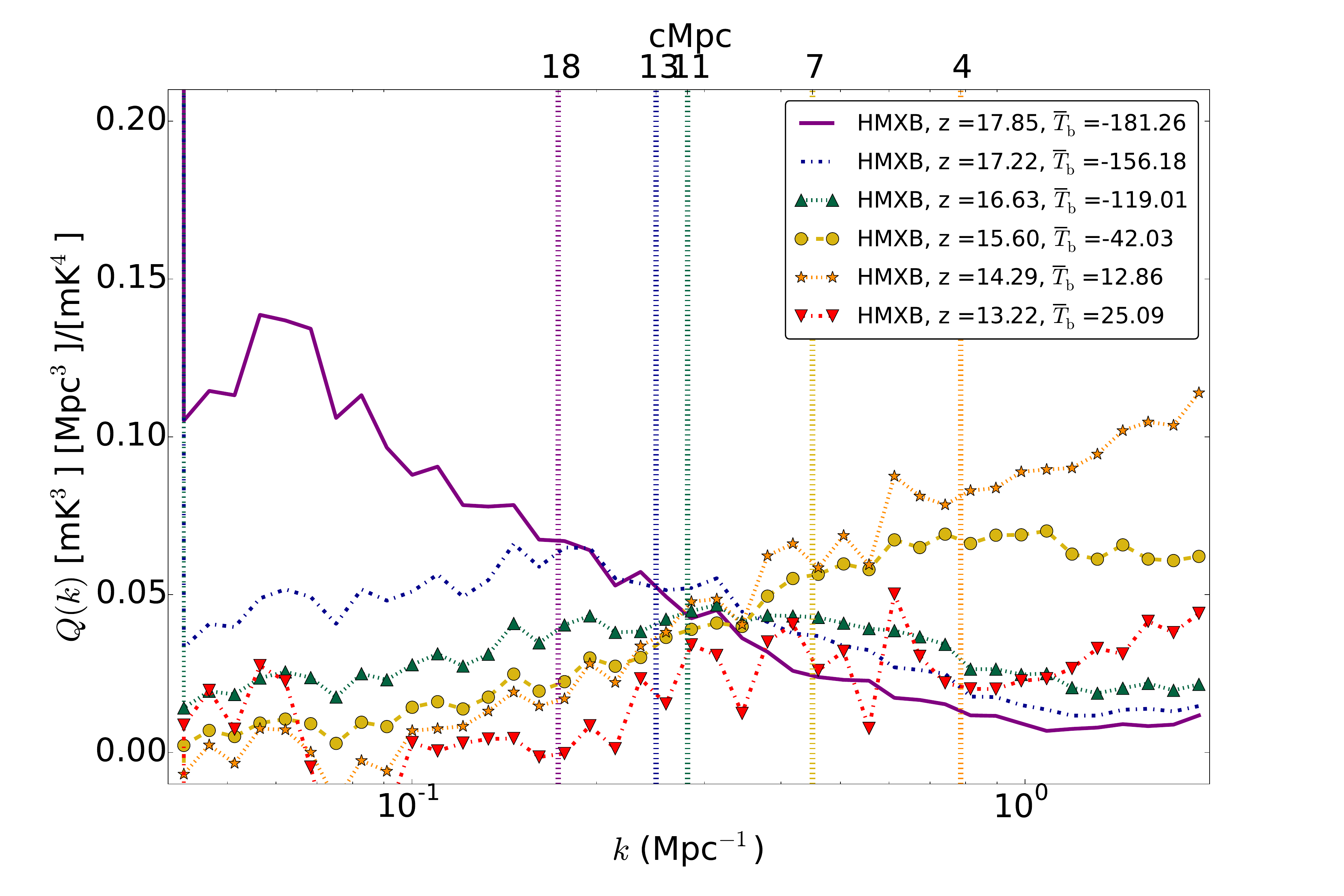}\\
    \includegraphics[trim=1.5cm 0.0cm 0.0cm 2.3cm, clip=true, scale=0.25]{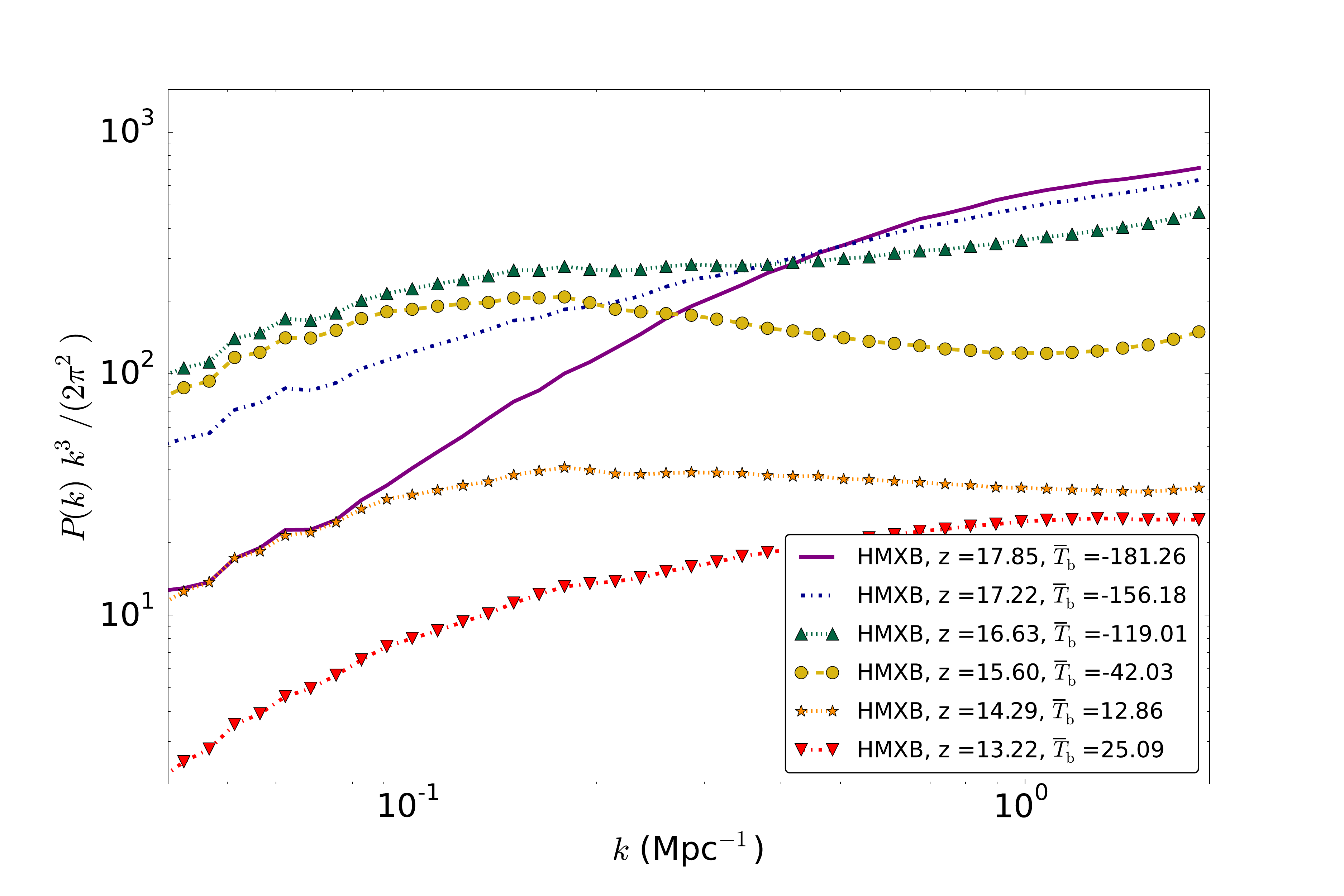}\\
  \end{array}$
  \caption{
  \textbf{Top} - Spherically-averaged reduced bispectrum
  with $k$ for the equilateral configuration of $k$ vectors for the \textit{HMXB} simulation..
  \textbf{Bottom} - power spectrum from the same simulation.
Q(k) retains a scale-dependent contribution from the power spectrum, especially relevant on larger scales.
  }
  \label{fig:Qvsk_contrast}
\end{figure}

The motivation for measuring $Q(k_1, k_2, k_3)$ from the brightness-temperature
field is less clear cut.
If we could measure the dimensionless brightness-temperature, i.e
$\delta_T = (T - \overline{T})/\overline{T}$, then it would obviously be useful
to identify when the bispectrum of the brightness-temperature is being driven
solely by the density field.
However, we measure the dimensional brightness-temperature, i.e. $(T - \overline{T})$,
and so $Q(k_1, k_2, k_3)$ is no longer dimensionless, it instead has units of
inverse brightness temperature (mk$^{-1}$ for the high-$z$ 21-cm signal).
This temperature dependence is particularly confusing during the epoch of heating
as the temperature will become very small as the field passes into emission,
and therefore $Q(k_1, k_2, k_3)$ can blow up during this phase due to division
by very small numbers.
Also, because of the brightness-temperature dependence of the 21-cm $Q(k)$, a contribution from the power spectrum remains in the statistic, the level of which is scale-dependent.
This is seen by comparing the top plot of Fig. \ref{fig:Qvsk_contrast} in which we plot the equilateral $Q(k)$-vs-$k$ for various redshifts with the bottom plot which shows $P(k)$-vs-$k$ for the same redshifts.
There is a evidence of a turnover that shifts from large to small scales, however, on larger scales it is not possible to concretely connect this with any physical scales in the map.
It is also clear that the drop in large-scale $Q(k)$ is strongly correlated with the increase in the power spectrum with decreasing redshift.

\bsp
\end{document}